\documentclass[article,11pt,nofootinbib,singlecolumn,superscriptaddress,preprintnumbers,floatfix]{revtex4-2} 
\usepackage[nodisplayskipstretch]{setspace}
\usepackage{cancel}
\usepackage{mathtools}
\usepackage{soul}
\usepackage{physics}
\usepackage{amssymb}
\usepackage{braket}
\usepackage{amsmath}
\usepackage{epsfig}
\usepackage{hyperref}
\usepackage{cleveref} 
\usepackage{breakurl}
\usepackage{bm}
\usepackage{color}
\usepackage{slashed}
\usepackage{tikz}
\usepackage{enumerate}
\usepackage[mathcal]{euscript}
\usetikzlibrary{decorations.markings}
\usepackage[english]{babel}
\usepackage{xcolor}
\makeatletter
\newcommand{\cz}{
  \mathord{\mathpalette\vaggelis@z{z}}%
}
\addto\captionsenglish{}

\usepackage[bottom=1in]{geometry}
\newcommand\beq{\begin{equation}}
\newcommand\eeq{\end{equation}}
\def\bea{\begin{eqnarray}}
\def\eea{\end{eqnarray}}

\DeclareSymbolFont{matha}{OML}{txmi}{m}{it}
\DeclareMathSymbol{\varv}{\mathord}{matha}{118}

\DeclareRobustCommand{\SkipTocEntry}[4]{}

\newcommand\beal{\begin{aligned}}
\newcommand\eeal{\end{aligned}}

\makeatletter
\renewcommand{\p@subsection}{}
\makeatother
\hypersetup{
    colorlinks=true,
    linkcolor=black,
    citecolor=black,
    urlcolor=black
}

\newcommand{\bn}{{\boldsymbol n}}

\newcommand{\bR}{{\boldsymbol R}}

\newcommand{\E}{\mathrm{e}}

\newcommand{\altchi}{{\mathsf{x}}}      
\newcommand{\altiota}{{\mathsf{I}}}   
\newcommand{\altupsilon}{{\mathsf{Y}}}


\definecolor{darkorange}{rgb}{1,0.549,0}

\begin{document}

\preprint{DESY-25-189\\\phantom{~}}
\title{Trails of clouds in binary black holes} 
\author{Mateja Bo\v{s}kovi\'c}
\affiliation{Deutsches Elektronen-Synchrotron DESY, Notkestr. 85, 22607 Hamburg, Germany}
\author{Rafael A. Porto}
\affiliation{Deutsches Elektronen-Synchrotron DESY, Notkestr. 85, 22607 Hamburg, Germany}
\author{Matthias Koschnitzke}
\affiliation{Deutsches Elektronen-Synchrotron DESY, Notkestr. 85, 22607 Hamburg, Germany}
\affiliation{II. Institut für Theoretische Physik, Universit\"{a}t Hamburg, Luruper Chaussee 149, 22761 Hamburg, Germany}
\begin{abstract}

Superradiant instabilities of rotating black holes can give rise to long-lived bosonic clouds, offering natural laboratories to probe ultralight particles across a wide range of parameter space. The presence of a companion can dramatically impact both the cloud’s evolution and the binary’s orbital dynamics, generating a trail of feedback effects that require detailed modelling. Using a worldline effective field theory approach, we develop a systematic framework for binaries on generic (eccentric and inclined) orbits, capturing both resonant and non-resonant transitions without relying solely on balance laws. We demonstrate the existence of ``co-rotating'' floating orbits that can deplete the cloud prior to entering the detector’s band, triggering eccentricity growth towards a sequence of fixed points. Likewise, we show that ``counter-rotating'' orbits can also deplete the cloud, driving (unbounded) growth of eccentricity. Furthermore, we uncover novel features tied to orbital inclination. Depending on the mass ratio, equatorial orbits can become unstable, and fixed points may arise not only for aligned or anti-aligned configurations but, strikingly, also at intermediate inclinations. We derive flow equations governing spin-orbit misalignment and eccentricity and identify distinctive signatures that can reveal the presence of boson clouds in the binary's history, as well as key features of possible in-band transitions. These results refine and extend earlier work, yielding a more faithful description of the imprints of ultralight particles in gravitational-wave signals from binary black holes, signatures that are within reach of future detectors such as LISA, Cosmic Explorer, and the Einstein~Telescope.

\end{abstract} 
 
\maketitle
\newpage
\tableofcontents

\newpage
\section{Introduction}
The ultralight and weakly interacting frontier of particle physics is difficult to probe with traditional collider and tabletop experiments. Yet, accessing it is well motivated by many theoretical considerations. The axion is a leading candidate to solve the strong CP problem~\cite{Hook:2018dlk}, while string theory generically predicts a plethora of ultralight particles~\cite{Arvanitaki:2009fg,Demirtas:2018akl,Mehta:2021pwf}, which at the same time also serve as potential sources of  dark matter in the universe~\cite{Marsh:2015xka,Hui:2016ltb,Hui:2021tkt}. 
It is thus not surprising that the direct detection of gravitational waves (GWs) by the LIGO-Virgo collaboration \cite{LIGOScientific:2016aoc} was quickly recognized to also open a window to physics beyond the Standard Model that may only interact gravitationally. For instance, the authors of \cite{Arvanitaki:2010sy} pointed out that measurements of  black hole (BH) spins allow for indirect constraints on axion-like particles. As is well known, for ultralight scalars propagating in the vicinity of spinning BHs superradiant instabilities can extract the BH's rotational energy, leading to the formation of boson clouds \cite{1971JETPL..14..180Z,1972JETP...35.1085Z,Press:1972zz,East:2018glu,Brito:2015oca}.
Thus, by virtue of superradiance, BHs cannot retain a high spin in the presence of ultralight bosons with Compton wavelengths comparable to the Schwarzschild radius. Moreover, the end state of the superradiant evolution is also observationally relevant, as the boson clouds themselves are sources of monochromatic GWs that depend only on the mass of the new particle, e.g., \cite{Arvanitaki:2010sy,Arvanitaki:2014wva,Brito:2014wla}.

\vskip 4pt  Several properties of these so-called {\it gravitational atoms} are controlled by 
\beq \alpha \equiv \frac{GM 
\mu}{\hbar c} = 0.07 \left(\frac{M}{10M_\odot}\right)\left(\frac{\mu}{10^{-12}\rm eV}\right),\eeq
the ``structure constant," with $\mu,M,$ the masses of the ultralight boson and BH, respectively. In the non-relativistic limit, the (quasi-stable) states of the cloud can be described in terms of hydrogenic-type spectrum~\cite{Detweiler:1980uk,Dolan:2007mj,Arvanitaki:2009fg,Arvanitaki:2010sy}. Although strong self-interactions, controlled by the axion decay constant $f_a$, tend to inhibit the growth of the cloud, superradiance can probe up to ten orders of magnitude in $f_a\,(\lesssim m_\mathrm{Pl})$ and $\mu$, which in turn correspond to  stellar-mass to supermassive astrophysical BHs in nature~\cite{Gruzinov:2016hcq, Baryakhtar:2020gao, Witte:2024drg}. Constraining boson clouds through measurements of the BH's spin and/or observations of the GWs emitted by boson clouds with future GW detectors such as LISA~\cite{LISA:2022kgy}, the Cosmic Explorer~\cite{Reitze:2019iox}, the Einstein Telescope \cite{Maggiore:2019uih,Abac:2025saz}, and others~\cite{Kawamura:2020pcg,Berlin:2021txa,Baum:2023rwc}, thus provides a window towards testing the existence of new particles in nature, e.g. \cite{Arvanitaki:2014wva,Brito:2014wla,Arvanitaki:2016qwi,Brito:2017zvb,Palomba:2019vxe,Baryakhtar:2020gao, Zhu:2020tht, Ng:2020ruv,Khalaf:2024nwc,Witte:2024drg,Caputo:2025oap,Gavilan-Martin:2026zzw}.\vskip 4pt 
Exploiting these features, characteristic of isolated gravitational atoms, remained the predominant approach towards probing the ultralight frontier until the work of \cite{Baumann:2018vus} revealed a much richer picture: ultralight particles can produce a non-trivial imprint in the GW signals from binaries where a boson cloud interacts with a companion. Crucially, resonant transitions between different states of the gravitational atom can radically modify the (cloud-free) vacuum evolution. Since then, a large body of work has been devoted to understanding the orbital dynamics of binary systems with BHs carrying a boson cloud, e.g. \cite{Baumann:2019ztm,Baumann:2021fkf,Tong:2022bbl,Takahashi:2023flk,Tomaselli:2023ysb,Brito:2023pyl,Boskovic:2024fga,Tomaselli:2024bdd,Dyson:2025dlj,Kim:2025wwj}. For instance, the work of~\cite{Baumann:2019ztm} demonstrated the existence of ``Landau-Zener" (LZ) transitions, inducing floating and sinking orbits, while ionization effects were discussed in~\cite{Baumann:2021fkf}. More recently, the existence of fixed points in the dynamics of the orbital eccentricity was demonstrated in~\cite{Boskovic:2024fga, Tomaselli:2024bdd}, and a first attempt at describing generic orbits, although limited to binaries with a large mass ratio, appeared also in \cite{Tomaselli:2024bdd,Tomaselli:2024dbw}.\vskip 4pt

The goal of this paper is to provide a more complete analysis of the dynamics of gravitational atoms in binary systems and associated GW signatures.\footnote{We focus on scalar clouds. While vectors exhibit rich phenomenology \cite{Baryakhtar:2017ngi,East:2017ovw,East:2017mrj}, the dynamics rests on the same foundation~\cite{Baumann:2019ztm}; accordingly, our analysis also provides a blueprint for the vector case.} In~particular, we extend (and correct) previous results in the literature regarding their chronological history and orbital backreaction. This is achieved by introducing a systematic treatment of the cloud and orbital sectors, via an Effective Field Theory (EFT) framework \cite{Goldberger:2004jt,Goldberger:2005cd,Porto:2005ac,Porto:2007qi,Porto:2016pyg,Goldberger:2022rqf}. In the EFT approach a gravitational atom interacting with long-wavelength perturbations, generated by a point-particle perturber, is described via a worldline action endowed with a series of (time-dependent) multipole moments~\cite{Baumann:2018vus}. The latter are determined by matching to the microphysics of the boson cloud. The worldline EFT then allows us to study the evolution of generic binary orbits directly at the level of the Hamiltonian equations, without relying entirely on balance laws. Our formalism is broadly applicable across the full range of mass ratios and captures both resonant and non-resonant mixing effects. This~enables the systematic study of possible transitions as well as a thorough assessment of their impact on the cloud's fate and orbital evolution.\vskip 4pt  

As anticipated in~\cite{Boskovic:2024fga, Tomaselli:2024bdd}, we demonstrate the existence of fixed points governing the evolution of the eccentricity for floating orbits. We further reveal analogous fixed points in the spin-orbit misalignment angle (\textit{obliquity}) $\beta \equiv \angle(\bm{{\cal S}},\bm{L})$, where $\bm{L}$ and $\bm{{\cal S}}$ denote the orbital angular momentum and spin of the gravitational atom, respectively. In doing so, we refine and correct several aspects of the eccentricity/obliquity flow diagrams previously depicted~in~\cite{Tomaselli:2024bdd}. Notably, depending on the mass ratio, we find that equatorial orbits may become unstable, and fixed points can emerge not only for co-rotating ($\beta=0$) and counter-rotating $(\beta=\pi)$ configurations, but also at intermediate obliquities. Consistent with previous findings~\cite{Tomaselli:2024bdd}, we confirm that for $\beta \simeq 0$ clouds in binaries that form at low orbital frequencies (corresponding to GW signals below the LISA bandwidth) are typically disrupted in the intermediate-mass to extreme-mass-ratio inspiral (IMRI/EMRI) limits. We extend this conclusion to the comparable-mass regime, and show that disruption persists across a broader region of parameter space, even for $\beta \simeq \pi$.\vskip 4pt 

In addition to the issue of stability, we also uncover a series of striking features that emerge from the presence of boson clouds in binary systems. Most remarkably, we find faster-than-vacuum evolution of the orbital eccentricity and spin-orbit obliquity, driven by resonant (and non-resonant) transitions. The correlated off-band dynamics acts as a forensic marker of a pre-existent cloud, while in-band resonances can significantly reshape the GW phase evolution. While a full assessment requires more detailed studies, our findings provide a more solid foundation for waveform modelling. Beyond the expected finite-size \cite{Baumann:2018vus,Baumann:2019ztm,Chia:2023tle} and ionization \cite{Baumann:2021fkf} effects, the features reported here define concrete targets that motivate dedicated search pipelines for third-generation detectors such as LISA \cite{LISA:2022kgy}, Cosmic Explorer \cite{Reitze:2019iox}, and the Einstein Telescope \cite{Abac:2025saz}.

\vskip 4pt {\bf Outline.} In \S\ref{sec:framewrk} we introduce the worldline effective theory and derive the associated equations of motion. In \S\ref{sec:match} we perform the matching to the microphysics of the gravitational atom and extract the value of the (time-dependent) multipoles. In \S\ref{sec:dynamics} we study the  dynamics of the cloud and binary orbit, including spin effects. We detail the floating conditions and demonstrate the existence of fixed points---for equatorial and misaligned orbits---in both IMRI/EMRI and comparable-mass~regimes. We also present flow diagrams for the eccentricity and obliquity. In \S\ref{sec:pheno} we turn to phenomenology, focusing on possible transitions and highlighting two observational cases: orbital relics and concrete scenarios yielding in-band GW phase evolution. We conclude in \S\ref{sec:concl} with a summary, comparison with previous literature, and future directions. A series of appendices collect technical details. Throughout this paper we use conventions from~\cite{Boskovic:2024fga} (hereafter referred to as the Letter).

\section{Framework} \label{sec:framewrk}

We start by discussing the basic features of the systematic framework we introduce to describe the orbital dynamics of gravitational atoms in binary systems without relying on balance laws. 

\subsection{Worldline theory}
For a gravitational atom composed of a BH surrounded by a cloud, for which the mass-energy density peaks around a few orders of magnitude away from the horizon, at $r_\mathrm{c} \simeq M/\alpha^2$ ( with $\alpha < 1$)~\cite{Arvanitaki:2009fg}, an EFT worldline approach may be implemented  to describe the interaction with a companion during the inspiral regime of a binary system (see also~\cite{Baumann:2018vus}). In the early stages their separation obeys $R \gg r_\mathrm{c}$,\footnote{In order to simplify the notation in this paper we depart from the choice in \cite{Baumann:2019ztm} and use $R$ instead of $R_\star$ for the separation, and similarly later on for the orbital elements.} such that we can treat the gravitational atom as a point-like object endowed with a series of multipole moments \cite{Goldberger:2004jt,Goldberger:2005cd,Goldberger:2022rqf,Porto:2005ac,Porto:2007qi,Porto:2016pyg,Goldberger:2020fot}. At leading (quadrupolar) order, it can be described by the worldline action
\begin{eqnarray}
S_\mathrm{WEFT} = -  \int d \tau \left[{\cal M}(\tau) + \frac{1}{2} \omega_\mu^{ab} {\cal S}_{ab}(\tau) v^\mu(\tau) + {\cal Q}_{L}(\tau) \nabla_{L-2} E^{i_{\ell-1}i_{\ell}} + \cdots \right].\label{weft}
\end{eqnarray}
where $\omega^{ab}_\mu$ is the spin connection, and $\cal M(\tau)$ denotes the total mass, $v^\mu$ the four-velocity, and ${\cal S}_{ab}(\tau)$ the spin tensor of the gravitational atom---including both the BH and the cloud. The gravitational field also enters through $E_{l m}=E_{\mu\nu} e_l^\mu e_m^\nu$, the electric component of the Weyl tensor projected onto a local frame satisfying $g_{\mu\nu} e^l_\mu e^m_\nu =\eta_{lm}$, and $\nabla$ is the covariant derivative. The associated ${\cal Q}_L(\tau)$ coefficients, with $L=\{i_1\cdots i_\ell\}$, describe the (time-dependent) mass-type multipole moment of the object (in the Fermi frame) and the dots include magnetic-type couplings which are not relevant for our purposes. See \cite{Goldberger:2004jt,Goldberger:2005cd,Goldberger:2022rqf,Porto:2005ac,Porto:2007qi,Porto:2016pyg,Goldberger:2020fot} for more details.\footnote{The description via \eqref{weft} is valid in the regime where the companion is {\it outside} of the cloud, which is the main focus of our work here. We will briefly comment on dipolar-type couplings later on.}\vskip 4pt

\begin{figure}
    \centering
    \includegraphics[width=0.6\textwidth]{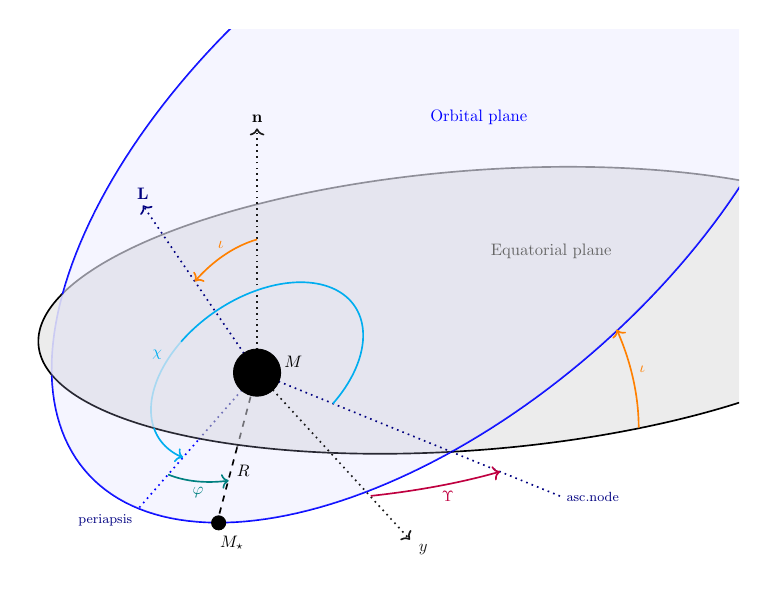}
    \caption{Euler-angle  rotation $\frak{R}(\bm{n},\bm{\hat{L}})$ from the reference frame, defined by the fixed axis $\bm{n}$, to the (non-inertial) \textit{orbital} frame, whose $z$-axis is aligned with the orbital angular momentum $\bm{L}$, while the $y$-axis points in the direction orthogonal to the periapsis, with the orbital elements $\mathbb{E}$ indicated.}
    \label{fig:orb_ref}
\end{figure}

 Our task is then to solve for the orbital dynamics of a point-like object with time-dependent mass, ${\cal M}(t)$, spin vector, $\bm{{\cal S}}(t)$, and  mass multipole moments, ${\cal Q}_L(t)$, interacting with a perturber of mass $M_\star$, which (for simplicity) we assume has negligible spin and internal structure. We will consider the non-relativistic regime, incorporating also Post-Newtonian (PN) corrections such as radiation-reaction and spin effects.\vskip 4pt  We move to the center-of-mass frame of the binary system and choose a unit vector, $\bm{n}$, as a fixed axis. We introduce, as is common in celestial mechanics, e.g.,~\cite{Tremaine_Dynamics}, the semi-major axis, $a$, eccentricity, $e$, inclination, $\iota$, longitude of the ascending node, $\Upsilon$, and argument of the periapsis $\chi$ as the parameters to describe the orbit (see Fig.~\ref{fig:orb_ref}).  We complete the set by adding the {\it mean anomaly}, $\vartheta$, obeying $\dot{\vartheta}=\Omega$ for the Kepler problem, with $\Omega=\sqrt{{\cal M}(1+q)/a^3}$, and $q=M_\star/\cal M$ the mass ratio. We will often find it convenient to use the {\it eccentric} anomaly, $E$, related to the {\it true anomaly}, $\varphi$, related to the relative distance via $R=a(1-e^2)/(1+e\cos{\varphi})$ (see App.~\ref{app:cel_mech}).\vskip 4pt 

We will refer to $\{a,e,\iota\}$ as \textit{principal} and to $\{\vartheta,\chi,\Upsilon\}$ as \textit{positional} elements. For spin degrees of freedom, on the other hand, we  will use the Euler angles associated with the rotation $\frak{R}(\bm{n},\bm{\hat{{\cal S}}})$, that will allow us to express the dynamics of $\bm{{\cal S}}$ in the terms of the parameters in the $\bm{n}$-frame. The flow of orbital elements $\mathbb{E}\equiv\{a,e,\iota,\vartheta,\chi,\Upsilon\}$ can be  written in the form of Hamiltonian equations,
\begin{equation} \label{eq:orb_evo_H}
\frac{d\mathbb{E}}{dt} = \hat{\mathbb{M}} \frac{\partial}{\partial \mathbb{E}} (H_\mathrm{K} + H_\mathrm{I}) \,,
\end{equation}
where $H_\mathrm{K} = - q{\cal M}^2/(2a)$ is Kepler's Hamiltonian, and $ H_\mathrm{I}$ representing the corrections beyond the point-particle limit, due to the extendedness of the gravitational atom, as well as PN effects, that follow from our worldline action \cite{Porto:2016pyg}. The $ \hat{\mathbb{M}}$ matrix takes into account the transformation between $\mathbb{E}$ and Delaunay's canonical basis (see App.~\ref{app:cel_mech}). The resulting orbital equations take the form,
\begin{eqnarray}
\frac{d a}{dt} &=& -\frac{2 \partial_\vartheta H_\mathrm{I}}{a \text{\textmu} \Omega } \,, \label{eq:a_gen} \\
\frac{d e^2}{dt}  &=& 2 \sqrt{1-e^2} \frac{ \left(\partial_{\chi} H_\mathrm{I}- \partial_\vartheta H_\mathrm{I}\right)}{a^2 \text{\textmu} \Omega} \,,  \label{eq:e_gen} \\
\frac{d\iota}{dt}  &=& \frac{\partial_{\Upsilon} H_\mathrm{I}- \partial_{\chi} H_\mathrm{I} \cos \iota}{a^2 \sqrt{1-e^2} \text{\textmu} \Omega  \sin \iota}  \,, \label{eq:i_gen} \\
\frac{d\vartheta}{dt}  &=& \Omega + \frac{2 a e \partial_a H_\mathrm{I}+\left(1-e^2\right) \partial_e H_\mathrm{I}}{a^2 e \text{\textmu} \Omega } \,, \label{eq:tht_gen} \\
\frac{d\chi}{dt}  &=& -\frac{\left(1-e^2\right) \partial_e H_\mathrm{I}-e \partial_{\iota} H_\mathrm{I}  \cot \iota }{a^2 e \sqrt{1-e^2} \text{\textmu} \Omega } \,,  \label{eq:chi_gen} \\
\frac{d\Upsilon}{dt}  &=& -\frac{\partial_{\iota} H_\mathrm{I}}{a^2 \sqrt{1-e^2} \text{\textmu} \Omega  \sin \iota} \label{eq:ups_gen} \,,
\end{eqnarray}
with $\text{\textmu}\equiv q{\cal M}/(1+q)$, the reduced mass.\vskip 4pt

The evolution of the spin also follows from the worldline theory in \eqref{weft}~\cite{Porto:2016pyg}. However, the derivation of the contribution from the multipolar terms is slightly trickier, since it involves varying the action with respect to the internal angular variables~\cite{Porto:2005ac}. For instance, the  equations of motion can be found in \cite{Goldberger:2020fot} for the leading quadrupolar case.  See also the Newtonian example in \cite{Tremaine_Dynamics,poissonwillbook} and App.~\ref{app:spin} for more details. \vskip 4pt

We find three types of perturbations in our problem: conservative, mixing, and dissipative. The conservative type does not change the orbital energy and angular momentum, and therefore they can only influence the positional elements (see, e.g.,~\cite{1976AmJPh..44..944B}). Mixing perturbations are those that exchange energy and angular momentum between the orbit and the gravitational atom, while dissipative processes emit energy-momentum to infinity. Although mixing perturbations are intrinsically dissipative, the rearrangement of the cloud's degrees of freedom during a resonant transition occurs, in general, on different time scales compared to the absorption of angular momentum and energy by the parent BH. We are interested in tracking the dynamics of the system as a result of these perturbations on \textit{secular} timescales, much longer than an orbital period. In principle, this would allow us to work within an {\it adiabatic approximation}, averaging over orbital timescales. For radiation-reaction effects, for instance due to GW emission or BH absorption, we will proceed in this fashion and add orbit-averaged quantities to the right-hand-side of the above equations. For the terms that drive orbital resonances, on the other hand, we will solve the full set of equations across the resonant dynamics without resorting to orbital averages.\vskip 4pt

\subsection{Interacting Hamiltonian} 
After adding a companion at a relative distance $R$, the mass-multipole couplings in the worldline action in \eqref{weft} produce the following Hamiltonian interaction
\beq \label{eq:H_int_multi}
H_\mathrm{I} \supset -M_\star \sum_{l \geq 2} \frac{(-1)^l}{l!} {\cal Q}^L e^A_{\;\!L} \partial_A \left( \frac{1}{R} \right)\,,
\eeq
with $\bm{R}=R\bm{\hat R}$, and the vierbein, $e^A_{\;\!L}$, is responsible for mapping from the orbital to the local frame  in which we will (shortly) perform the matching in~\S\ref{sec:match}. Contracting $e^A_{\;\!L} \partial_A$ and using the standard relations between the symmetric trace-free (STF) tensors and spherical harmonics we find 
\beq
H_\mathrm{I} \supset V_{\cal Q} \equiv - M_\star \sum_{l \geq 2} \sum^{l}_{m=-l} \frac{ 4 \pi }{2l+1} \frac{Y_{lm}(\bm{\hat{R}})}{R^{l+1}} {\cal Q}_{lm} \,, \quad {\cal Q}^{ L } = \frac{4 \pi l!}{(2l+1)!!} \sum^m_{m=-l} \left( \mathcal{Y}^{\langle L \rangle }_{lm} \right)^\ast {\cal Q}_{lm} \,, \label{eq:ir_multi} 
\eeq
where $\mathcal{Y}^{ L}_{lm} $ are STF tensor harmonics, and
\begin{eqnarray} \label{eq:wigner}
Y_{lm} (\hat{\bR}) &=& \sum^{l}_{g=-l} [D^{(l)}_{m g}(\bm{{\cal S}},\bm{L})]^\ast Y_{lg} \left(\frac{\pi}{2},\varphi \right) \,,\\
\left[ D^{(l)}_{m g} \right]^\ast &\equiv& \E^{i g \xi +i m \kappa} [d^{(l)}_{m g}(\beta)]^\ast \nonumber \,,
\end{eqnarray}
with $D \, (d)$ are Wigner (small) matrices~\cite{1987AuJPh..40..465M} depending on the obliquity, associated with the rotation, $\frak{R}(\bm{{\cal S}},\bm{L}) \equiv \{\kappa,\beta,\xi\}$, connecting the local frame of the gravitational atom and the orbital frame.\footnote{Notice that the $g=m$ term has support around $\beta=0$ (co-rotating limit), while $g=-m$ has support around $\beta=\pi$ (counter-rotating).} In general, we describe this rotation through an intermediate (inertial) reference frame $\bn$, such that $\frak{R}(\bm{{\cal S}},\bm{L}) = \frak{R}(\bm{n},\bm{L}) [\frak{R}(\bm{n},\bm{{\cal S}})]^{-1} $, see Fig.~\ref{fig:orb_ref} and App.~\ref{app:spin}.
\vskip 4pt 

From the interacting Hamiltonian we derive the evolution of the orbital parameters. We can also derive the equations of motion for the spin, yielding (to all orders in the multipole expansion)
\begin{eqnarray}
\dot{{{\cal S}}}^j_{\cal Q} 
&=&  \left( \frak{R}(\bm{n},\bm{{\cal S}})^{-1} \right)^j_a \epsilon^{abc} \sum^\infty_{l \geq 2} \frac{M_\star (-1)^l}{R^{l+1}} \sum^l_{m=-l} {\cal Q}_{lm} \left[\int d \Omega_{\bm{\hat{k}}} Y_{lm}(\bm{\hat{k}}) \hat{k}^b \frac{d P_l(x)}{d x} \Big|_{x=\bm{\hat{k}} \cdot \bm{\hat{R}}} \right] \hat{R}^c \,,  \label{eq:spin_dynamics}
\end{eqnarray}
where the $P_l$'s are Legendre polynomials. (See App.~\ref{app:spin} for more details.)\vskip 4pt While the above equation includes the exchanges of energy and angular momentum between the orbit and the cloud, the contribution from the multipolar couplings to the external flux of total angular momentum turns out to be suppressed with respect to the leading (point-like) term \cite{Apostolatos:1994mx}. Hence, at leading order in the radiation-reaction (RR) forces, we have 
\begin{eqnarray} \label{eq:spin_balance}
\frac{d\bm{\mathcal{S}}}{dt}\Big|_\mathcal{Q} +\frac{d\bm{L}}{dt}\Big|_\mathcal{Q} \simeq 0\,.
\end{eqnarray}

More generally, to model RR effects we will introduce to the right-hand-side of \eqref{eq:a_gen}-\eqref{eq:ups_gen} the following terms,
\begin{eqnarray}
\frac{da}{dt} \Big|_\mathrm{RR} &=& -\frac{64  {\cal M}^3 q (q+1)}{5 a^3} f(e)\,, \label{eq:RR_a} \\
\frac{de^2}{dt} \Big|_\mathrm{RR} &=& \frac{64 {\cal M}^{3}q(1+q)}{a^4} \sqrt{1-e^2}\left( g(e)-f(e)  \sqrt{1-e^2}\right) \label{eq:RR_e}\,,
\end{eqnarray}
with $f(e) = \tfrac{1+\frac{73 e^2}{24}+\frac{37 e^4}{96}}{(1 - e^2)^{7/2}}$ and $g(e) = \tfrac{1+\frac{7 e^2}{8}}{(1-e^2)^2}$, upon adiabatic averaging~\cite{Peters:1963ux,Peters:1964zz}.\vskip 4pt

Other than the multipole moments, additional terms present in $H_I$ include mass and spin PN corrections, which we will consider up to ${\cal O}(\Omega a)^5$, or 2.5PN order, when the leading radiation-reaction term kicks in, see, e.g.,~\cite{Porto:2016pyg}. For simplicity, we ignore the spin of the companion, and therefore the possibility of a second cloud in the binary system. This assumption can be easily relaxed in the worldline approach by enlarging the parameter space.

\section{Matching} \label{sec:match}

In what follows we describe the matching procedure that allows us to read off the value of the ${\cal Q}_{lm}(t)$ coefficients from the knowledge of the short-distance physics of the gravitational atom, and subsequently the interacting Hamiltonian governing the binary dynamics. Moreover, as explained later in~\S\ref{sec:pheno}, many of the phenomenologically relevant cases can be studied in terms of an effective two-level model, which we describe below. The discussion generalizes straightforwardly to a multi-level system (see App.~\ref{app:atom} for details).\vskip 4pt

\vskip 4pt For the matching, we choose a frame where the $z$-axis is aligned with $\bm{\mathcal{S}}$, the total spin of the gravitational atom. The latter includes both  the cloud's and BH's spin, i.e. $\bm{\mathcal{S}}=\bm{S}_\mathrm{c}+\bm{S}$, which we will assume remain parallel, $\bm{S}_\mathrm{c} \parallel \bm S$, throughout the evolution of the system\footnote{While this is a valid approximation for the type of transitions we consider here (where coriolis-type effects are suppressed) the misalignment may be relevant near the Bohr regime.}.
As the choice of Boyer-Lindquist coordinates is adapted to the asymptotic observer, unlike the analysis in \cite{Goldberger:2020fot}, we perform the matching in a frame that is not fully co-rotating with the BH. 

\subsection{Clouds in external fields} \label{sec:micro}

Following~\cite{Baumann:2018vus}, we perform a field redefinition, $\Psi = (\psi \E^{-i \mu t} + \mathrm{c.c.})/\sqrt{2\mu}$, average over the (high-frequency) $\mu$ terms, and expand in the regime $\alpha \ll 1$, obtaining,
\begin{eqnarray} \label{eq:schrod}
i \dot{\psi}  + \mathcal{I} &=& \left( - \frac{1}{2\mu}\nabla^2 - \frac{\alpha}{r} + V_\mathrm{R} +  V_\star + V_\mathrm{sg}  \right) \psi \,.
\end{eqnarray}
The potential in this equation includes: $V_\mathrm{R}$, the ($\alpha$-suppressed) relativistic corrections to the bound potential~\cite{Baumann:2018vus,Baumann:2019eav}; $V_\mathrm{sg}$, accounting for the self-gravity of the cloud~\cite{Baryakhtar:2020gao,Kim:2025wwj}; and $V_\star$, induced by the presence of a perturber~\cite{Baumann:2018vus,Baumann:2019ztm}. The remaining term, parameterized on the left-hand-side with $\mathcal{I}$, include all the non-Hermitian contributions due to decaying modes~\cite{Detweiler:1980uk,Dolan:2007mj,Arvanitaki:2010sy,Baumann:2019eav}, GW emission~\cite{Arvanitaki:2010sy,Yoshino:2013ofa,Arvanitaki:2014wva,Brito:2017zvb,Siemonsen:2022yyf} and ionization~\cite{Baumann:2021fkf}.\vskip 4pt 
 
 Ignoring dissipative and self-gravity effects, in isolation the cloud is described by bound states resembling the spectrum of the hydrogen atom, $ \psi^\mathrm{H}_a \equiv \langle r | n_al_am_a \rangle$ ~\cite{Detweiler:1980uk}, with various  (fine and hyperfine) energy splittings~\cite{Baumann:2018vus,Baumann:2019eav}, scaling as 
 \beq \label{eq:spectrum_R}
 \epsilon_{n_a l_a m_a}=\mu\left(1-\tfrac{\alpha^2}{2n_a^2}+f_{n_a l_a} \, \alpha^4 +h_{n_a l_a}\, \tilde a\, m_a \,\alpha^5 + \mathcal{O}(\alpha^6)\right) \,,\eeq 
 where $\tilde a$ is the (dimensionless) spin of the BH. While many of these states are long-lived, they are ultimately unstable, with a decaying width,\footnote{The coefficients in~\eqref{eq:spectrum_R} and \eqref{eq:spectrum_I} can be found in App. A of the Letter.} 
 \begin{eqnarray}  \label{eq:spectrum_I}
\Gamma_{n_a l_a m _a} \simeq 2\tilde{r}_+C_{n_a l_a}\,g_{l_a m_a} \alpha^{4l_a+5}(\epsilon_{n_a l_a m_a} - m_a \Omega_H) \,,
\end{eqnarray}
with $M \Omega_H = \tilde a/[2 (1+\sqrt{1-\tilde a^2})]$. Moreover, the cloud can also deplete through GW emission, where the typical decay width scales as (with $M_\mathrm{c}$ the instantaneous mass of the cloud)~\cite{Siemonsen:2022yyf}
\begin{eqnarray}
\Gamma^\mathrm{GW}_{n_a l_a m_a} \simeq \frac{M_\mathrm{c}}{M^2} G_{n_a l_a m_a} \alpha^{4m_a + 10} \,, \quad G_{211} \simeq 0.025 \,, \quad G_{322} \simeq 2 \cdot 10^{-7} \,. \label{eq:gw_from_cloud}
\end{eqnarray}

For the case of an orbital companion outside of the cloud ($R > r_\mathrm{c}$), the interacting potential is given by~\cite{Baumann:2018vus,Baumann:2019ztm}
\begin{eqnarray} \label{eq:cloud_pot}
V_\star(r) &=& - \mu M_\star \sum_{l  \geq 2 } \sum^{l}_{m=-l} \frac{4\pi}{2 l +1}  \frac{r^{l}}{R^{l+1}} Y_{lm} (\bm{\hat{r}}) Y^\ast_{lm} (\bm{\hat{R}}) \,, 
\end{eqnarray}
whereas for $R < r_c$ the dipole coupling ($l=1$)  also contributes, see~\cite{Detweiler:2003ci,Brito:2023pyl,Duque:2023seg,Tomaselli:2024bdd}. The $V_\star$ term leads to both corrections to the binding energies as well as resonant mixing~\cite{Baumann:2018vus,Baumann:2019ztm}.\vskip 4pt It is instructive to separate the contribution from the $m=0$ and $m \neq 0$ terms. Because of the orthogonality of spherical harmonics, the former is the only part that can contribute to the ``diagonal" terms. In addition, since (upon orbital averaging) these corrections vary slowly with time, we can use an adiabatic approximation to absorb them into the wavefunctions. We can then estimate the corrections to the binding energy and radial function using (time-independent) perturbation theory as follows,
\begin{eqnarray} \label{eq:cons_shift}
\mathcal{E}_a \simeq \epsilon_a +\langle \psi^{\rm H}_a |V^{m=0}_\star|\psi_a^{\rm H}\rangle \,, \quad \mathcal{R}_a \simeq \mathcal{R}_a^{\mathrm{H}} + \sum_{b \neq a} \frac{\bra{\psi^{\rm H}_b}V^{m=0}_\star\ket{\psi^{\rm H}_a}}{\epsilon_a-\epsilon_b} \mathcal{R}^\mathrm{H}_a   \,,
\end{eqnarray}
where $\mathcal{R}^\mathrm{H}_a$ are the Hydrogenic radial functions. Within this extended Hilbert space, the effect due to the remaining mixing (``off-diagonal") terms can be then parameterized through the following ansatz 
\begin{eqnarray} \label{eq:psi_expansion}
|\psi\rangle = \sum_a c_a(t) |a\rangle \,,\quad  \langle r|a\rangle = \psi_a =\mathcal{R}_{a} Y_{a}   e^{-i (\mathcal{E}_a - \mu) t} \,. 
\end{eqnarray}
The task is reduced to {\it non-perturbatively} solving for the time-dependent $c_a(t)$ coefficients, as in \cite{Baumann:2019ztm,Takahashi:2023flk,Boskovic:2024fga,Tomaselli:2024bdd}. We start by using \eqref{eq:wigner} to rewrite the overlap due to the multipole expansion in \eqref{eq:cloud_pot} (see also~\cite{Baumann:2019ztm,Boskovic:2024fga,Tomaselli:2024bdd}), and by performing an expansion into (a finite number of obliquity) $g$-overtones and (an infinite sum of eccentric) $k$-overtones,\footnote{Technically speaking, the sum over overtones converges only for moderate values of the eccentricity. See App.~\ref{app:ecc} for more details.} yielding $(a\neq b)$
\begin{eqnarray}
\bra{a}V_\star\ket{b}_{lm} &=& \sum_{g,k} \eta^{(ab)}_{l,m,g,k} \E^{-i \left( \Sigma^{(ab)}_{g,k}  + \Delta \mathcal{E}_{ab}  t \right)} \,, \, \Sigma^{(ab)}_{g,k}  =   (g-k)\vartheta + g\xi + m \kappa  \label{eq:eta} \\
\eta^{(ab)}_{l,m,g,k} &\approx&  \eta^{(ab)}_{l,m,m,0}  d^{(l)}_{mg}(\beta) H_{l, g,k}
   \left(\frac{\Omega}{\Omega_0} \right)^{\frac{2}{3}(l+1)}   \frac{Y_{lg}\left(\frac{\pi}{2},0\right)}{Y_{lm}\left(\frac{\pi}{2},0\right)}\,, \nonumber \\
H_{l, g,k} &\equiv& \tfrac{(-\mathrm{sgn}[k] g e)^{|k|}}  {|k|!} \left(1  + \tfrac{(l + 1) k }{2 (-g)} + \mathcal{O}(e^{2})\right) , \nonumber \\
\eta^{(ab)}_{l,m,m,0}  & = & - \frac{\mu M_\star}{r_\mathrm{c}} \left( \frac{r_\mathrm{c}}{a_0} \right)^{l+1} \frac{4\pi}{2l + 1} Y_{lm}\left(\frac{\pi}{2},0\right) \left( I_r  I_\Omega \right)^{(ab|lm)} \,,   \nonumber \\
I^{(ab|lm)}_r &\equiv& \int^{\mathsf{R}\to \infty} d\mathsf{r} \mathsf{r}^2 \hat{\mathcal{R}}_b \hat{\mathcal{R}}_a \mathsf{r}^{l}\,,\quad I^{(ab|lm)}_\Omega \equiv \int d\Omega_{\hat{r}} Y^\ast_a(\hat{r})   Y_{lm} (\hat{r}) Y_{b}(\hat{r})\nonumber , 
\end{eqnarray}
where $\mathsf{r} \equiv r/r_\mathrm{c}$, $\mathsf{R} \equiv R/r_\mathrm{c}$,
$\hat{\mathcal{R}}=r_\mathrm{c}^{3/2} \mathcal{R}$, and $a_0 = [{\cal M}(1+q)/\Omega_0^2]^{1/3}$ is a semi-major axis associated with a reference frequency, $\Omega_0$, which we define below, cf.~\eqref{eq:frakf}. 
\vskip 4pt Mixing occurs provided the selection rules are satisfied~\cite{Baumann:2018vus}: $m=-\Delta m_{ab}$, $l+l_a+l_b=2\mathbb{Z}$, $|l_a-l_b| \leq l \leq l_a+l_b$, and it is resonantly enhanced when  (see~\cite{Baumann:2019ztm} and \S\ref{sec:dynamics})
\begin{eqnarray}
\Delta^{(ab)}_{g,k}=0  \,, \quad  \Delta_{m,g,k} \equiv   \Delta \mathcal{E}_{ab} + \dot{\Sigma}^{(ab)}_{g,k} \label{eq:resonance} \,,
\end{eqnarray}
where $\Delta \mathcal{E}_{ab} = \mathcal{E}_b-\mathcal{E}_a$ (and similarly for $n,l,m$ differences).  The resonances can be further classified as being of the Bohr ($\mathcal{B}$; $\Delta n_{ab} \neq 0$), fine ($\mathcal{F}$; $\Delta n_{ab} = 0$, $\Delta l_{ab} \neq 0$) and hyperfine ($\mathcal{H}$; $\Delta n_{ab} = \Delta l_{ab} =0$, $\Delta m_{ab} \neq 0$) type~\cite{Baumann:2018vus,Baumann:2019ztm}.\vskip 4pt 

Using \eqref{eq:cons_shift}, we can estimate the relative error in the binding energy due to the tidal interaction at a given resonant transition, $\Omega \simeq \Delta {\cal E}_{ab}$; upon orbital averaging, this yields
\begin{eqnarray} \label{eq:energy_pert_shift}
\left(\frac{\Delta\mathcal{E}_{ab} - \Delta\epsilon_{ab}}{\Delta\epsilon_{ab}}\right)_{\rm res}  & \simeq & 
\frac{\Delta \epsilon_{ab}/\mu}{ \alpha^2 }  \frac{q}{1+q} \frac{2-3 \sin^2 \beta}{(1-e^2)^{3/2}}  \,, 
%
\end{eqnarray}
which remains small away from the Bohr regime [see \eqref{eq:spectrum_R}], as well as for small eccentricities. The (leading) scaling in $\alpha$ for the allowed transitions is then given by
\begin{eqnarray} \label{eq:res_positions}
\Delta \mathcal{E}_{ab} \sim \frac{\alpha^p}{M}   \left[ 1 + \delta_{p,7} \left( \frac{\tilde{a}}{\alpha} -1 \right) \right] \,, \quad  p=\{3,5,7\} \,\,,\, \mathrm{for} \,\, \{\mathcal{B}\,,\mathcal{F} \,,\mathcal{H}\} \,.
\end{eqnarray}

In order to monitor different overtones, %
and the shift of the energy split $\Delta \mathcal{E}_{ab}$ due to changes in~$\alpha$, it is useful to introduce the following parameters (adapted from the Letter):%
\begin{eqnarray} \label{eq:frakf}
\Omega^{(ab)}_0 &\equiv& \left| \frac{\Delta \mathcal{E}_{ab}}{\Delta m_{ab}} \right| \,, \quad \frak{s} \equiv \mathrm{sgn}\left( \frac{\Delta \mathcal{E}_{ab}}{\Delta m_{ab}} \right) \,,  \quad 
\frak{f}^{(ab)}\equiv \frac{\Omega}{[\Omega^{(ab)}_0]^\mathrm{sat}}\,, \quad \frak{f}^{(ab)}_{g,k} \equiv \frac{-\frak{s}\Delta m_{ab}}{g- k} \nonumber\,,  
\end{eqnarray}
where $\alpha_\mathrm{sat} \equiv \mu M_\mathrm{sat}$ is the initial value at birth, when the cloud saturates the superradiant condition. (In general, we will use `sat' to refers to initial values at saturation of the superradiance growth.) For simplicity, we will ignore the impact of non-zero values for $\{\dot{\vartheta}-\Omega,\dot{\xi},\dot{\kappa}\}$, such that (see \S\ref{sec:dynamics} for more details) 
\begin{equation}
\Delta^{(ab)}_{g,k} \simeq \frak{s}\Delta m_{ab}\, [\Omega^{(ab)}_0]^{\mathrm{sat}} \left[\frac{\Omega_0}{\Omega_0^\mathrm{sat}} - \frac{\frak{f}}{\frak{f}_{g,k}}\right]^{(ab)} \,.  \label{eq:f_k}
\end{equation}
The existence of resonances requires $\frak{f}^{(ab)}_{g,k}>0$. For instance, in the equatorial limit, we have the resonance condition $k/(-\frak{s} \Delta m_{ab}) < \pm \frak{s}$ for $g= \pm m$.\vskip 4pt

To set some of the relevant timescales of the LZ phenomena, it is useful to linearize the frequency evolution around a reference overtone, $\Omega_{g,k}$, such that \cite{Baumann:2019ztm}
\beq \label{eq:Omega_linearized}
\Omega \approx [\Omega^{(ab)}_{g,k}]_\mathrm{sat} + \gamma^{(ab)}_{g,k} t\,, \quad \gamma^{(ab)}_{g,k}\equiv \frac{96}{5} 
\left( \frac{ q {\cal M}^{5/3}}{(1+q)^{1/3}} \right)_\mathrm{sat} \left[\Omega^{11/3}_{g,k}\right]^{(ab)}_\mathrm{sat}\,,
\eeq
and, as we did in the Letter, introduce the following rescaled variables, 
\begin{eqnarray}
\label{eq:rescaling_LZ}
 \quad z^{(ab)}_{l,m,g,k} &\equiv& \frac{\left(\eta^{(ab)}_{l,m,g,k}\right)^2}{\gamma^{(ab)}_{g,k}} \,,  \,\, v_{g,k}^{(ab)\pm} \equiv \frac{|\Gamma^\pm_{ab}|}{\sqrt{\gamma^{(ab)}_{g,k}}} \,,\,\, \Gamma^{\pm}_{ab} \equiv \Gamma_a \pm {\Gamma}_b\,, \nonumber \\ 
 w^{(ab)}_{g,k} &\equiv& \frac{\Omega^{(ab)}_{g,k}}{\sqrt{\gamma^{(ab)}_{g,k}}}\,,\,\, \tau^{(ab)}_{g,k} = \sqrt{\gamma^{(ab)}_{g,k}}t\,. 
\end{eqnarray} 
 Modulo the orbital backreaction we discuss momentarily in \S\ref{sec:dynamics}, the inspiral is mostly driven by radiation-reaction effects. The evolution near a resonance transition then mimics a LZ-type problem~\cite{landau,Zener:1932ws}, and generalizations thereof~\cite{PhysRevA.46.4110, 1996PhRvA..53.4288V,1997PhRvA..55.2982V}. See \cite{Baumann:2019ztm,Takahashi:2023flk,Boskovic:2024fga} for more~details.

 \vskip 4pt  In the remainder of this paper we concentrate on hyperfine, fine, and (early) Bohr transitions that are within the regime of validity of multipole-expanded effective theory. Because of this,  we will not consider Bohr transitions that happen {\it inside} the cloud nor ionization effects~\cite{Baumann:2021fkf}. In addition, for simplicity, we will ignore self-interactions~\cite{Gruzinov:2016hcq, Baryakhtar:2020gao, Witte:2024drg} and self-gravity effects~\cite{Kim:2025wwj}. See \S\ref{sec:concl} for comments on this point.

\subsection{Two-level atom}  \label{sec:1l_atom}

In the non-relativistic limit, the density of the cloud in a $\psi_{a}$ state is given by $\rho_a = \mu N_\mathrm{c} \mathcal{R}^2_a Y_a Y^\ast_a$, where $N_\mathrm{c}$ is the occupancy, which determines the mass, $M_\mathrm{c} \simeq \mu N_\mathrm{c}$ and spin $S_\mathrm{c} \simeq m_a N_\mathrm{c}$, of the cloud. 
The multipolar decomposition is obtained as a projection of the density profile on the spherical harmonics.  In particular, the $Q^c_{lm}$'s of the cloud (in the local frame) vanish for $m \neq 0$, and for the rest we find  
\begin{eqnarray} \label{eq:quad_1lvl}
 Q^c_{lm} &=& \int d\bm{r} r^l \rho(\bm{r}) Y^\ast_{lm} (\hat{r})=\delta_{m0}\, M_c  r^l_c  \left[I_r   I_\Omega \right]^{(aa|l0)} \quad (\rm one-level) \,,
\end{eqnarray}
which dominates over the quadrupole moment of the BH, e.g.,
\begin{eqnarray} \label{eq:Q_c_vs_bh}
\frac{Q_\mathrm{c}}{Q_\mathrm{BH}} \Big|_{20} \sim \frac{M_\mathrm{c}}{M} \frac{ \tilde{a}}{\alpha^6}   \,,
\end{eqnarray}
such that we can approximate ${\cal Q}_{lm} \simeq Q^c_{lm}$.\vskip 4pt

Introducing a second state, $\psi_b$,  and  allowing for level mixing, we find
\begin{eqnarray}
\rho(t)= M_c \left|c_a(t) \mathcal{R}_a Y_a + c_b(t) \mathcal{R}_b Y_b \right|^2 \,, 
\end{eqnarray}
where the $c_i$'s are dimensionless occupancies of the two states $\{\ket{a}, \ket{b}\}$, obeying $|c_a|^2+|c_b|^2=1$. In this scenario, the multipole moments become (with $\bar m = -m$)
\begin{eqnarray}  \label{eq:Q_2_state}
Q^c_{lm} &=& \frac{1}{2} M_c r^{l}_c \times \Big[ \sqrt{1-\sigma^2} I^{(ab|lm)}_r  (-1)^{m} \left( e^{i\delta}    I^{(ab|l\bar{m})}_\Omega +e^{-i\delta}     I^{(ba|l\bar{m})}_\Omega \right) \quad (\rm two-level) \nonumber \\
&& \qquad \quad +  \left([I_r   I_\Omega]^{(aa|l0)}(1+\sigma) +  [I_r   I_\Omega]^{(bb|l0)} (1-\sigma) \right) \delta_{m0} \Big]  \,, 
\end{eqnarray}
where we introduced the canonically-conjugate variables $(\delta,\sigma)$ through the relation~\cite{PhysRevA.59.620}:
\begin{align} \label{eq:sgm_delta}
c_a = \sqrt{\frac{1+\sigma}{2}}  \exp \left[-i  \Delta \mathcal{E}_{ab} \, t/2  \right]   \,, \quad
c_b = \sqrt{\frac{1-\sigma}{2}}  \exp \left[ i (\delta +  \Delta \mathcal{E}_{ab} \, t/2)  \right],  
\end{align}
and, as before, we distinguished the diagonal contribution (due to the $m=0$ part) from the mixing terms. (Note that \eqref{eq:Q_2_state} reduces to \eqref{eq:quad_1lvl} when $\sigma \to 1$.)\vskip 4pt  
 It is straightforward to compute the effects from the diagonal piece. Upon orbital averaging we find, e.g. for the dominant $l=2$ mode, 
\begin{eqnarray} \label{eq:VQm0}
\langle V^{\mathrm{diag}}_{\cal Q}  \rangle_{l=2} =  \frac{1}{4} \frac{N_\mathrm{c} }{\sqrt{1-e^2}^3}  \left[ 2-3\sin^2 \beta \right]  \left( (1+\sigma)  \eta^{(aa)}_{2,0,0,0} +  (1-\sigma)  \eta^{(bb)}_{2,0,0,0} \right) \,.
\end{eqnarray}
In contrast, the contribution from the mixing ($a \leftrightarrow b$) terms in~\eqref{eq:ir_multi}, can be written as\footnote{Notice that, unlike diagonal terms, the presence of the factor of $(\delta - \Sigma^{(ab)}_{g,k} )$ in the argument prevents us from implementing an orbital averaging.} 
\begin{eqnarray} \label{eq:Vq_mixing}
V^{a \leftrightarrow b}_{\cal Q} = N_\mathrm{c} \sqrt{1-\sigma^2} \sum_l \sum^l_{m=0} \sum_{g,k} \eta^{(ab)}_{l,m,g,k} \cos{\left(\delta - \Sigma^{(ab)}_{g,k}  \right)} \,.
\end{eqnarray}

To conclude this section, it is instructive to examine the Hamiltonian governing the cloud’s mixing evolution between the $\ket{a},\ket{b}$ states under the tidal field in terms of the $(\delta,\sigma)$ canonical variables, which takes the form (see App.~\ref{app:atom})
\begin{align} \label{eq:hamiltonian_psi}
     H_\mathrm{c}  &= \frac{N_\mathrm{c}}{2} \left( -   \Delta \mathcal{E}_{ab} \sigma + 2  \sqrt{1-\sigma^2} \sum_l \sum^l_{m=0} \sum_{g,k} \eta^{(ab)}_{l,m,g,k} \cos(\delta - \Sigma^{(ab)}_{g,k} )  \right) \,.
    \end{align}
From here we notice that, although the expression in \eqref{eq:ir_multi} is in principle different than \eqref{eq:cloud_pot}, the evolution of the cloud and  backreaction on the binary system is described by the same interacting terms, cf.~\eqref{eq:Vq_mixing} and \eqref{eq:hamiltonian_psi}.  This is expected, since changes in the cloud are compensated by changes to the orbit. Let us emphasize, nonetheless, the energy and angular-momentum redistribution is now described in terms of instantaneous laws rather than relying on balance equations.

\section{Dynamics}  \label{sec:dynamics}

We are now in a position to describe the most salient features of the gravitational atom and orbital dynamics. For simplicity, throughout this section we consider the basic ingredients to describe two-state level mixing. (See App.~\ref{app:atom} for additional details.) We will use the results from this section later on in \S\ref{sec:pheno} to describe phenomenologically relevant scenarios, including a discussion on multi-state level mixing and other effects. 

\subsection{Gravitational atom} \label{sec:overview_GA}

The phase-space evolution equations follow from the Hamiltonian in \eqref{eq:hamiltonian_psi}. There are, however, several ingredients that need to be incorporated to fully describe the dynamics. As we mentioned earlier, we must include dissipative effects, e.g., cf.~\eqref{eq:spectrum_I}. The latter modify the right-hand-side of the evolution equations, which then take the form,
\begin{align}
\frac{d\sigma}{dt} &=  -2 \sum_{l,m,g,k} \eta^{(ab)}_{l,m,g,k} \nu^{(ab)}_{g,k} - \Bar{\Gamma}^{-}_{ab} (1-\sigma^2) \,, \label{eq:d_sgm_decay}\\
\frac{d\delta}{dt}  &=   - \Delta \mathcal{E}_{ab} - \frac{2\sigma}{1-\sigma^2}  \sum_{l,m,g,k}  \eta^{(ab)}_{l,m,g,k}  u^{(ab)}_{g,k} \,, \label{eq:del_Vq} \\
u^{(ab)}_{g,k} &\equiv  \sqrt{1-\sigma^2} \cos{(\delta - \Sigma^{(ab)}_{g,k} )}   \,,\quad
 \nu^{(ab)}_{g,k} \equiv - \sqrt{1-\sigma^2} \sin{(\delta - \Sigma^{(ab)}_{g,k} )} \label{eq:u_def} \,,
\end{align}
where $\bar \Gamma_{a(b)}$ is associated with both the reabsorption into the BH $(\Gamma_{a(b)})$, c.f. \eqref{eq:spectrum_R}, and the emission of GWs by the cloud itself $(\Gamma^{\rm GW}_{a(b)})$, c.f.~\eqref{eq:gw_from_cloud}. 
\vskip 4pt

To close the system, we also need to track the (instantaneous) occupancy of the cloud 
\begin{equation}
\frac{d N_\mathrm{c}}{dt}  = -  (\Bar{\Gamma}^{+}_{ab} +\Bar{\Gamma}^{-}_{ab} \sigma) N_\mathrm{c} \,, \label{eq:d_occup}
\end{equation}
together with the (adiabatic) co-evolution of the BH mass and spin\footnote{A more accurate description may be achieved by balancing the change in the BH parameters with the flux at the horizon from the near-zone solution of the Klein-Gordon equation, co-evolved with the orbital dynamics.  See also~\cite{Takahashi:2023flk,Hui:2022sri,Kim:2025wwj}.}
\begin{eqnarray}
\frac{d\alpha}{dt}   &\simeq& \mu^2  N_\mathrm{c}  (\Gamma^{+}_{ab} +\Gamma^{-}_{ab} \sigma)   \,, \label{eq:D_alpha} \\
 \frac{d(\tilde{a} \alpha^2)}{dt}  &\simeq&  - \frac{d}{dt} \left[ \frac{\mu^2 N_\mathrm{c}}{2} \left(m_a + m_b + \left(m_a-m_b \right) \sigma \right) \right]_{\bar{\Gamma}_{a(b)}\to \Gamma_{a(b)}}\,, \label{eq:D_a_tilde}
\end{eqnarray}
where the dependence on the decay width in \eqref{eq:D_a_tilde} enters upon taking the time derivative and replacing $\dot N_c$ by \eqref{eq:d_occup} with $\bar{\Gamma}_{a(b)}=\Gamma_{a(b)}$. Notice that this set of equations is coupled, and therefore also $\Gamma_{a(b)}$, $\Delta {\cal E}_{ab}$, etc., become dynamical variables through a varying coupling constant.\vskip 4pt  For some of the (short) narrow  transitions that may occur in the $\mathcal{H}$ to early $\mathcal{B}$  regime, the effects due to GW depletion of the cloud (as well as ionization) are negligible. This allows us to approximate $\bar{\Gamma}_{ab} \simeq \Gamma_{ab}$~in~\eqref{eq:d_occup}, and eliminate~\eqref{eq:D_alpha}, yielding
\begin{eqnarray} \label{eq:alpha_via_n}
    \alpha \simeq \alpha_\mathrm{sat} \left[1+\frac{\alpha_\mathrm{sat}}{M^2_\mathrm{sat}}\left(N^{\mathrm{sat}}_\mathrm{c}-N_\mathrm{c} \right) \right] \,.
\end{eqnarray}
From here, and via their $\alpha$-dependence [see \eqref{eq:spectrum_I} and \eqref{eq:res_positions}], we obtain the following shifted values,
\beq \label{eq:change_alpha_G}
\Delta \Omega^{(ab)}_0 \simeq p [\Omega^{(ab)}_0]^{\rm sat} \, \alpha_\mathrm{sat}  \frac{N_\mathrm{c}}{M^2_\mathrm{sat} } \left(1- \frac{N_\mathrm{c}}{N^{\mathrm{sat}}_\mathrm{c}} \right)\,,
\quad \Delta \Gamma_{a(b)} \simeq (4l+5)\Gamma^{\rm sat}_{a(b)} \alpha_\mathrm{sat}  \frac{N_\mathrm{c}}{M^2_\mathrm{sat} } \left(1- \frac{N_\mathrm{c}}{N^{\mathrm{sat}}_\mathrm{c}} \right)\,,
\eeq 

In what follows, we consider a few useful limits that can help us understand some of the features of the general case. We first look at the equations in~\eqref{eq:d_sgm_decay} and~\eqref{eq:del_Vq} when a single overtone dominates. We then consider perturbative mixing. We study the possibility of overlapping overtones in the low-frequency limit $(\Omega \ll \Omega_{g,k})$, and also examine the decoupling limit which mimics a LZ-type transition~\cite{Baumann:2019ztm}.\vskip 4pt

{\it \underline{Dominant overtone}.} When the dynamics is dominated by a single $(g,k)$ overtone, a convenient description is given by the Feynman-Vernon-Hellwarth representation~\cite{1957JAP....28...49F, 1996PhRvA..53.4288V, 2023PhRvE.107f4211M}. The phase-space dynamics takes place on the unit (``Bloch'') sphere spanned by $\{\sigma, u^{(ab)}_{g,k}, \nu^{(ab)}_{g,k}\}$, where the variables $\{|u^{(ab)}_{g,k}| \leq 1, |\nu^{(ab)}_{g,k}| \leq 1\}$ are defined in~\eqref{eq:u_def}, and their corresponding evolution equations are (see~\cite{1996PhRvA..53.4288V} and App.~\ref{app:atom})
\begin{eqnarray}
\frac{du^{(ab)}_{g,k}}{dt} &=& \bar{\Gamma}^{-}_{ab} u^{(ab)}_{g,k} \sigma - \nu^{(ab)}_{g,k} \Delta^{(ab)}_{g,k} \,, \label{eq:du_decay} \\
\frac{d \nu^{(ab)}_{g,k}}{dt} &=&   \sigma\big(2  \eta^{(ab)}_{l,m,g,k}  + \bar{\Gamma}^{-}_{ab}  \nu^{(ab)}_{g,k}\big) + u^{(ab)}_{g,k} \Delta^{(ab)}_{g,k} \label{eq:dv_decay}\,.
\end{eqnarray}

{\it \underline{Perturbative mixing}.} In the regime where the dynamics is controlled by a single overtone and the population of the $\ket{b}$-state is small (i.e. $|\sigma - 1| \ll 1$), the evolution equations can be solved perturbatively. We start by using~\eqref{eq:d_sgm_decay} and~\eqref{eq:dv_decay} and expressing $\nu^{(ab)}_{g,k}$ and $u^{(ab)}_{g,k}$ in terms of $\sigma$ and its derivatives,
\begin{eqnarray}
\nu^{(ab)}_{g,k} &=& - \frac{\bar{\Gamma}^{-}_{ab} - \bar{\Gamma}^{-}_{ab}  \sigma^2 + \dot{\sigma}}{2  \eta^{(ab)}_{l,m,g,k} } \,, \label{eq:nu_via_sigmaD} \\
u^{(ab)}_{g,k} &=& \frac{-1}{\Delta_{g,k}} \left(2  \eta^{(ab)}_{l,m,g,k}  \sigma + \bar{\Gamma}^{-}_{ab} \nu^{(ab)}_{g,k} - \dot{\nu}_{g,k}  \right) \,.
\end{eqnarray}
Introducing the parametrization 
\begin{equation}\label{eq:eps}
\sigma = 1 - \sum_n \varepsilon^{2n} F^{(2n)}
\end{equation} 
the constraint $[u^{(ab)}_{g,k}]^2+[\nu^{(ab)}_{g,k}]^2 = 1-\sigma^2$ turns into an equation for the $F^{(2n)}$ coefficients and their derivatives, where  we look for perturbative solutions at leading order, i.e., $\eta^{(ab)}_{l,m,g,k} \sim \mathcal{O}(\varepsilon)$. Furthermore we assume the adiabatic condition $\dot{F}^{(2)} \sim \dot{\eta}^{(ab)}_{l,m,g,k} \ll 1$, which applies to a large number of relevant cases (see App.~\ref{app:atom}). Keeping terms up to ${\cal O}(\varepsilon^2)$ we then find
\begin{eqnarray}
F^{(2)} &=& \frac{2 [ \eta^{(ab)}_{l,m,g,k} ]^2}{[\Delta^{(ab)}_{g,k}]^2 + (\bar{\Gamma}^{-}_{ab})^2} \,, \nonumber \\
\quad \nu^{(1)}_{g,k}  &=& \frac{-2 \bar{\Gamma}^{-}_{ab}  \eta^{(ab)}_{l,m,g,k}  }{[\Delta^{(ab)}_{g,k}]^2 + (\bar{\Gamma}^{-}_{ab})^2} \,, \quad u^{(1)}_{g,k}  =  \frac{-2 \Delta^{(ab)}_{g,k} \eta^{(ab)}_{l,m,g,k}  }{[\Delta^{(ab)}_{g,k}]^2 + (\bar{\Gamma}^{-}_{ab})^2}  \,, \label{eq:pert_mix}
\end{eqnarray}
which implies, $\dot{\delta} = \dot{\Sigma}^{(ab)}_{g,k} + \mathcal{O}(\varepsilon)$. These expressions allow us, in the regime of validity of perturbative mixing, to {\it integrate out} part of the cloud sector and to concentrate only on the remaining $\{N_\mathrm{c}, \alpha, \tilde{a}\}$ variables. The results agree with the ``adiabatic following'' approximation in~\cite{1997PhRvA..55.2982V}.\vskip 4pt 

The above approximation is valid provided $F^{(2)}\ll 1$, or equivalently $\sqrt{{[\Delta^{(ab)}_{g,k}]^2 + (\bar{\Gamma}^{-}_{ab})^2}} \gg  \eta^{(ab)}_{l,m,g,k}$, which justifies our power-counting rules. Extending the perturbative solution towards the resonance (when $\Delta^{(ab)}_{g,k} = 0$), we find that the validity of the approximation is linked to the strong decay regime, i.e., $|\bar{\Gamma}^{-}_{ab}|\gg \eta^{(ab)}_{l,m,g,k}$. In contrast, in the weak-decay regime, perturbative results can be applied only in the early stages of the transition, in particular to set up the initial conditions for numerical evolution.\vskip 4pt

{\it \underline{Low-frequency limit}.} For a single overtone, the perturbative mixing approximation relies on the Bloch-sphere formulation, which cannot be extended to multipole overtones. We can, nonetheless, consider instead the low-frequency limit $\frak{f}^{(ab)} \ll 1$, where $\Delta^{(ab)}_{g,k} \simeq \Delta \mathcal{E}_{ab} \gg \dot{\Sigma}^{(ab)}_{g,k}$ [see~\eqref{eq:frakf}] takes on the same value for all overtones. Using~\eqref{eq:d_sgm_decay} and~\eqref{eq:del_Vq}, we start by writing the generalization of the Bloch constraint 
\begin{eqnarray}
\label{eq:sgm_del_contraint}
\left(\dot{\delta} + \Delta \mathcal{E}_{ab}  \right) &=& \left(\frac{2\sigma}{\sqrt{1-\sigma^2}}\right)^2 \left[|W|^2 - \left(\frac{\dot{\sigma}+(\bar{\Gamma}^{-}_{ab})^2 (1-\sigma^2)}{2\sigma} \right) \right]\,, \\ 
 W &\equiv& e^{i\delta} \sum_{g,k}  \eta^{(ab)}_{l,m,g,k} e^{-i \Sigma^{(ab)}_{g,k}} \,, \quad U \equiv \mathrm{Re}[W] \,, \quad V \equiv - \mathrm{Im}[W]  \,,   \nonumber 
\end{eqnarray}
such that,
\begin{eqnarray}
|W|^2 &=& \sum_{d} S_d^2 + 2\sum_{d<d'} S_d S_d' \cos{\left[\Sigma^{(ab)}_{g,k}-\Sigma_{m,g',k'} \right]} \,, \quad S_d \equiv   \sum_{(g,k|d)} \eta^{(ab)}_{l,m,g,k}  \,,  \label{eq:abs_W}
\end{eqnarray}
where $d \equiv g - k=g'-k'= \cdots$, includes in the same class a series of (degenerate) overtones, and $\sum_{(g,k|d)}$ indicates a sum over $(g,k)$ pairs with $g-k=d$. We resort again to a perturbative expansion of the sort, $\sigma = 1 + \sum_n \varepsilon^{2n} F^{(2n)}$, using the same  scaling rules and assuming adiabaticity as before, i.e,  $\eta^{(ab)}_{l,m,g,k} \sim \mathcal{O}(\varepsilon)$, $\dot{F}^{(2)} \sim \dot{\eta}^{(ab)}_{l,m,g,k} \ll 1$. We assume the solution does not depart significantly from the one obtained for a single overtone in~\eqref{eq:pert_mix}, such that $\dot{\delta} \sim \dot{\Sigma}^{(ab)}_{g,k}$. This allows us, in the low-frequency limit 
($\frak{f}^{(ab)} \ll 1$), to ignore the factor of $\dot \delta$ on the left-hand side of~\eqref{eq:sgm_del_contraint}. We can then perform an orbit-average on both sides, yielding
\begin{eqnarray}
F^{(2)} = \frac{2  \langle |W|^2 \rangle }{(\Delta \mathcal{E}_{ab})^2 + (\bar{\Gamma}^{-}_{ab})^2} \,, \,\, V^{(1)} =  \frac{-2 \bar{\Gamma}^{-}_{ab}  \langle |W|^2 \rangle  }{(\Delta \mathcal{E}_{ab})^2 + (\bar{\Gamma}^{-}_{ab})^2}  \,, \,\, U^{(1)} = \frac{-2 (\Delta \mathcal{E}_{ab})  \langle |W|^2 \rangle  }{(\Delta \mathcal{E}_{ab})^2 + (\bar{\Gamma}^{-}_{ab})^2} \,, \label{eq:low_f_limit}
\end{eqnarray}
where, away from the resonance, we have $\langle |W|^2 \rangle  \to \sum_{d} S_d^2 $ (see \S\ref{sec:equatorial}). Notice that, restricting the above result to the single-overtone case, i.e., $U=\eta^{(ab)}_{l,m,g,k} u^{(ab)}_{g,k}$ etc., the result agrees with the low-frequency limit of~\eqref{eq:pert_mix}, as expected.\vskip 4pt

{\it \underline{Decoupling limit}.}  Another important limit is when $N_\mathrm{c} \ll M^2$, which {\it linearizes} the Schr\"odinger problem in ~\eqref{eq:schrod}. Using~\eqref{eq:Omega_linearized}, and returning to the dominant-overtone case, the solution can be written in terms of the parabolic cylinder functions (cf. App. B of the Letter) and the adiabaticity of the resonant transition is then controlled by the typical size of the $z^{(ab)}_{l,m,g,k}$ parameters, as $\sigma(\infty) \sim 2 \exp{\left(-2 \pi z^{(ab)}_{l,m,g,k}/|g+k|\right)} - 1$, while the temporal dynamics of the transition depends on the ratio $v_{g,k}^{(ab)-}/\sqrt{z^{(ab)}_{l,m,g,k}}$~\cite{1997PhRvA..55.2982V, Baumann:2019ztm, Boskovic:2024fga}. For weak decay, $v^{(ab)-}_{g,k} \ll \sqrt{z^{(ab)}_{l,m,g,k}}$, in the adiabatic regime ($z^{(ab)}_{l,m,g,k} \gtrsim 1$) we find \textit{narrow} resonances taking place over a time $\tau_\mathrm{LZ} \simeq 4\sqrt{z^{(ab)}_{l,m,g,k}}$~\cite{Vitanov:1998yn}. A resonance becomes \textit{wide} either in the super-adiabatic $(z^{(ab)}_{l,m,g,k} \gg 1)$ weak-decay regime or for strong decay, $v^{(ab)-}_{g,k} \gg \sqrt{z^{(ab)}_{l,m,g,k}}$. In the latter case one finds [via~\eqref{eq:RR_a}, \eqref{eq:rescaling_LZ}, \eqref{eq:d_occup}, \eqref{eq:pert_mix}]
\begin{eqnarray} \label{eq:nc_pert}
\frac{1}{N_\mathrm{c}}\frac{dN_\mathrm{c}}{d\frak{f}^{(ab)}} \simeq - [\frak{f}^{(ab)}]^{(4l-7)/3} \times  \frac{2  \, [v^{-} \hat{z}_{l,m} w]^{(ab)}_{g,k}}{f(e) \left[(v^{-}_{g,k} )^2+\left(1- \tfrac{\frak{f}}{\frak{f}_{g,k}} \right)^2 \left(\Delta m_{ab}\, w_{g,0}\right)^2 \right]^{(ab)}} \,,
\end{eqnarray}
where $\hat{z}_{l,m, g,k} \equiv  z_{l,m, g,k} / [\frak{f}^{(ab)}]^{(4l+1)/3}$. This illustrates that the cloud can experience significant decay even when $\frak{f}^{(ab)} \ll \frak{f}^{(ab)}_{g,k}$. \vskip 4pt

\subsection{Orbit \& Spin} \label{sec:overview_orbit}

 It is often customary to include the inclination, $\iota$, among the principal elements, together with $\{e,\Omega\}$. However, a self-consistent description of the dynamics can be achieved by tracking the obliquity, $\beta$, instead. We summarize the main equations below and refer the reader to App.~\ref{app:cel_mech} and App.~\ref{app:spin} for further details.\vskip 4pt 

{\underline{\it Principal elements \& Obliquity}.} The evolution of these parameters receives contributions both from $V^{a\leftrightarrow b}_{\cal Q}$ [via~\eqref{eq:a_gen}-\eqref{eq:e_gen} and~\eqref{eq:Vq_mixing}] and from radiation reaction [cf.\ \eqref{eq:RR_a}-\eqref{eq:RR_e}]. Combining these contributions, we find
\begin{eqnarray}
\frac{d \Omega}{dt} &=& f(e)\, \mathcal{G}\, \gamma^{(ab)}_{m,0}\,
[\frak{f}^{(ab)}]^{11/3} 
- \sum_{l,m,g,k} \left(\frac{\frak{f}^{(ab)}}{\frak{f}^{(ab)}_{g,k}} \right)^{4/3}  (g-k)   \left(\sqrt{\gamma} \, b \,  \eta_{l,m}  \, \nu \right)^{(ab)}_{g,k} \,, \label{eq:a_Vq} \\
\frac{de^2}{dt}   &=& \frac{2\sqrt{1-e^2}}{3[\Omega^{(ab)}_0]^\mathrm{sat}} \Bigg[f(e) \, \mathcal{G}\, \gamma_{m,0}\,
\frak{f}^{8/3} \left( \frac{g(e)}{f(e)}-\sqrt{1-e^2} \right) + \sum_{l,m,g,k} \frac{1}{\frak{f}_{g,k}} \left(\frac{\frak{f}}{\frak{f}_{g,k}}\right)^{1/3} \times \nonumber \\
&&  (g-k) \left(  \frac{g}{\frak{s} \Delta m} \frak{f}_{g,k} + \sqrt{1-e^2}  \right) \left(\sqrt{\gamma} \, b \, \eta_{l,m} \, \nu  \right)_{g,k} \Bigg]^{(ab)} \,, \label{eq:e_Vq} \\
b^{(ab)}_{g,k} &\equiv& \frac{3(1+q)^{1/3}  (M \Omega_{g,k}^{(ab)})^{1/3}_\mathrm{sat}}{q(\mathcal{M}/M_\mathrm{sat})^{5/3}} [w^{(ab)}_{g,k}]^\mathrm{sat} \frac{N_\mathrm{c}}{M^2_\mathrm{sat}} \,, \quad \mathcal{G} \equiv \left( \tfrac{\mathcal{M}^5 q^3/(1+q)}{\left[\mathcal{M}^5 q^3/(1+q)\right]_\mathrm{sat}} \right)^{1/3} \label{eq:bk_zk} \,,
\end{eqnarray}
where $b^{(ab)}_{g,k}>0$ parametrises the orbital backreaction.\vskip 4pt The derivation of the obliquity flow involves a few additional steps, most notably establishing the relationship between $\beta$ and the other relevant parameters in a generic $\bm{n}$-frame. After several manipulations, we arrive at (see App.~\ref{app:spin}) 
\begin{equation}
\frac{d\cos \beta }{dt}   = 
\sum_{l,m,g,k} \left( \eta_{l,m} \, \nu \right)^{(ab)}_{g,k}
\left[  \frac{\left(\frak{f}/\frak{f}_{g,k}\right)^{1/3}\sqrt{\gamma_{g,k}}\, b_{g,k}}{3\Omega^{\mathrm{sat}}_{g,k}\sqrt{1-e^2}}  \left( m - g \cos{\beta} \right) -\frac{ (g- m \cos{\beta})}{s_\mathrm{c}} \right]^{(ab)} ,  \label{eq:i_Vq}  
\end{equation}
with
\begin{equation}
s_\mathrm{c} \equiv \frac{\mathcal{S}(t)}{N_\mathrm{c}(t)} \simeq \tilde{a} \left( \frac{\alpha}{\alpha_\mathrm{sat}} \right)^2 \left( \frac{N_\mathrm{sat}}{M^2_\mathrm{sat}} \right)^{-1}  + \left(m_a + m \frac{\sigma-1 }{2} \right)\,. 
\end{equation}

It is instructive to note that~\eqref{eq:i_Vq} simplifies in two relevant limits,
\begin{equation}
\frac{d\cos \beta }{dt}   \simeq  
\sum_{l,m,g,k}\left(\sqrt{\gamma} \, b \, \eta_{l,m} \, \nu  \right)^{(ab)}_{g,k} \left[ \frac{\left(\frak{f}/\frak{f}_{g,k}\right)^{1/3}}{3\Omega^\mathrm{sat}_{g,k}\sqrt{1-e^2}}  \right]^{(ab)}\left( m - g \cos{\beta} \right) 
 \,, \quad  \frac{b^{(ab)}_{g,k}}{[w^{(ab)}_{g,k}]_\mathrm{sat}} \gg \left(\frac{\frak{f}^{(ab)}_{g,k}}{\frak{f}^{(ab)}}\right)^{1/3}  \label{eq:i_Vq_emri} \end{equation}
 and
\begin{equation}
\frac{d\cos \beta }{dt}  \simeq  -
\sum_{l,m,g,k}\left( \eta_{l,m} \, \nu  \right)^{(ab)}_{g,k}  \frac{(g- m \cos{\beta})}{s_\mathrm{c}}  \,, \quad  \frac{b^{(ab)}_{g,k}}{[w^{(ab)}_{g,k}]_\mathrm{sat}} \ll \left(\frac{\frak{f}^{(ab)}_{g,k}}{\frak{f}^{(ab)}}\right)^{1/3}  \,,  \label{eq:i_Vq_q1} 
\end{equation}
which we may loosely refer to as the IMRI/EMRI ($S_\mathrm{c} \gg L$) and the comparable-mass ($S_\mathrm{c} \ll L$) limits, respectively, after noticing
\begin{eqnarray} \label{eq:spin_est}
\frac{S_\mathrm{c}}{L}  \simeq  \frac{N_\mathrm{c}}{M^2} \alpha^{p/3} \frac{(1+q)^{1/3}}{q} \simeq \frac{b^{(ab)}_{g,k}}{[w^{(ab)}_{g,k}]_\mathrm{sat}} \,,
\end{eqnarray}
near a resonant transition ($M\Omega \simeq \alpha^p$).\vskip 4pt

The alert reader will recognize that~\eqref{eq:a_Vq},~\eqref{eq:e_Vq} and~\eqref{eq:i_Vq_emri} differ non-trivially from their counterparts in~\cite{Tomaselli:2024bdd} (see App.~\ref{app:flux_balance} for more details). As we shall see, this has important phenomenological consequences.\vskip 4pt  

{\it \underline{Positional elements}.}  For the region of parameter space on which we focus here, the positional angles do not play a leading role. However, they do affect various details of the resulting dynamics (see App.~\ref{app:spin}). The positional elements evolve not only due to $V^{a\leftrightarrow b}_{\cal Q}$ in \eqref{eq:Vq_mixing}, but also due to $V^{\rm diag}_{\cal Q}$ in \eqref{eq:VQm0} and other conservative PN terms. Below we focus on the periapsis precession, $\chi$, which is analogous to the evolution of $\vartheta$ and $\Upsilon$. We find, [via~\eqref{eq:a_gen}–\eqref{eq:ups_gen},\eqref{eq:Vq_mixing},\eqref{eq:VQm0}]
\begin{eqnarray}
\label{eq:chi_Vq}
\frac{d\chi }{dt} \Big|_{\cal Q}  &= &  - \frac{[\frak{f}^{(ab)}]^{1/3} }{[\Omega^{(ab)}]^{\rm sat}_0} \Biggl\{\frac{\left(\sqrt{\gamma} b \right)^{(ab)}_0}{8 (1-e^2)^2}  \big( 1 + 3 \cos (2 \beta) +  2 \sin (2 \beta) \partial_\iota \beta \cot \iota \,   \big) \times \\ && \quad \times\left[ (1+\sigma)  \eta^{(aa)}_{2,0,0,0} +  (1-\sigma)  \eta^{(bb)}_{2,0,0,0} \right]  
+ \mathcal{O}\left( \eta^{(aa)}_{l >2 ,0,0,0}, \eta^{(bb)}_{l >2 ,0,0,0} \right)\nonumber  \\  && +\sum_{l,m,g,k} 
  \frac{\left(\sqrt{\gamma}\, b \right)^{(ab)}_{g,k}}{3 \,[\mathfrak{f}^{(ab)}_{g,k}]^{4/3} \sqrt{1-e^2}}  
  \Biggr[ 
    u \left(  \cot{\iota} \,  \partial_\iota \beta 
      \partial_\beta \eta_{l,m}
      - \tfrac{(1-e^2)}{e} \partial_e \eta_{l,m}
    \right)  \nonumber \\
&&   + \nu \cot{\iota} \partial_\iota \left(g  \kappa + m  \xi \right)  \eta_{l,m}
  \Biggr]^{(ab)}_{g,k}
 \Biggr\} \,, 
\nonumber
\end{eqnarray}
where in the first line we concentrated on the dominant $l=2$ mode in $V^{\rm diag}_{\cal Q}$.\vskip 4pt

Let us add a few observations. First, in addition to inducing shifts in the energy and wavefunction corrections [cf.~\eqref{eq:energy_pert_shift}], tidal deformations also act as a conservative effect that manifests as the precession of positional elements in the orbital sector. Second, in the equatorial limit, $\chi$ and $\Upsilon$ become degenerate, and the divergent term proportional to $\cot{\iota}$ (in the third row) is cancelled by a corresponding term from $\dot{\Upsilon}$. A similar cancellation occurs between $\dot{\chi}+\dot{\Upsilon}$ and $\dot{\vartheta}-\Omega$ in the $e \to 0$ limit (see~\eqref{eq:tht_gen}-\eqref{eq:ups_gen} and~\cite{Tremaine_Dynamics} for further discussion).

 \vskip 4pt \subsection{Equatorial limit} \label{sec:equatorial}

A phenomenologically relevant situation is that of equatorial orbits, with the spin aligned (or anti-aligned) with the angular momentum. For co-rotating ($\beta = 0)$ or counter-rotating ($\beta=\pi$) orbits the tidal overlap $\eta^{(ab)}_{l,m,g,k} \propto d^{(l)}_{mg}(\beta)$ is supported only for $g = + m$ and $g=-m$, respectively. In what follows we describe the main features of the dynamics in the vicinity of a particular $(g,k)$ overtone. For concreteness, let us take $\bm{n} = \bm{\hat{L}}$ as the reference $z$-axis. The residual angular dynamics is then restricted to  $SO(2) \cross SO(2)$ angular variables. We will describe the evolution of the system ignoring at first the effects due to precession, which we will return to at the end of this subsection. \vskip 4pt

{\it \underline{Floating fixed points}.} %
Backreaction effects are most pronounced in the weak-decay regime, and become increasingly prominent the narrower the resonance. In this situation, when the transfer of energy-momentum between the cloud and the orbit happens on short time scales, the orbital dynamics encounters a number of fixed points in the frequency, $\Omega \simeq \Omega_{g,k}$, yielding \textit{floating} orbits~\cite{Baumann:2019ztm, Boskovic:2024fga, Tomaselli:2024bdd}. Remarkably, during floating, the evolution equations in~\eqref{eq:a_Vq} and~\eqref{eq:e_Vq} also feature a series of fixed points in the  eccentricity~\cite{Boskovic:2024fga, Tomaselli:2024bdd}, for which $d\Omega/dt=de^2/dt=0$, when the following condition is met
\begin{eqnarray} \label{eq:flt_ecc}
 \frac{g(e)}{f(e)} = \frac{g \frak{s}}{-\Delta m_{ab}}  \frak{f}^{(ab)}_{g,k} \,.
\end{eqnarray}

Let us focus first on transitions with $\Delta m_{ab}<0$, i.e. $\frak{s}=1$. On the one hand, for co-rotating orbits we have $g=m=-\Delta m_{ab}>0$ and early overtones, for which $0<\frak{f}^{(ab)}_{g,k} <1$, fixed points occur at $e_{\rm cr}$ between $[0.3,0.6]$, whereas for $\frak{f}^{(ab)}_{g,k} \geq 1$ and $k < -\Delta m_{ab}$ we have $e_{\rm cr}=0$, and the eccentricity decays faster than in vacuum~\cite{Boskovic:2024fga}. 
On the other hand, for counter-rotating orbits, with $g=-m<0$, all resonant overtones, which are only present for $k < \Delta m_{ab}$, lead to growth of eccentricity. Remarkably, since $de^2/dt$ remains positive, there is no fixed point (nor upper bound) in the counter-rotating case, although the conditions that sustain floating $(\dot \Omega=0)$ break down as $e \to 1$ (see below).  As we explained in the Letter, the growth of eccentricity for co-rotating orbits can be understood from the balance laws. While the orbital frequency remains fixed, the loss of orbital angular momentum yields $\tfrac{d}{dt}(L^2) \propto - \frac{d}{dt}(e^2)$. Since $\dot{L} \propto \left(\frak{s}\tfrac{g}{m} - \frak{f}^{(ab)}_{g,k}\right)$, for $g=m$ we find that the eccentricity grows (decays) for early (late) resonances. A similar argument applies also to the counter-rotating case. 
The flux obeys $\dot{L} \simeq - (g/m) \dot{\mathcal{S}} + \mathcal{T}_\mathrm{GW}$, with $\mathcal{T}_\mathrm{GW}<0$ (see the Letter for the explicit expression). As the spin of the gravitational atom decreases ($\Delta m_{ab}<0 \to \dot{\mathcal{S}}<0$), we find the eccentricity grows for $g=-m$, provided the conditions for floating are met.\vskip 4pt 

The situation is reversed for transitions with $\Delta m_{ab}>0$. The main overtone is available only on counter-rotating orbits, with early(late) overtones producing growth(decay) of eccentricity; whereas for the case of co-rotating orbits, all  overtones with $k < - \Delta m_{ab}$ lead to growth of eccentricity.%

The evolution of the eccentricity in (rescaled) time can be obtained by replacing the floating condition, $\dot{\Omega}=0$, in~\eqref{eq:a_Vq}, allowing us to eliminate $\nu^{(ab)}_{g,k}$ in~\eqref{eq:e_Vq}, yielding
\begin{eqnarray}
\label{eq:flt_ecc_growth}
e (\tau^{(ab)}_{g,k}) &\simeq& e_{\mathrm{cr}} \sqrt{1-\left(1- \frac{e^2_{\mathrm{res}^{-}}}{e^2_\mathrm{cr}} \right) \exp\left[-\frac{\tau^{(ab)}_{g,k}}{w^{(ab)}_{g,k}} \Xi'(e_\mathrm{cr})\right]} \,, \\
\Xi &\equiv& - \frac{2\sqrt{1-e^2}f(e) \left(  \frac{g(e)}{f(e)} - 
\frac{g}{g-k}\right)}{3} \nonumber \,,
\end{eqnarray}
where $e_{\mathrm{res}^-}$ is the eccentricity at the onset of the resonance. In order to identify the associated timescale, using~\eqref{eq:d_sgm_decay} and~\eqref{eq:a_Vq} (neglecting the cloud decay and taking $f(e) \simeq 1$) we find 
\begin{eqnarray} \label{eq:flt_time}
\sigma^{\mathrm{float}} \simeq 1 - \frac{2 \tau^{(ab)}_{g,k}}{(g-k)b^{(ab)}_{g,k}} \,, \quad [\tau^{(ab)}]^\mathrm{float}_{g,k} \simeq  (g-k) b^{(ab)}_{g,k} \,,
\end{eqnarray}
such that the eccentricity growth in~\eqref{eq:flt_ecc_growth} is controlled by the ratio $[b/w]^{(ab)}_{g,k} \simeq S_\mathrm{c}/L$. Expectedly, the largest growth of eccentricity occurs when the spin of the cloud dominates over the orbital angular momentum ($S_\mathrm{c} \gg L$). \vskip 4pt

\begin{figure*}[t!]
\centering

\resizebox{\textwidth}{!}{%
  \includegraphics[height=6cm]{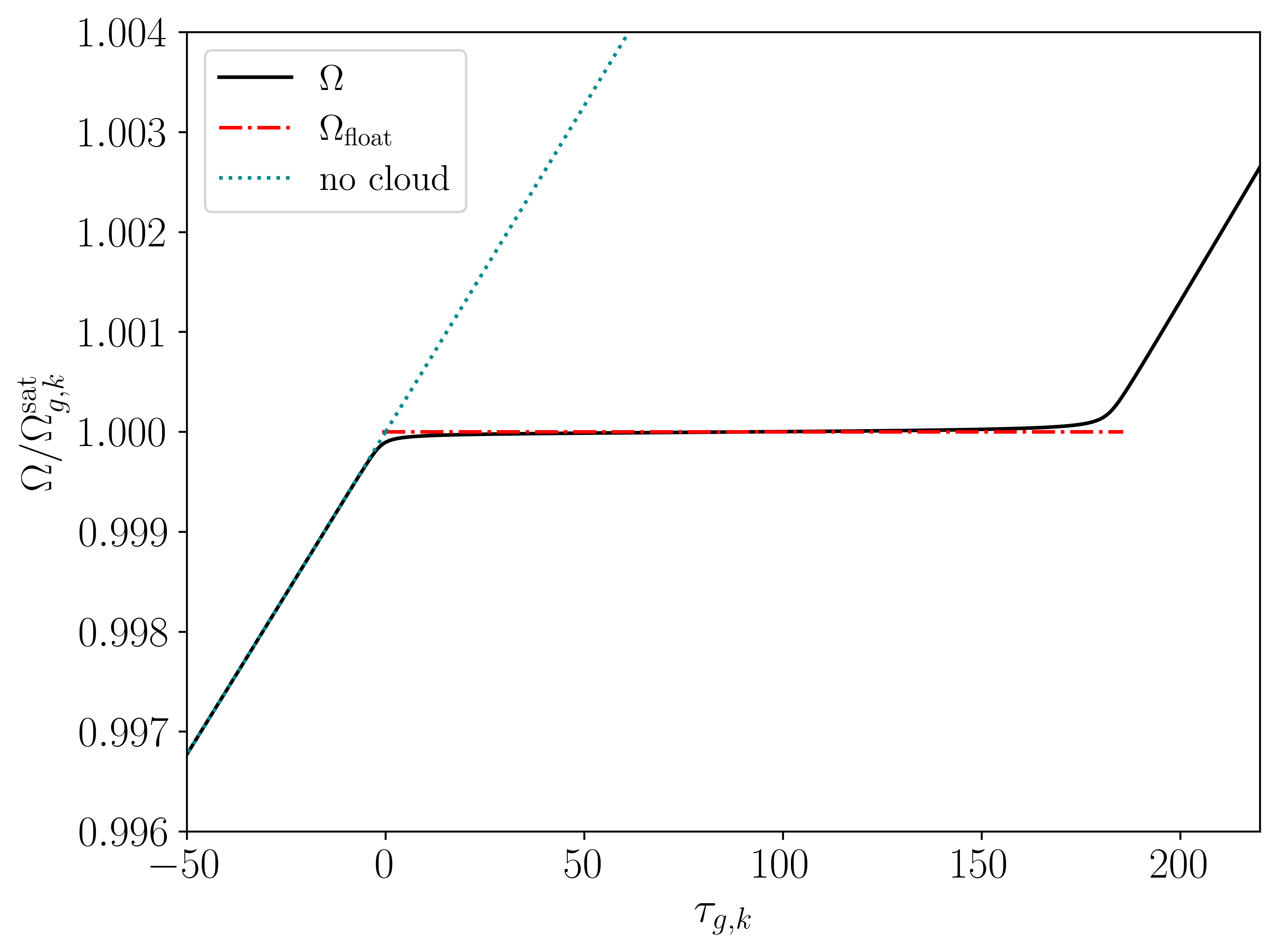}\qquad
  \includegraphics[height=6cm]{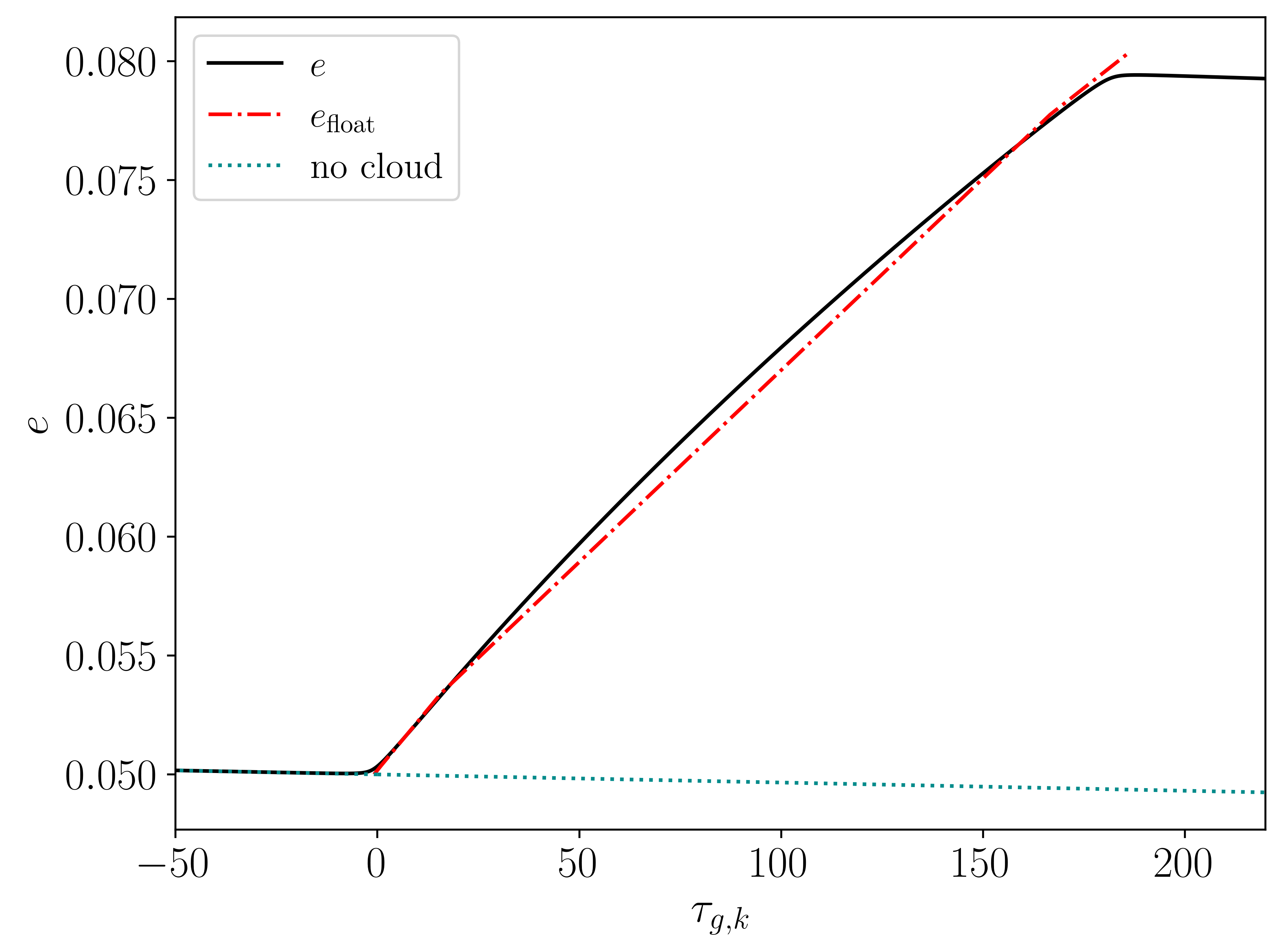}%
}

\caption{Evolution of the orbital frequency [{\it left}] and eccentricity [{\it right}] through the $(k=-1)$ $\ket{544} \to \ket{533}$ transition, with $(q,\alpha)_\mathrm{sat}=(1,0.3)$, $(N_\mathrm{c}/M^2)_\mathrm{sat}=0.13$, $e(\Omega^\mathrm{sat}_{1,-1})=0.05$, and in the weak-decay regime. Numerical solution (solid black), floating period and eccentricity growth [via~\eqref{eq:flt_time},\eqref{eq:v_flt}] (red, dot-dashed), and standard radiation-reaction (RR) vacuum evolution (cyan, dotted).}
\label{fig:ideal_flt}

\end{figure*}

{\it \underline{Floating criteria}.} From the dynamical equations we can readily find out what is necessary to sustain floating orbits. 
Following \eqref{eq:d_sgm_decay},\eqref{eq:du_decay}-\eqref{eq:dv_decay}, the dynamics of $\nu^{(ab)}_{g,k}$ obeys a parabolic-like trajectory from $  \nu^{(ab)}_{g,k}(-\infty)=0$ to $ \nu^{(ab)}_{g,k}(+\infty)=0$. From \eqref{eq:a_Vq}, the floating condition, $\dot \Omega\simeq 0$, implies
\begin{eqnarray} \label{eq:v_flt}
 [\nu^{(ab)}_{g,k}]^{\mathrm{float} } \simeq \frac{f(e) }{ (g-k) [b^{(ab)}]^{\mathrm{sat}}_{g,k} \sqrt{z^{(ab)}_{l,m,g,k}} }   \,,
\end{eqnarray}
constrained by $| [\nu^{(ab)}_{g,k}]^{\mathrm{float} }| \leq 1$ [cf.~\eqref{eq:u_def}]. In addition, since the sign of $\dot{\nu}^{(ab)}_{g,k}$ is determined by the sign of $\sigma$, this implies that the bracket in the first term in  \eqref{eq:dv_decay} must be positive during floating (since the second term becomes subdominant at the resonance). These two constraints yield the following condition for the critical amount of cloud required to maintain a floating orbit, 
\begin{eqnarray} \label{eq:flt_res_break}
N_\mathrm{c} > N^{\mathrm{cr}}_\mathrm{c} \equiv N_\mathrm{c, sat} \frac{f(e)}{ (g-k)  [b^{(ab)}]^{\mathrm{sat}}_{g,k}\sqrt{z^{(ab)}_{l,m,g,k}}} \times \mathrm{max}\left[ 1 \,,\, \frac{v^{(ab)-}_{g,k}}{2\sqrt{z^{(ab)}_{l,m,g,k}}} \right] \,,
\end{eqnarray}
which should be satisfied both at the beginning and at any point during the floating transition. The criteria in~\eqref{eq:flt_res_break} can be satisfied by increasing either $b^{(ab)}_{g,k}$ or $z^{(ab)}_{l,m,g,k}$. However, a large value for $z^{(ab)}_{l,m,g,k}$ would broaden the resonance, thus weakening the counter-balance to radiation reaction. By comparing the floating timescale~\eqref{eq:flt_time} to the typical (unperturbed) duration of a LZ transition, $\tau_\mathrm{LZ} \simeq 4 \sqrt{z^{(ab)}_{l,m,g,k}}$, the condition $\tau_\mathrm{LZ} < \tau_\mathrm{float}$, requires
\begin{eqnarray} \label{eq:flt_over_adia}
b^{\mathrm{sat}}_{g,k}  \gtrsim 4\sqrt{z^{(ab)}_{l,m,g,k}} /(g-k) \,,
\end{eqnarray}
which further restricts the values that support floating. An example highlighting the agreement between numerical and analytic results is depicted in Fig.~\ref{fig:ideal_flt}.\vskip 4pt 

The above requirements delineate the conditions under which floating orbits can exist, and, in turn, determine the circumstances in which floating ceases before the population transfer is complete. Notice that for early overtones, provided $e_{\mathrm{res}^-} < e_\mathrm{cr}$, the fact that $z^{(ab)}_{l,m,g,k} \propto e^{2|k|}$ grows during the transition will tend to decrease the value of $N^{\mathrm{cr}}_\mathrm{c}$, thus allowing for a self-sustained floating condition. In contrast, the decrease of eccentricity that occurs for the late overtones increases the value of $N^{\mathrm{cr}}_\mathrm{c}$, moving in the direction of the {\it resonance breaking}. Furthermore, in the counter-rotating case, as $e \to 1$ we have $f(e) \sim (1-e)^{-7/2}$, which would also break the floating conditions. These findings are in agreement with the limits discussed in~\cite{Boskovic:2024fga, Tomaselli:2024bdd}, within the overlapping realm of validity.\vskip 4pt 

\begin{figure*}[t!]
\centering

\resizebox{\textwidth}{!}{%
  \includegraphics[height=6cm]{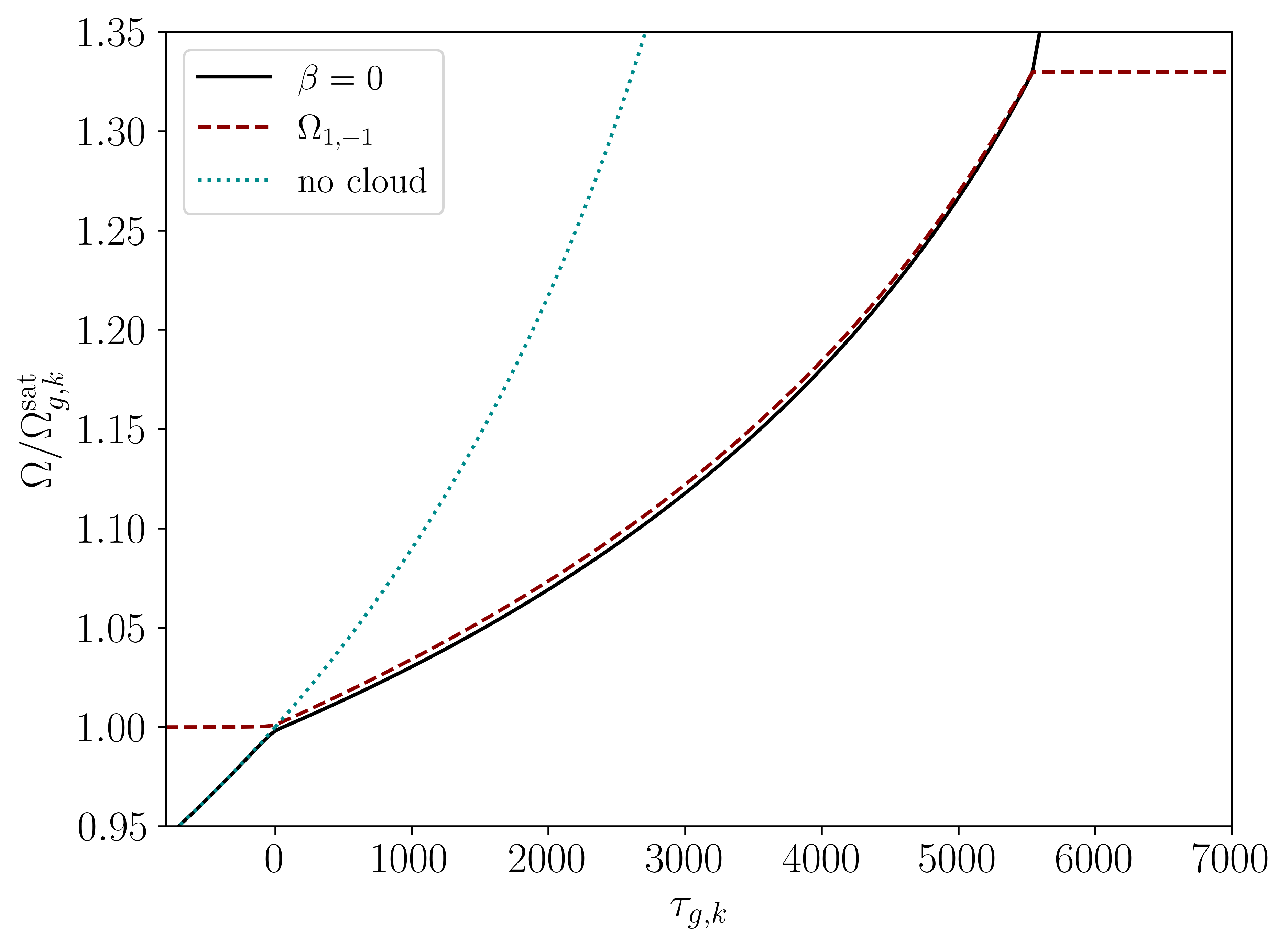}\qquad
  \includegraphics[height=6cm]{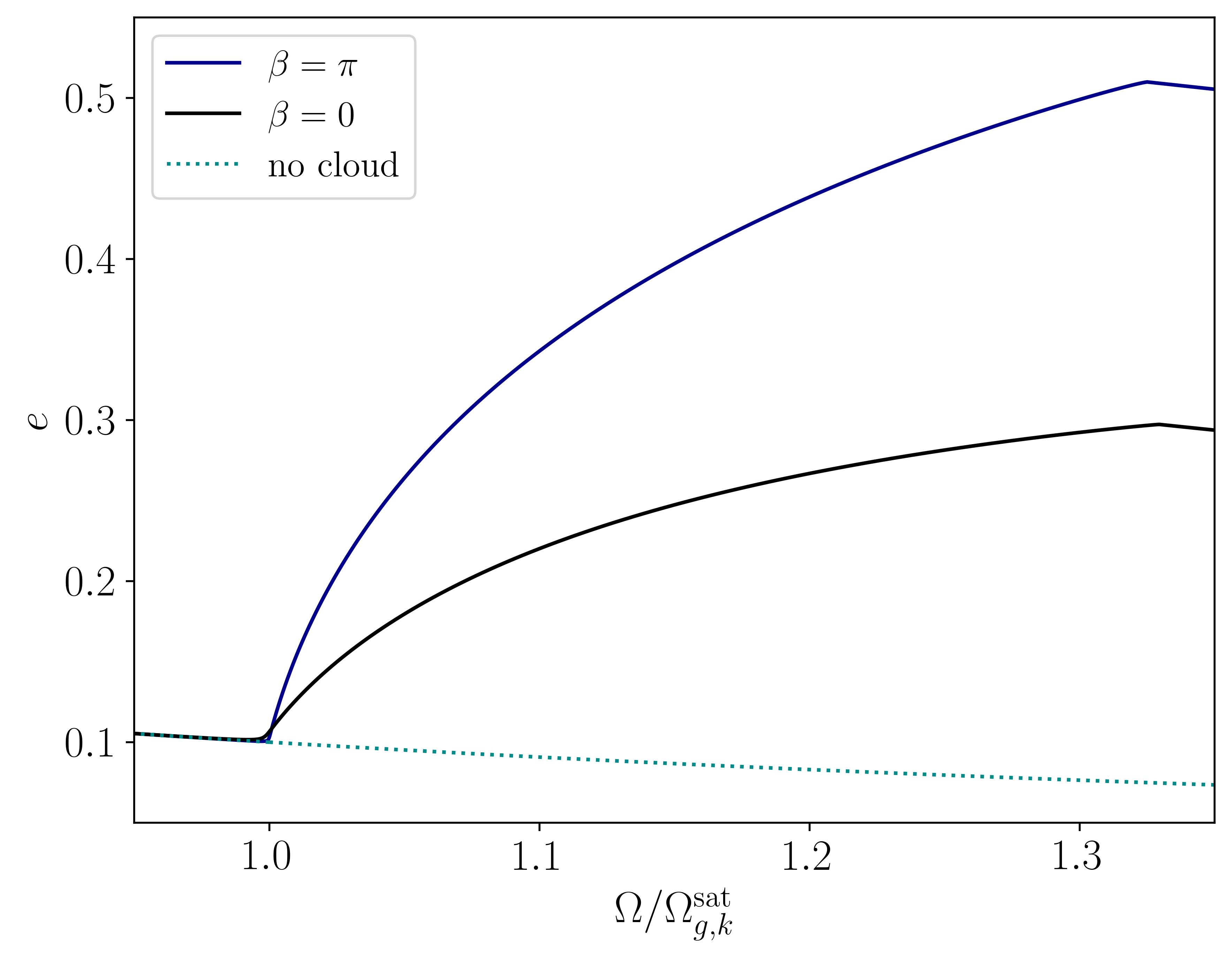}%
}
\caption{Time evolution of the orbital frequency [{\it left}] and frequency evolution of the eccentricity [{\it right}] through the $\ket{322} \to \ket{311}$ transition, with $(q,\alpha)_\mathrm{sat}=(0.1,0.22)$, $(N_\mathrm{c}/M^2)_\mathrm{sat}=0.33$, and $e(\Omega^\mathrm{sat}_{g,k})=0.1$. Numerical solution (solid black) and radiation-reaction vacuum evolution (cyan, dotted) are shown in both cases for the co-rotating case ($g=1,k=-1)$. We also illustrate the growth of eccentricity for the $k=-2$ counter-rotating overtone (blue).}
\label{fig:quasi_flt}
\end{figure*}

{\it \underline{Quasi-Floating}.} The floating conditions, as previously stated, can only apply in the weak-decay regime $v^{(ab)-}_{g,k} \ll \sqrt{z^{(ab)}_{l,m,g,k}}$ and $b_{g,k} \ll 1/v^{(ab)-}_{g,k}$. In the moderate- and strong-decay cases, the change of the BH mass, via absorption, unavoidably leads to a shift in the resonant frequency $\Omega^{(ab)}_{g,k} \sim \alpha^p$ towards higher values. Provided that the resonance is not too wide, the orbital frequency will follow the shift in the condition $\Omega \simeq \Omega^{(ab)}_{g,k}(\alpha)$ as $\alpha$ evolves [cf.~\eqref{eq:change_alpha_G}]. Following~\eqref{eq:d_occup} and \eqref{eq:pert_mix}, we can show that this {\it quasi-floating} scenario is characterized by an evolution of the orbital frequency which is slower than in vacuum, $\dot{\Omega}^{(ab)}_{g,k}/\dot{\Omega}|_\mathrm{RR} \ll 1$, provided 
\begin{eqnarray}
\alpha \frac{N_\mathrm{c}}{M^2}  \left( \frac{w_{g,k} z_{l,m,g,k}}{v^{-}_{g,k}} \right)^{(ab)} \ll 1 \,.
\end{eqnarray}
From~\eqref{eq:a_Vq} we can also solve for $\nu^{(ab)}_{g,k}$ in this regime, yielding
\begin{eqnarray}
 \nu^\mathrm{quasi-float}_{m,g,k} = \frac{f(e) \left(\frak{f}^{(ab)}/\frak{f}^{(ab)}_{g,k} \right)^{7/3}-p [w^{(ab)}_{g,k}]^{\mathrm{sat}} \left(\dot{\alpha}/\alpha\right) 
 \left(\frak{f}^{(ab)}/\frak{f}^{(ab)}_{g,k} \right)^{-1/3}}{(g-k) b^{(ab)}_{g,k} \left(N_\mathrm{c}/N^{\mathrm{sat}}_\mathrm{c} \right) \sqrt{z^{(ab)}_{l,m,g,k}}}  \,.
\end{eqnarray}
Since $[\nu^{(ab)}_{g,k}]^\mathrm{quasi-float} \lesssim [\nu^{(ab)}_{g,k}]^\mathrm{float}$ [c.f.~\eqref{eq:v_flt}], the impact on the eccentricity evolution will be somewhat weaker compared to the floating scenario, while the quasi-floating lasts longer than~\eqref{eq:flt_time}. However, quasi-floating alleviates at the same time some of the resonance breaking conditions. We illustrate the previous discussion with a specific example in Fig.~\ref{fig:quasi_flt}. 
\vskip 4pt

\begin{figure*}[t!]
\begin{tabular}{cc}
\includegraphics[width=.5\textwidth]{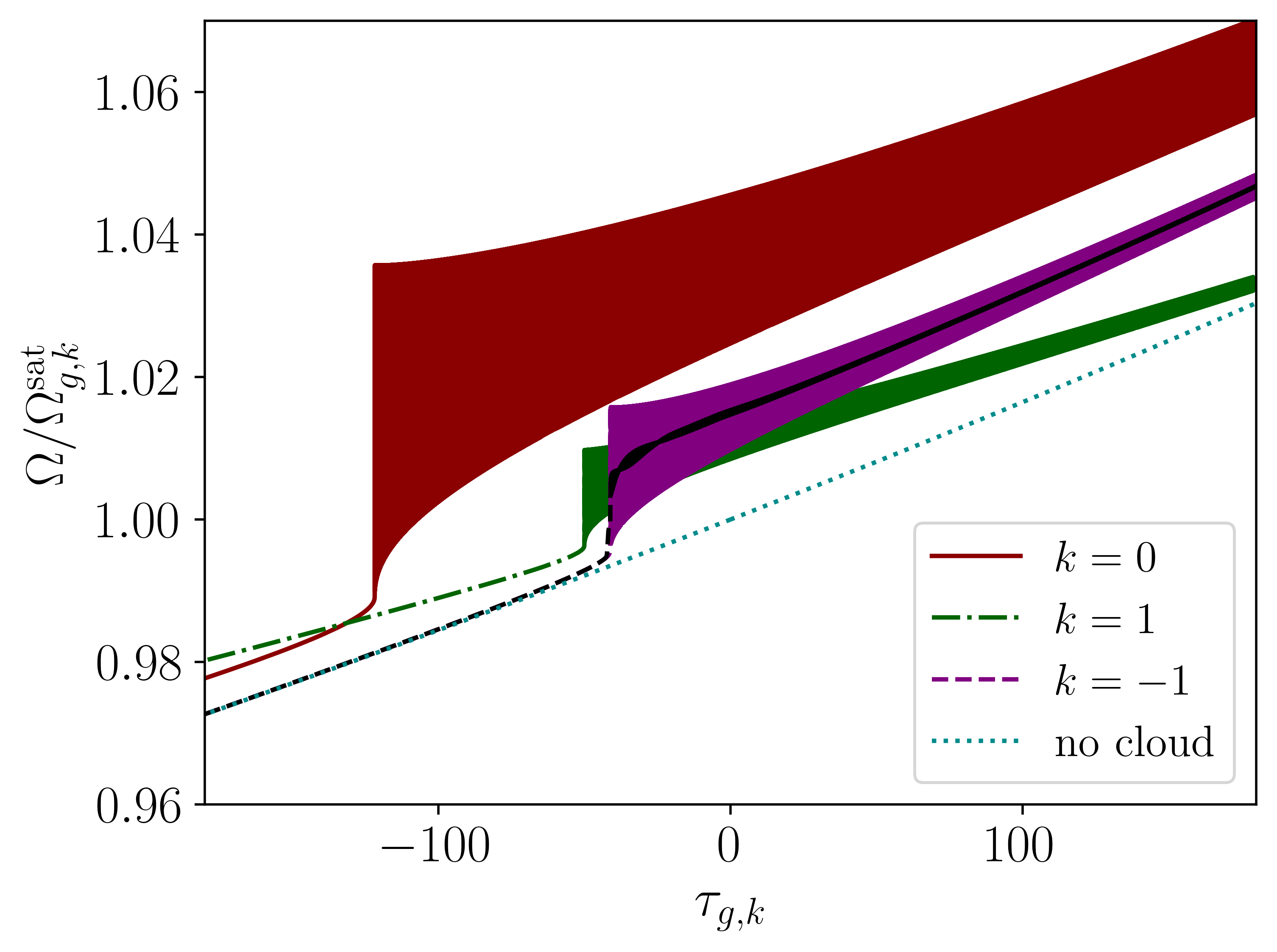}
\qquad
\includegraphics[width=.5\textwidth]{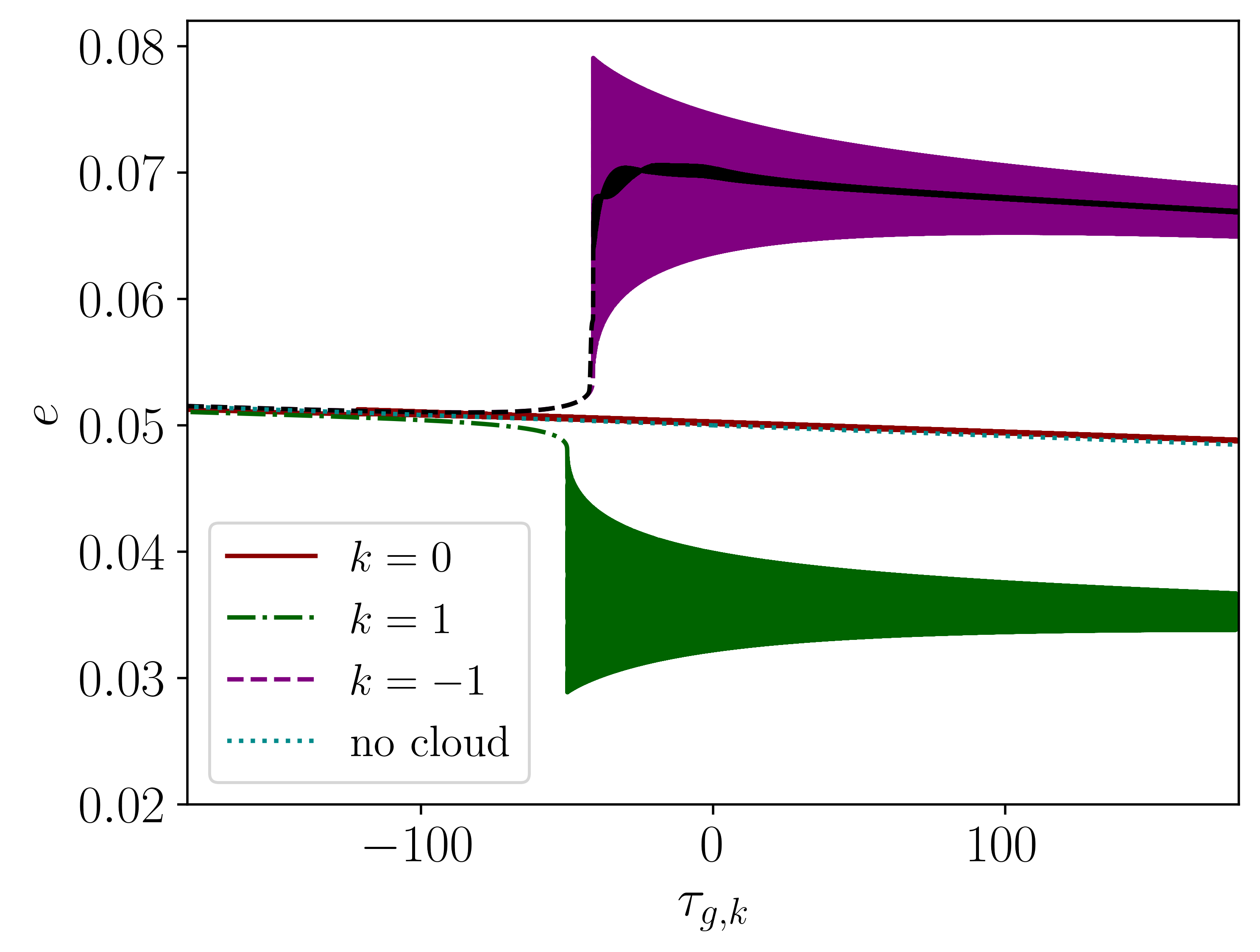}
\end{tabular}
    \caption{Time evolution of the orbital frequency [{\it left}] and eccentricity [{\it right}] during a sinking (early Bohr) transition $\ket{211} \to \ket{54-4}$ on the counter-rotating orbit for $(g=-5,k=-1,0,1)$ overtones (dashed purple, solid red, and dot-dashed green), with $(q,\alpha)_\mathrm{sat}=(0.05,0.1)$, $(N_\mathrm{c}/M^2)_\mathrm{sat}=0.33$, $e(\Omega^\mathrm{sat}_{g,k})=0.05$,  respectively. The appearance of the shaded band is due to the rapidly oscillating nature of the solution. Therefore, we also plot the averaged solution in the case of $k=-1$ resonance (black). The corresponding vacuum evolution is indicated in dotted cyan.} 
    \label{fig:sinking}
\end{figure*}

{\it \underline{Sinking}.} In contrast to floating, transitions with $\Delta {\cal E}_{ab} > 0$ can drain energy from the orbit, thus accelerating the inspiral~\cite{Baumann:2019ztm}. This \textit{sinking} of the orbit leads to transitions which reduce adiabaticity, limiting the backreaction effects. This can be seen, in the weak-decay regime, by introducing a renormalized LZ parameter, i.e., $z^{(ab)}_{l,m,g,k} \to \zeta^{(ab)}_{l,m,g,k}\equiv (\eta^{(ab)}_{l,m,g,k})^2/\dot{\Omega}$, which can be shown to scale as \cite{Boskovic:2024fga} 
\begin{eqnarray}
\zeta^{(ab)}_{l,m,g,k} \simeq \left(\frac{z^{(ab)}_{l,m,g,k}}{\pi b^2_{g,k}} \right)^{1/3} \,.
\end{eqnarray}
Hence, as the backreaction parameter becomes large, $b^{(ab)}_{g,k} \gg 1$, the duration of the transition also shrinks, thus reducing the impact on the orbit. Sinking resonances have so far been studied only in the weak-decay regime on the main $(g,k)=(m,0)$ overtones, and shown to produce, when $b^{(ab)}_{g,k} \gg 1$, a kick in the orbital frequency and eccentricity followed by transient oscillations~\cite{Baumann:2019ztm, Tomaselli:2024bdd}. However, these types of resonances are a more general phenomena, for instance when $\Delta m_{ab}>0$, sinking occurs for co(counter)-rotating orbits whenever $k>-|\Delta m_{ab}|$ ($k>|\Delta m_{ab}|$). Although the evolution of the eccentricity during sinking is less predictable, we can show that in the $b^{(ab)}_{g,k} \gg  1$ limit, the evolution equation in~\eqref{eq:e_Vq} implies that a drop (or jump) in eccentricity is possible for early (or later) overtones on co-rotating orbits, whereas on counter-rotating orbits the eccentricity always decreases. The criteria is reversed,  between co- and counter-rotating orbits, when $\Delta m_{ab} <0$. Furthermore, the stronger the decay, the smoother the transition, shifting the onset of mixing to earlier times. In contrast to floating, the presence of strong decay can therefore have a larger impact on the orbit for sinking transitions. We illustrate a few dynamical features of sinking transitions in Fig.~\ref{fig:sinking}.\vskip 4pt

{\it \underline{Non-resonant mixing}.} So far we have assumed the cloud survives until the resonance. However, for sufficiently wide transitions, the cloud may be depleted long before the resonance frequency is reached. Nonetheless, even in such circumstances, the orbital backreaction can imprint a significant departure from the standard vacuum evolution.
The depletion of the cloud operates irrespectively of the sign of $\Delta {\cal E}_{ab}$, and even if the resonance conditions are not met. For this non-resonant behaviour, the strongest overlap turns out to be with the main overtone, with $k=0$, even for moderate eccentricities. In the case of non-resonant mixing, we solve the evolution of the orbital elements and spin angles in frequency via perturbative mixing using~\eqref{eq:pert_mix}. Regardless of the sign of $\Delta {\cal E}_{ab}$, from~\eqref{eq:a_Vq} and~\eqref{eq:e_Vq} we find that for $\Delta m_{ab}<0$ on co-rotating ($g=-\Delta m_{ab}$) orbits, the backreaction to non-resonant mixing tends to make the transition more adiabatic, while the eccentricity depletes faster than in vacuum. These features are reversed for counter-rotating ($g= \Delta m_{ab}$) orbits with $\Delta m_{ab}<0$. The exact opposite occurs for transitions with $\Delta m_{ab}>0$.\vskip 4pt

{\it \underline{Precession effects on resonances.}} The precession of the orbit can influence the position and duration of the resonances. The non-Keplerian flow of $\dot{\vartheta}$, together with $\{\dot{\xi},\dot{\kappa}\}$, compete with the orbital frequency~$\Omega$, affecting the resonant condition [cf.~\eqref{eq:resonance}]. For instance, consider the apsidal precession, which involves an interplay between PN, $V_\mathrm{1PN}$, and the multipolar, $V_\mathcal{Q}$, corrections. We can estimate the rate of precession at the resonance ($M\Omega \sim \alpha^p$) as follows,\footnote{The periapsis precession originating from the 1PN potential obeys $\dot{\chi}_{\mathrm{1PN}} > 0$, while the sign of $\dot{\chi}_Q$ depends on the angular profile of the cloud and the mixing dynamics.}
\begin{eqnarray}
\frac{\dot{\chi}}{\Omega} \Big|_\mathrm{1PN} &\sim & \alpha^{2p/3} \frac{(1+q)^{2/3}}{1-e^2} \,, \label{eq:Dchi_1pn}
\\
\frac{\dot{\chi}}{\Omega}\Big|_{Q_{20}} &\sim&\alpha^{4 p/3-3} \frac{N_\mathrm{c}/M^2}{(1+q)^{2/3}} \frac{1}{(1-e^2)^3} 
\,. \label{eq:Dchi_qm0} 
\end{eqnarray}
It follows that for all $\mathcal{H}/\mathcal{F}$ resonances $(p=5,7)$, the correction due to precession is negligible for $q \lesssim 1$ and $e \lesssim 0.5$. Hence, for our purposes in this paper, we can neglect the impact of precession on the resonance conditions. However, that is no longer the case deep into the Bohr regime, nor for highly eccentric orbits.\vskip 4pt

{\it \underline{Stability.}}  Finally, as a prelude to the general situation, let us analyse the dynamical equations in the vicinity of the equatorial case and for the dominant $g=\pm m$ resonant transitions. Using that $(\eta^{(ab)}_{l,m} \nu)_{g,k}>0$ [c.f. \eqref{eq:nu_via_sigmaD}], from~\eqref{eq:i_Vq_emri} we can read off the following condition 
\begin{eqnarray}
{\rm sgn}\left(\partial_t \cos{\beta} \right) = {\rm sgn}\left( m  - g \cos{\beta}\right) \,, \qquad (S_c \gg L)\,, \label{eq:overt_stab_emri}
\end{eqnarray}
which can be used to study the stability of planar orbits in the IMRI/EMRI limit. In particular, for allowed transitions with $\Delta m_{ab} < 0$ ($m>0$) near co-rotating orbits ($g=m$, $|\beta| \simeq 0$, $\cos\beta \lesssim 1$) $\beta$ decreases, yielding a stable configuration. In contrast, near counter-rotating orbits ($g=-m$, $|\beta| \simeq \pi$, $\cos\beta \gtrsim -1$), an increasing $\beta$  makes the equatorial orbit unstable. The situation is reversed for allowed transitions with $\Delta m_{ab}>0$ $(m<0)$.\vskip 4pt

For comparable masses, from~\eqref{eq:i_Vq_q1} we find
\begin{eqnarray}
{\rm sgn}\left(\partial_t \cos{\beta} \right) = {\rm sgn}\left(m \cos{\beta} - g\right) \,,  \qquad (S_c \ll L)\,.\label{eq:overt_stab_q1}
\end{eqnarray}
The above condition implies that for all $\Delta m_{ab}>0$ allowed transitions ($m<0$), both co-rotating and counter-rotating equatorial orbits are stable. However, transitions with $\Delta m_{ab}<0$ $(m>0)$ along equatorial configurations become unstable to small perturbations.\vskip 4pt

Let us emphasize two important points regarding the above discussion. First of all, it applies exclusively to resonant transitions with $g = \pm m$, which dominate in the vicinity of $\beta = 0$ or $\beta = \pi$. Depending on the initial conditions, subdominant resonant transitions (having a small but nonzero overlap) may still render quasi-planar orbits unstable. This is the case when an attractive fixed point emerges, away from the equatorial plane. As we shall see shortly, such scenarios arise for transitions with $g>m$, and more prominently in the $q \ll 1$. Secondly, the rate of growth (decrease) of $\beta$ towards the fixed points depends on various parameters, including the type of transition,  mass ratio, and the cloud's density. Therefore, although in theory some equatorial orbits are unstable, in practice they may not all experience large perturbations. Both these features have relevant phenomenological implications to which we return in the following section.

\subsection{Misaligned orbits} \label{sec:incl}

We move now onto the more general case of {\it misaligned} orbits with $\beta \in [0,\pi]$, focusing on the phenomenologically relevant cases of (quasi)-floating and non-resonant mixing. As a side product, we will also provide flow diagrams highlighting the dynamics near the equatorial limit. As before, we restrict our attention to the region of parameter space in which precession effects do not modify the resonance behavior (see App.~\ref{app:spin}).\vskip 4pt

{\it \underline{Floating fixed points on inclined circular orbits.}} Let us consider first an idealized floating scenario, i.e., $\dot{\Omega}^{(ab)}_{g,0} \simeq 0$ and $\bar{\Gamma}^{-}_{ab} \simeq 0$, on a circular orbit.\footnote{Contrary to the claims in~\cite{Tomaselli:2024bdd}, the circular limit is consistent with $\beta \neq 0$ (see App.~\ref{app:flux_balance}).} Resonances are possible provided $\frak{f}^{(ab)}_{g,0}>0$, i.e., $g>0$, with both $\frak{s}=\pm 1$. In these circumstances, and using~\eqref{eq:v_flt}, we can find a closed-form solution to~\eqref{eq:i_Vq}:
\begin{eqnarray} \label{eq:i_emri_circ}
\cos\beta &\simeq& \left(\frac{m}{g} - \left[\frac{m}{g} - \cos{\beta(0)} \right] \exp{-\frac{b_{g,0}}{w_{g,0}} \frac{t}{b_{g,0}}} \right) \,, \qquad (S_c \gg L) \,,  \label{eq:beta_circ_emri} \\
\cos{\beta} &\simeq&  \frac{\cos{\beta(0)}\left(1+\tfrac{m_a}{(\tilde{a}/N_\mathrm{c})} \right) - \tfrac{t}{b_{g,0} (\tilde{a}/N_\mathrm{c})} }{1+\tfrac{m_a}{(\tilde{a}/N_\mathrm{c})}  - \tfrac{g}{m}\tfrac{t}{b_{g,0} (\tilde{a}/N_\mathrm{c})}}  \,, \qquad (S_c \ll L) \,.
\label{eq:beta_circ_q1}
\end{eqnarray}

Let us consider the IMRI/EMRI limit first. For $g=m$ overtones $(\Delta m_{ab} <0)$ we find that $\beta(t) \to 0$, consistently with our previous discussion on stability. In contrast, for $g=-m>0$ $(\Delta m_{ab} >0)$, the situation is reversed and $\beta_\mathrm{cr} = \pi$ becomes a fixed point. Let us emphasize, however, that other fixed points are also possible. For instance, for $m/g < 1$, the condition $\cos{\beta}_\mathrm{cr} = m/g$ determines a series of attractor points; whereas, for $m/g > 1$, there is no real solution, and therefore $\beta(t) \to 0$.\vskip 4pt  

In the comparable-mass limit, the obliquity is governed primarily not by the transfer of momentum from the cloud to the orbit, but by the cloud's internal dynamics. This yields the milder behavior shown in~\eqref{eq:beta_circ_q1}. For example, in the diluted-cloud limit, $\tilde{a}/N_\mathrm{c} \to \infty$, one recovers the expected $\beta(t)\simeq \beta(0)$. The more interesting regime turns out to be when $\tilde{a}/N_\mathrm{c}\simeq 1$. As noted in~\eqref{eq:overt_stab_q1}, transitions with $g=m>0$ make the critical point at $\beta_{\rm cr}=0$ unstable.\footnote{The resulting instability in~\eqref{eq:beta_circ_q1} grows slowly for $\beta(0)\simeq 0$, as long as $m_a>m$. By contrast, for $m_a<m$ the weak-decay approximation breaks down. We discuss representative examples in \S\ref{sec:pheno}.} More generally, for resonances with $g\neq m$, the fixed points are unstable when $g/m<1$, whereas for $g/m>1$ the would-be fixed point lies outside the real domain and the evolution instead drives $\beta(t)\to\pi$.
\vskip 4pt

Throughout the previous discussion we assumed the floating conditions~\eqref{eq:flt_res_break},~\eqref{eq:flt_over_adia} are satisfied. However, since the overlap depends on the obliquity $\eta^{(ab)}_{l,m,g,k} \propto d^{(l)}_{mg}(\beta)$, analogously to the arguments below~\eqref{eq:flt_over_adia} the sustainability of a floating configuration is not always guaranteed. As an example, let us consider the $g=m$ resonant overtone, first in the IMRI/EMRI limit. Here we have that $ \eta^{(ab)}_{l,m,m,k} \sim d^{(l)}_{mm}(\beta)$ is maximal at $\beta = 0$, and goes smoothly towards zero at $\beta=\pi$. For orbits starting at $\beta(0) \simeq \pi$, the floating conditions require that $b^{(ab)}_{g,k}$ is sufficiently large [cf.~\eqref{eq:flt_res_break}] to offset the angular suppression $d^{(l)}_{mm} \sim (\beta - \pi)^{2m}$. Yet, once floating begins, it can become progressively more sustainable as $\beta(t) \to 0$. This changes for the case of comparable masses, where a fixed point is present at $\beta_\mathrm{cr} = \pi$. In such a case, starting at $\beta(0) \simeq 0$,  the dynamical evolution would move towards resonance breaking instead. These two situations are somewhat similar to the differences in the evolution of the eccentricity on equatorial orbits for early and late overtones, respectively.\vskip 4pt

\begin{figure}[t!]
  \centering

  \makebox[\textwidth][c]{%
    \begin{minipage}{0.42\textwidth}
      \centering
      \includegraphics[width=\textwidth]{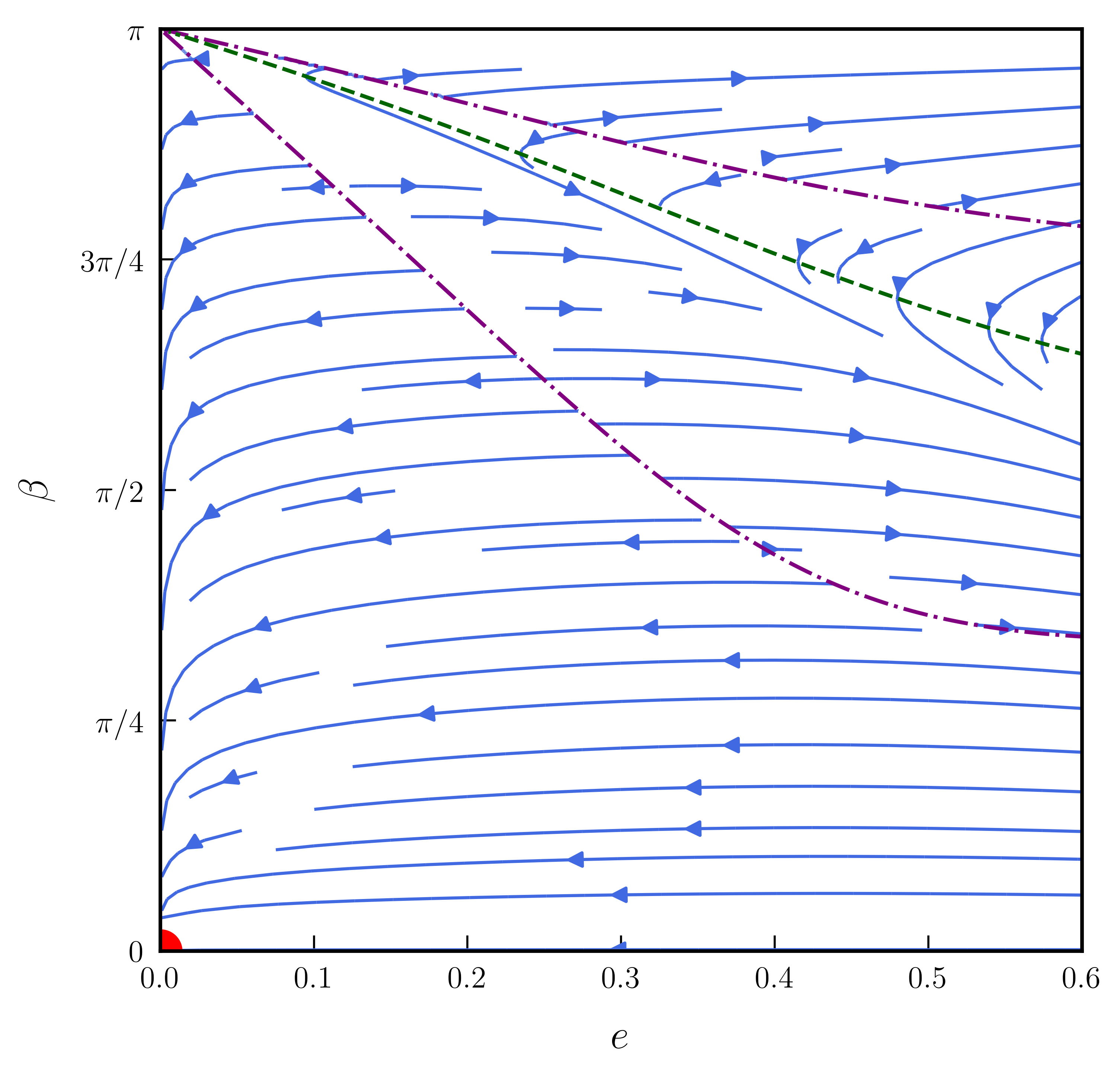}
    \end{minipage}\hspace{0.03\textwidth}%
    \begin{minipage}{0.42\textwidth}
      \centering
      \includegraphics[width=\textwidth]{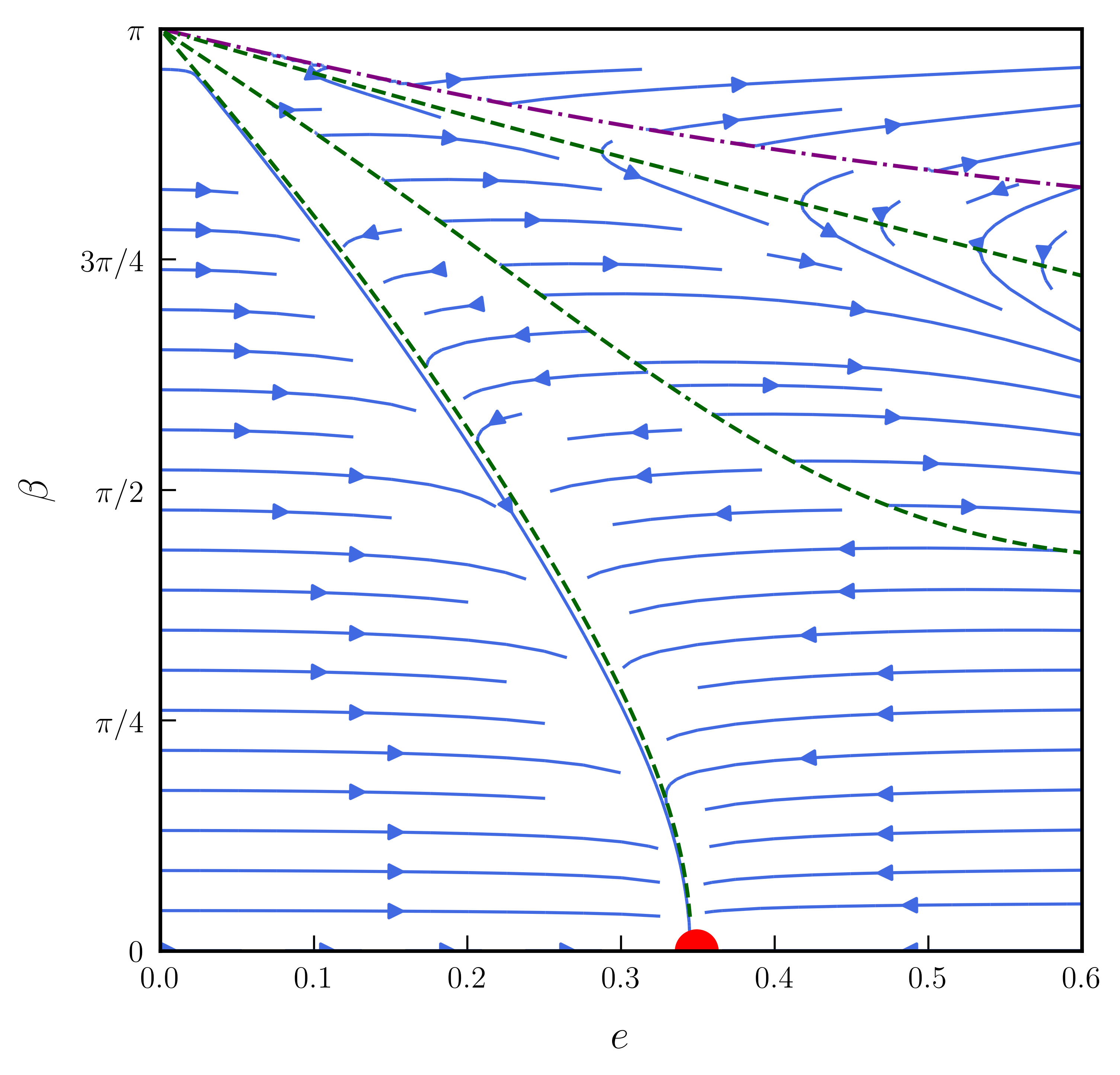}
    \end{minipage}%
  }

  \vspace{0.5em}

  \makebox[\textwidth][c]{%
    \begin{minipage}{0.42\textwidth}
      \centering
      \includegraphics[width=\textwidth]{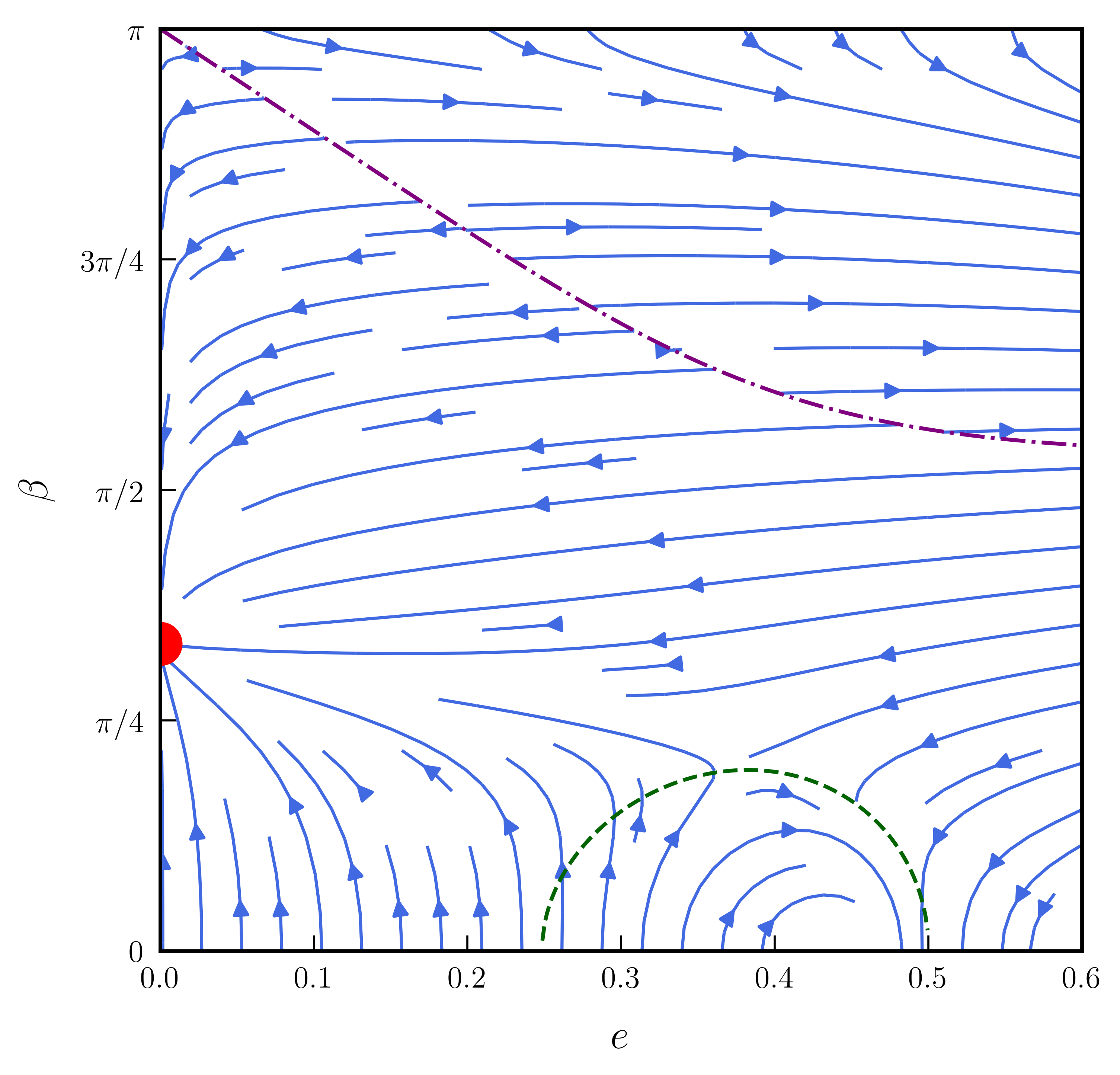}
    \end{minipage}\hspace{0.03\textwidth}%
    \begin{minipage}{0.42\textwidth}
      \centering
      \includegraphics[width=\textwidth]{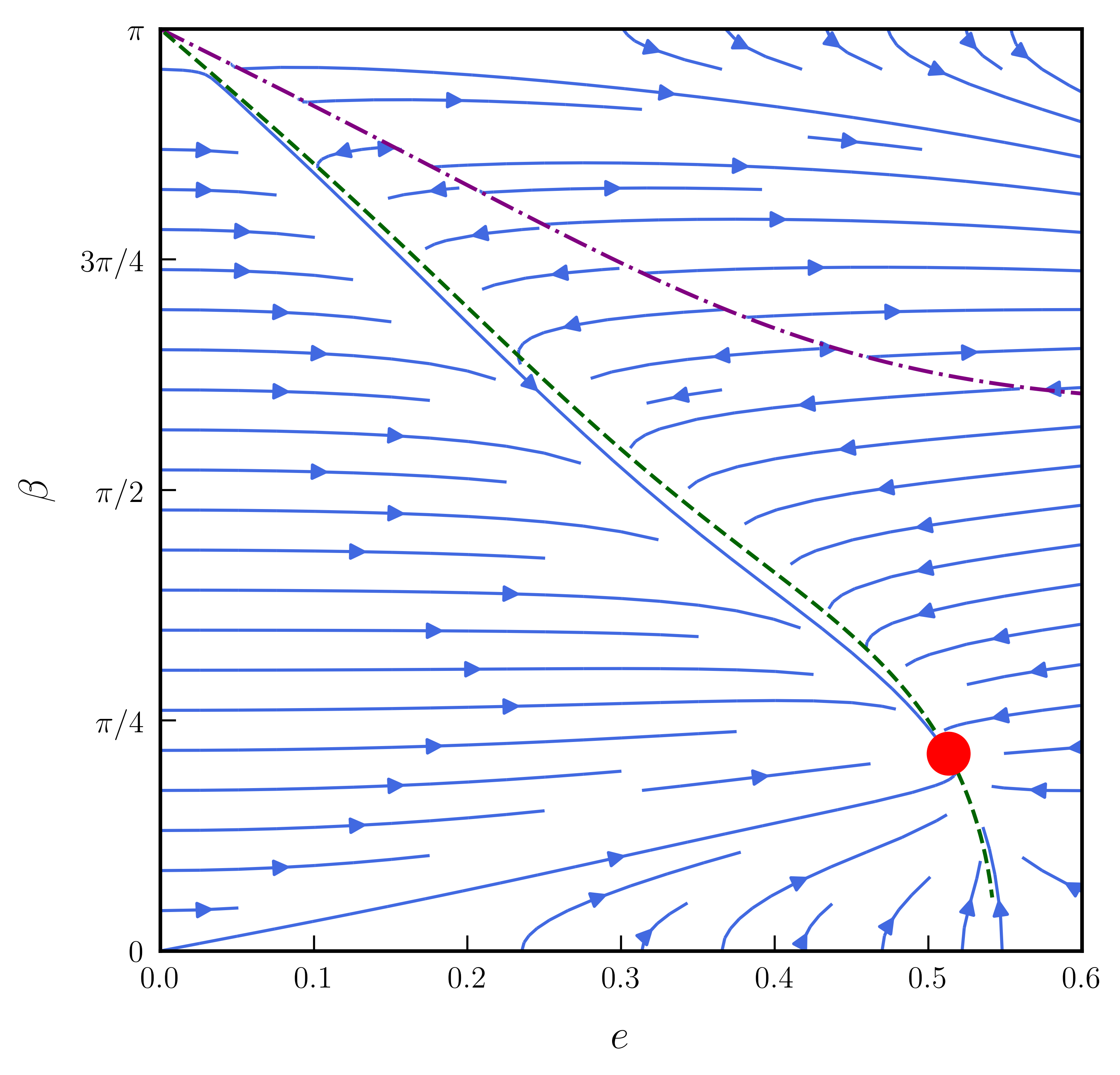}
    \end{minipage}%
  }
  \caption{Eccentricity/obliquity $\{e,\beta\}$ flow for degenerate overtones $\Omega^{(ab)}_{(g,k|d)} =  (m/d) \, \Omega^{(ab)}_0 $ in the limit $(b_{d}/w_d)^{(ab)} \gg 1$ [via~\eqref{eq:e_i_emri}] for $l=2$ and: $(m,d)=(2,2)$ [\textit{upper left}], $(2,3)$ [\textit{upper right}], $(1,2)$ [\textit{lower left}] and $(1,3)$ [\textit{lower right}]. Fixed points are indicated by red dots, and (non-equatorial) separatrices by purple and green dashed curves. The strongest (eccentric) overtone on the left diagrams corresponds to the main one, at $k=0$, while on the right the strongest one is at $k=-1$. The lower panel exhibits the non-equatorial fixed points at $(e,\beta_\mathrm{cr}) = (0, \pi/3)$ [{\it left}; cf.~\eqref{eq:beta_circ_emri}] and $(0.48, \pi/4)$ [{\it right}].}
  \label{fig:e_i_flow}
\end{figure}

{\it \underline{Floating on generic orbits in the IMRI/EMRI limit.}}  The most general situation involves orbits with both $\beta \neq 0$ and $e \neq 0$. In this scenario, there is an interplay between $g-$ and $k-$overtones that can lead overtone degeneracy. More concretely, to the existence of a tower of overtone pairs $(g,k)$, $(g',k'), \cdots $, that may resonate at the same orbital frequency, provided $g-k = g'-k' = \cdots = d$ [see \eqref{eq:abs_W}]. (Recall $\dot{\Sigma}^{(ab)}_{g,k} \approx (g-k)\dot{\vartheta}$ [see~\eqref{eq:f_k}].) This possibility arises whenever $Y_{lg}(\pi/2,0) \neq 0$, which ensures a non-zero  mixing overlap [cf.~\eqref{eq:eta}]. Although more involved, these degenerate overtones share the same Bloch variables $(\nu,u)^{(ab)}_{d}$,
which allows us to simplify the problem. Let us first consider the weak-decay regime and assume that the floating conditions are valid, i.e., $\dot{\Omega}^{(ab)}_{g,k} \simeq 0$, $\bar{\Gamma}^{-}_{ab} \simeq 0$. Under these assumptions we find, generalizing~\eqref{eq:v_flt}, 
\begin{eqnarray} \label{eq:v_flt_inc}
 [\nu^{(ab)}]^{\mathrm{float} }_{d}\simeq \frac{f(e) }{ (g-k) \, b_{d} \sum_{(g,k|d)} \mathrm{sgn}(\eta^{(ab)}_{l,m,g,k}) \sqrt{z^{(ab)}_{l,m,g,k}}  }  \,,
\end{eqnarray}
where $\sum_{(g,k|d)}$ indicates a sum over degenerate overtone pairs. Notice that in~\eqref{eq:v_flt} the factor of $\mathrm{sgn}(\eta^{(ab)}_{l,m,g,k})$ is inconsequential, since an overall sign does not alter the physics~\cite{Baumann:2019ztm}. However, as we shall see, for degenerate overtones the relative signs turn out to play an important role. Substituting~\eqref{eq:v_flt_inc} into~\eqref{eq:e_Vq} and~\eqref{eq:i_Vq_emri}, and performing a change of variables $dt \to -d\sigma/(-\dot{\sigma})$, the evolution equations for $\{e,\beta\}$ become 
\begin{eqnarray} \label{eq:e_i_emri}
\frac{de^2}{d(-\sigma)} & \simeq& \left( \frac{b_d}{w_d} \right)^{(ab)} \frac{\sqrt{1-e^2} d \sum_{(g,k|d)} \left(\tfrac{g(e)}{f(e)}- \frac{g}{d} \right) \mathrm{sgn}(\eta^{(ab)}_{l,m,g,k}) \sqrt{z^{(ab)}_{l,m,g,k}}}{3\sum_{(g,k|d)}  \mathrm{sgn}(\eta^{(ab)}_{l,m,g,k}) \sqrt{z^{(ab)}_{l,m,g,k}}} \,,  \\
\frac{d\cos \beta }{d(-\sigma)} &\simeq &  \frac{\sum_{(g,k|d)} 
\left[ \left( \frac{b_d}{w_d} \right)^{(ab)} \frac{\left( m - g \cos{\beta} \right)}{3\sqrt{1-e^2}}   - \frac{(g- m \cos{\beta})}{s_\mathrm{c}(\sigma)}  \right]  \mathrm{sgn}(\eta^{(ab)}_{l,m,g,k}) \sqrt{z^{(ab)}_{l,m,g,k}}}{2\sum_{(g,k|d)}  \mathrm{sgn}(\eta^{(ab)}_{l,m,g,k}) \sqrt{z^{(ab)}_{l,m,g,k}}} \,. \nonumber
\end{eqnarray}
As expected, the factors of $\mathrm{sgn}(\eta^{(ab)}_{l,m,g,k}) \sqrt{z^{(ab)}_{l,m,g,k}}$ cancel out for a single dominant overtone but remain in the more general degenerate case. Notice that, in the $b_d/w_d \gg 1$ limit, the evolution equations become autonomous [depending only on $(\beta,e)$] and scale-free [after absorbing the overall factor into a new ``time'' variable $\sigma \to (b_d/w_d)\sigma$], which allows us to construct a series of flow diagrams. We illustrate four different situations in~Fig.~\ref{fig:e_i_flow}.\footnote{Similar plots can be found in~\cite{Tomaselli:2024bdd}, which however differ significantly from ours. The discrepancy arises from missing terms in~\cite{Tomaselli:2024bdd}, as well as the authors’ omission of degenerate overtones (see \S\ref{sec:concl}).} In addition to fixed points in the eccentricity at $\beta=0$ (equatorial), and the obliquity at $e=0$ (circular), fixed points exist also away from these limits. Moreover, we also encounter separatrices, originating from zeroes of the numerator and denominator of~\eqref{eq:e_i_emri}, that delineate different regimes. Let us emphasise that, while these diagrams provide useful guidance, they are constructed under the assumption of an uninterrupted floating configuration.\footnote{Moreover, they also illustrate the approximate location of the true fixed points of the system. This is because, in the IMRI/EMRI limit, the term in~\eqref{eq:i_Vq_emri} vanishes by construction at the fixed points, promoting the (would-be subleading) contribution in~\eqref{eq:i_Vq_q1} to a leading role. In practice, this slightly shifts the value of the fixed points relative to those displayed in the figures.} 
Therefore, they omit regimes in which floating breaks down. In general scenarios, the evolution of the orbit will ultimately depend on the initial conditions of the full system and parameter space, which we will discuss in more detail in \S\ref{sec:pheno}.\vskip 4pt

{\it \underline{Non-resonant mixing on generic orbits.}} Over the secular evolution of $\beta$, the binary may, in principle, sweep across the entire tower of $g$-overtones, including non-resonant modes with $d\leq 0$. Since our focus is the co-evolution of the cloud and the host binary in the regime $\frak{f}^{(ab)} \ll 1$, we can use the low-frequency approximation in~\eqref{eq:low_f_limit} to follow multi-$(g,0)$ overtones (notice that, for non-resonant mixing, the $k=0$ contribution typically dominates for each value of $g$). The low-frequency limit faithfully captures the cloud's decay, but it does not fully incorporate the corresponding orbital backreaction. To restore this effect, we complement it with the substitution rule, implied by~\eqref{eq:a_Vq}-\eqref{eq:i_Vq} (restricted to $k=0$),
\begin{eqnarray} \label{eq:low_f_subst}
\sum_g g \,\eta^{(ab)}_{l,m,g,0}\,\nu^{(ab)}_{g,0}
\;\to\;
\left(\frac{\sum_g g \,\eta_{l,m,g,0}^2}{\sum_g \eta_{l,m,g,0}^2}\,
V^{(1)}\right)^{(ab)}\,,
\end{eqnarray}
where $V^{(1)}$ is defined in~\eqref{eq:low_f_limit}. This agrees with the low-frequency limit of the equatorial dynamics, when only one overtone has support.

\section{Phenomenology} \label{sec:pheno}

For a given BH mass at the saturation point of superradiant growth, we can probe a range of ultralight bosons---set by
$\mu =\alpha_{\rm sat}/M_{\rm sat}$---through their imprint on GW signals from compact binaries. To determine the cloud properties at saturation, we adopt a quasi-adiabatic model of superradiant evolution~\cite{Arvanitaki:2014wva,Brito:2014wla,East:2018glu}, which fixes the relevant parameters and the cloud's state $(\ket{a})$, given the initial mass of a rapidly spinning BH (see App.~\ref{app:lines_res} for details). Astrophysical and cosmological formation scenarios then provide priors for the compact binaries containing BHs capable of hosting a boson cloud, including distributions for the (initial) orbital elements $\{\Omega,e,\beta\}_{\rm in}$~\cite{Belczynski:2001uc,Sesana:2010qb,Kowalska_2011,Breivik:2016ddj,Nishizawa:2016jji,Nishizawa:2016eza,Rodriguez:2016vmx,Rodriguez:2017pec,Rodriguez:2018pss,Lower:2018seu,Randall:2019znp,Fang:2019dnh,Romero-Shaw:2020thy,Sedda:2020wzl,2021MNRAS.507.2659G,Zevin:2021rtf,Gualandris_2022,Garg:2023lfg,Saini:2023wdk,Dhurkunde:2023qoe,Stegmann:2025shr,Larsen:2025ayq,Mancieri:2025cmx}. We co-evolve the gravitational atom and its host binary---consistently tracking level-mixing and precession effects, as well as resonant and wide-state transitions---until either the cloud is depleted, or largely disrupted; or it survives into the detector’s band. Both branches yield testable predictions. Early depletion generates changes in the distribution of orbital and spin elements relative to cloud-free baselines, whereas long-lived clouds produce notable in-band signals.\vskip 4pt 

Following~\cite{Breivik:2016ddj}, the Letter presented a proof-of-concept population study under the assumption of co-rotating, equatorial field binaries carrying a boson cloud. Here we lift these restrictions and incorporate several new features into a unified model. Although a comprehensive analysis that encompasses all relevant effects lies beyond the scope of this paper---for example, during the early inspiral the binary may be influenced by relaxation and environmental effects, e.g.,~\cite{Milosavljevic:2002ht,Sesana:2010qb,Barausse:2014tra,Burke-Spolaor:2018bvk,Ishibashi:2020zzy,Gualandris:2022kxh}, and by dynamical friction~\cite{Baumann:2021fkf,Tomaselli:2023ysb,Tomaselli:2024bdd} at later stages---we highlight key elements and isolate robust signatures that furnish concrete targets for GW searches and set a clear path forward for future analyses.\vskip 4pt  

We highlight our main results for stellar-mass $\big\{ 5 \lesssim M_{\rm sat}/M_\odot \lesssim 150\,; \,0.1 \lesssim q \lesssim 10\big\}$ and IMRI/EMRIs  $\big\{10^{2} \ll M_{\rm sat}/M_\odot\, ; \, (10 M_\odot/M_\mathrm{sat}) \lesssim q \lesssim 10^{-2}\big\}$ separately.\footnote{The case for the intermediate-mass regime, $M \simeq 10^{2}$--$10^{4}\,M_\odot$, has strengthened in recent years, with multiple lines of observational evidence, e.g.,~\cite{2024Natur.631..285H}, including constraints on their spins~\cite{Wen:2021yhz}. The lower-mass region will be within reach of forthcoming LIGO--Virgo--KAGRA observing runs~\cite{Cheung:2025grp}, and further explored by third-generation detectors such as ET~\cite{Abac:2025saz}. LISA, in turn, is expected to survey intermediate-mass systems across a broad range of parameter space, while also accessing the $10^{4}$--$10^{5}\,M_\odot$ region thought to inhabit the nuclei of dwarf galaxies~\cite{Greene:2019vlv}.}  We defer the more technical details to App.~\ref{app:lines_res}. For the sake of notation, we will also drop the `sat' label in the remainder of this section.

\subsection{Stellar binaries} \label{sec:stellar}

In what follows we take $t_\mathrm{age} \simeq 10^8 \, \mathrm{yr}$ as a reference value for the $\ket{211}$ and $\ket{322}$ excited states, but consider also scenarios with $t_{\rm age} \gtrsim 10^{9}{\rm yr}$  that can populate the $\ket{433}$ state for higher values of $\alpha$ (see~\cite{Caputo:2025oap} and references therein for a discussion on $t_{\rm age}$ values). \vskip 4pt

{\it \underline{Chronology of the $\ket{211}$ state.}} The first growing mode remains viable, provided $0.01 \, (0.02) \lesssim \alpha \lesssim 0.09$ $(0.11)$ for $M \simeq 5 M_\odot \,(150 M_\odot)$. Possible transitions then include quadrupolar $\mathcal{H}$-transitions $\ket{211} \to \{\ket{210},\ket{21-1}\}$ and early $\mathcal{B}$-transitions ($l \geq 4$) $\ket{211} \to \{\ket{l \,\, (l-1) \,\, -(l-1)}$ , $\ket{(l+2) \,\,  (l+1) \,\,  (l+1)} \}$, before entering the deep $\mathcal{B}$ regime and ionization. All of the narrow $\mathcal{H}$-resonances are of the (quasi-)floating type, with $\Delta m_{ab}<0$, while all early narrow $\mathcal{B}$-resonances are of the sinking type. This is also the case for the excited states we will study shortly.\vskip 4pt

In the regime $\alpha \lesssim 0.06$, the $\mathcal{H}$-transitions occur so early in the inspiral that they do not affect binaries which merge within a Hubble time. For such small values of $\alpha$, only clouds that form \emph{after} the $\mathcal{H}$-transitions can grow and survive long enough to enter the detector band. Once there, they encounter early (sinking) $\mathcal{B}$-resonances which, while not sufficiently adiabatic to disrupt the cloud, can still imprint significant changes to the orbital evolution. Subsequently, the system is governed by the ionization-driven inspiral and the deeper $\mathcal{B}$ regime; see, e.g.,~\cite{DellaMonica:2025zby} for a preliminary study.\vskip 4pt

\begin{figure*}[t!]
\begin{tabular}{cc}
\includegraphics[width=.5\textwidth]{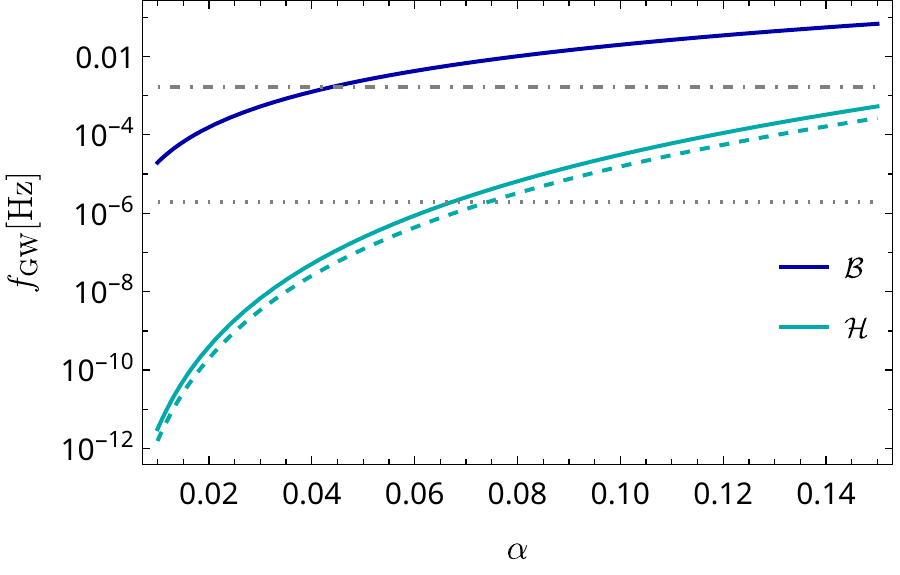}
\includegraphics[width=.5\textwidth]{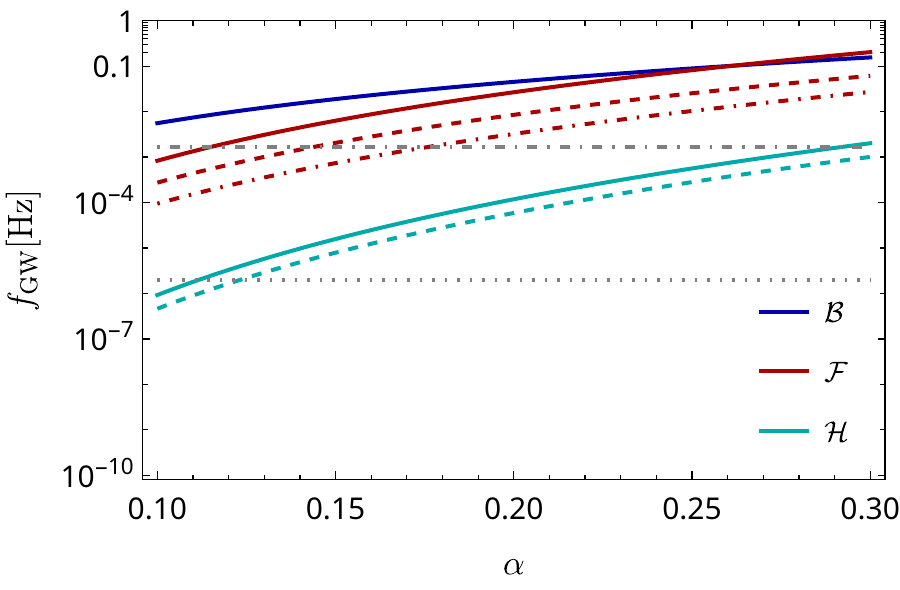}
\end{tabular}
    \caption{The value of $\Omega^{(ab)}_{g,0}$  for a selection of transitions with $M=70 M_\odot$. For the $\ket{211}$ state [\textit{left}], to $\ket{54-4}$ (blue) and to $\{\ket{21-1}, \ket{210} \}$ (cyan, full/dashed). For the $\ket{322}$ state [\textit{right}], to $\ket{54-4}$ (blue), to $\{\ket{300}, \ket{31m} \, (g=1), \ket{31m} \, (g=3)\}$ (red, full/dashed/dot-dashed) and to $\{\ket{320}, \ket{321} \}$ (cyan, full/dashed). The gray dot-dashed line indicates the lowest frequency for a non-monochromatic GW signal in the LISA band; while dotted line indicates the lowest frequency for a binary to coalesce within Hubble time (with $q=0.5, e_\mathrm{in} \ll 1$). }
    \label{fig:freq}
\end{figure*}

For larger values of $\alpha$, $0.06 \lesssim \alpha \lesssim 0.1$, boson clouds can undergo $\mathcal{H}$-transitions before the binary enters the detector band, see Fig.~\ref{fig:freq}. 
In this case, the strongest resonances [with $(g,k) = (2,0)$] efficiently deplete the cloud across most of parameter space, except within a narrow region around counter-rotating orbits ($\beta \simeq \pi$).\footnote{In principle, the cloud can also excite some of the earlier overtones with $k<0$; however, due to the width of the main overtone, for $\beta \lesssim \pi/2$ these are comparatively inefficient at growing the eccentricity.} From~\eqref{eq:flt_res_break}, we find that for $e \gtrsim 0.01\,(\alpha/0.1)^{8/9}$ and comparable masses, the strongest available ($k=-3$) transition disrupts the cloud.\footnote{Given the expected eccentricity distributions, see e.g.~\cite{1975MNRAS.173..729H,Breivik:2016ddj}, most binaries have moderate-to-high eccentricities in the early inspiral and therefore satisfy this condition.} By contrast, binaries on counter-rotating orbits with negligible eccentricity may survive the $\mathcal{H}$-regime, although up to $\simeq 50\%$ of the cloud's mass may be depleted.\vskip 4pt

{\it \underline{Chronology of the $\ket{322}$ state.}}  The next excited state is supported in the range $\alpha \gtrsim 0.1$, and it poses a richer resonant history. At first we find $\mathcal{H}$-resonances: $\ket{322} \to \{\ket{321},\ket{320},\ket{32-1},\ket{32-2}\}$, driven by $l=4$ transitions, with the first two also present with $l=2$ perturbations. Transitions to $\{\ket{320},\ket{32-2}\}$ are possible in the equatorial limit. Next, in the $\mathcal{F}$ regime, we find a (wide) quadrupolar transition to the (spherical) state $\ket{322} \to \ket{300}$, and three $(l=3)$-mediated narrower (quasi-floating) resonances: $\{\ket{311},\ket{310},\ket{31-1}\}$, where the middle one is only possible away from the equatorial plane.\vskip 4pt 

All of the $\mathcal{F}$-transitions occur in the LISA band in the lower range of BH masses, as well as 
$\mathcal{H}$-transitions for large values of $\alpha \gtrsim 0.3$. For high stellar masses, $\mathcal{H}$-resonances fall outside of the band, while $\mathcal{F}$-transitions typically occur inside for $\alpha \gtrsim 0.2$. Early $\mathcal{B}$-resonances, on the other hand, occur at comparable higher frequencies. (See Fig.~\ref{fig:freq}.)\vskip 4pt 

Depending on the birth frequency, the fate of the cloud is largely dictated by the wide mixing with the $\ket{300}$ state, located at
$\Omega_0^{(322,300)} \simeq (10^2 - 10^3)\,\Omega^{(322,32m)}_{g,k}$ and 
$\Omega^{(322,300)}_0 \simeq (3 - 15)\,\Omega^{(322,31m)}_{g,k}$, relative to other transitions (choosing the strongest eccentric overtones close to the main one at $k=0$). We then consider three basic scenarios for binary formation: \textit{(i)} prior to the $\mathcal{H}$-regime; \textit{(ii)} after the $\mathcal{H}$-regime; and 
\textit{(iii)} in the vicinity of the $\mathcal{F}$-transitions. These regimes are separated by roughly decades in frequency. \vskip 4pt

\begin{figure*}[t!]
\centering
\setlength{\tabcolsep}{0pt} 

\resizebox{\textwidth}{!}{%
  \includegraphics[width=.5\textwidth]{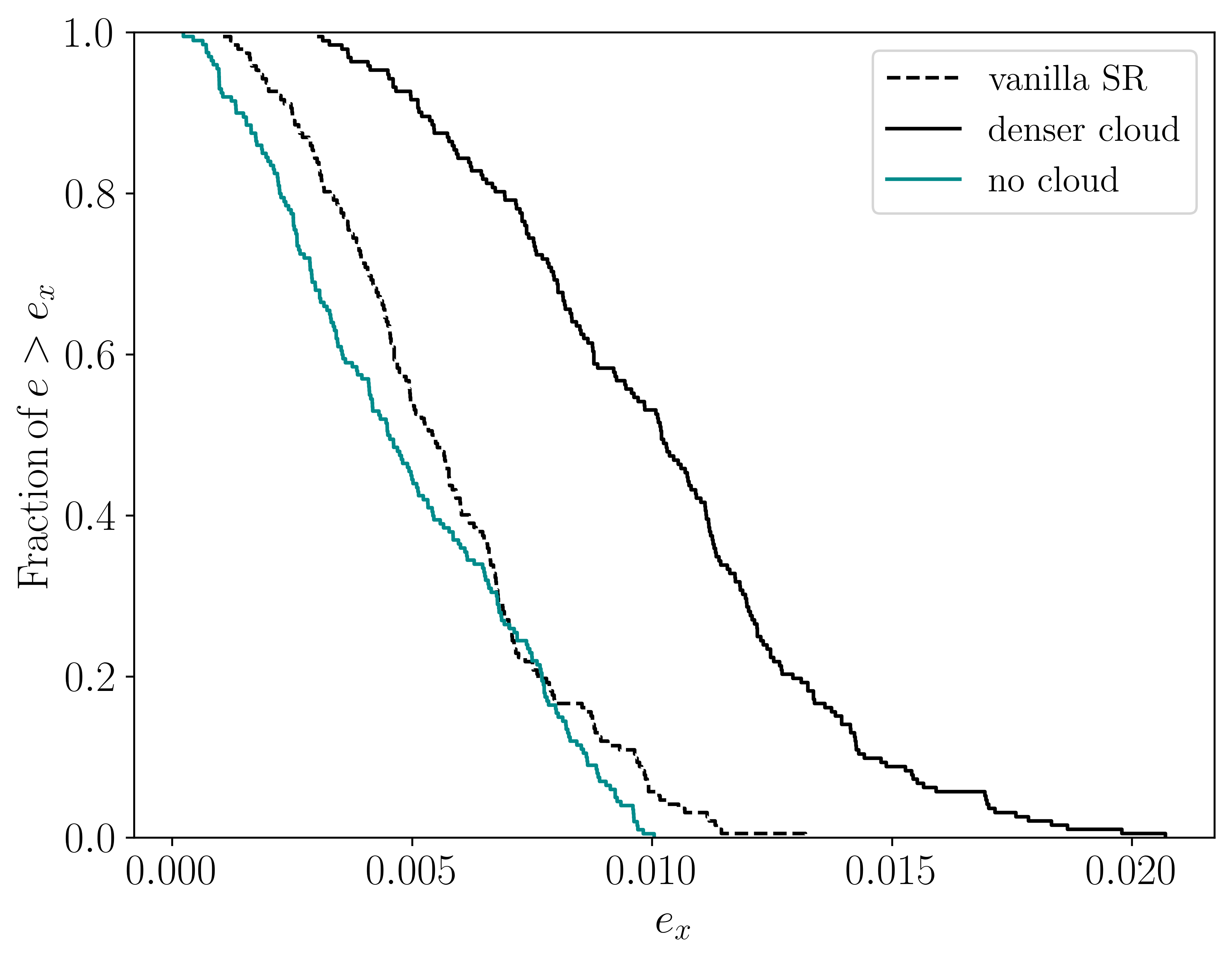}%
  \includegraphics[width=.52\textwidth]{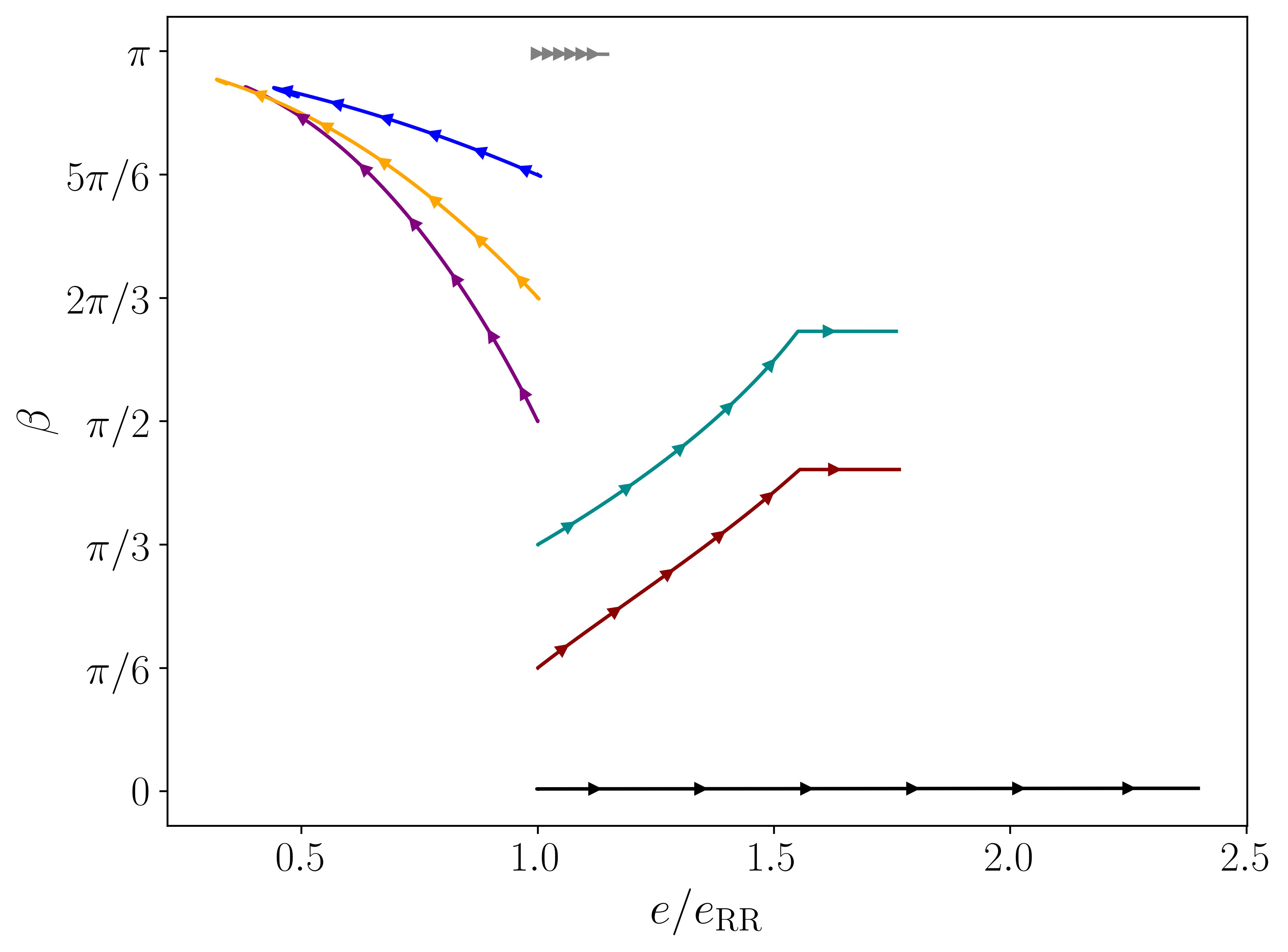}%
}
\caption{Features of the cloud-orbit co-evolution in scenario \textit{(i)} (see main text), due to the three-state mixing $\{\ket{322},\ket{32m},\ket{300}\}$, $m \in \{0,1\}$. Ratio of binaries observed at $f_{\rm GW} = 10^{-2} \, \mathrm{Hz}$ with eccentricities above a given value $e_x$, for $\beta_{\rm in} \simeq 0$ [\textit{left}]. We show both the vanilla case ($M_\mathrm{c}/M \simeq 0.05$, dashed black), as well as denser clouds ($M_\mathrm{c}/M \simeq 0.2$, solid black). Eccentricity/Obliquity $\{e/e_{\rm RR},\beta\}$ flow for vanilla clouds [\textit{right}], normalized to standard radiation-reaction (RR)-driven evolution in vacuum ($e_{\rm RR}$), for the specific case $\alpha=0.25$, $q=0.1$, $e_\mathrm{in}=0.3$, and seven $\beta_{\rm in}$ values.}
\label{fig:322_hyperfine}
\end{figure*}

In scenario~\textit{(i)}, mixing with the spherical state becomes relevant toward the upper end of the $\alpha$ range, since $\Omega^{(322,300)}_0/\Omega^{(322,32m)}_0 \sim \alpha^{-2}$, although the narrower $\mathcal{H}$-transitions may still be successfully excited.\footnote{Restricting to the dominant $(l=2)$-driven resonances, the strongest overtones are $(g,k) = (2,0)$, with the one connecting to $\ket{321}$ occurring first.} The quasi-floating conditions~\eqref{eq:flt_res_break} are easily satisfied even for modest eccentricities, so that early overtones are the first to be excited, leading to a moderate growth of eccentricity.
Let us consider a setup analogous to that discussed in the Letter, but with a fixed boson mass $\mu = 5 \times 10^{-13}\,\mathrm{eV}$, and a population of binary BHs with component masses $M/M_\odot \in [60,80]$ and initial eccentricities $e_{\rm in} \in [0.1,0.6]$, spanning a range of initial obliquities $\beta_{\rm in}$. We initialize the orbital frequency at $\Omega_{\rm in}/\pi \simeq 10^{-4}\,\mathrm{Hz}$ and evolve the systems up to the middle of the LISA band, corresponding to a (peak) GW frequency $f_{\rm GW} \simeq 10^{-2}\,\mathrm{Hz}$, where $f_{\rm GW} \simeq \frac{\Omega}{\pi} \frac{(1+e)^{1.1954}}{(1-e^2)^{3/2}}$ (cf.~\cite{Wen:2002km}). The results are shown in Fig.~\ref{fig:322_hyperfine}. The left panel displays the distribution of eccentricities for the {\it vanilla} superradiant scenario\footnote{Starting with a highly spinning BH in vacuum ($\tilde{a}_\mathrm{in} \lesssim 1$), the cloud occupancy can be computed for a given $(\alpha, \tilde{a})$ at the start of the each superradiant cycle $\ket{211} \to \ket{322} \to \cdots$. See App.~\ref{app:lines_res} for more details.}, together with the corresponding distribution obtained for a denser cloud. Consistent with~\cite{Boskovic:2024fga}, we find that a sizeable fraction of binaries (from $\approx 5\%$ and up to $50\%$ depending on the density of the cloud) are observed with eccentricities $e \gtrsim 0.01$ at $f_{\rm GW} \simeq 10^{-2}$Hz. The right panel shows the evolution of the obliquity for a representative configuration with $\alpha = 0.25$, $q=0.1$, and $e_{\rm in}=0.3$.\vskip 4pt

While in quasi-equatorial configurations the obliquity remains nearly unchanged, more generic setups tend to evolve towards counter-rotating orbits. Specifically, for $\beta \lesssim \pi/2$ the $k<0$ overtones are excited, driving an increase in eccentricity, whereas for $\pi/2 \lesssim \beta \lesssim \pi$ the dominant $k=0$ resonance typically operates, reducing the eccentricity. In nearly equatorial cases, $\beta \simeq \{ 0, \pi\}$, the eccentricity flow is instead governed by mixing with the $\ket{300}$ state.\footnote{The negligible change in eccentricity at $\beta_{\rm in}\simeq \pi$ in the plot (to the right) is a consequence of the given choice of parameters (which are meant to illustrate the radical changes in the obliquity). In fact, large deviations in the eccentricity distribution can occur also for counter-rotating configurations, and are due to a portion of binaries that  undergo a handful of strong $(g,k)=(-2,-3)$ resonances.}\vskip 4pt

\begin{figure*}[t!]
\centering

\makebox[\textwidth][c]{%
  \includegraphics[width=.5\textwidth]{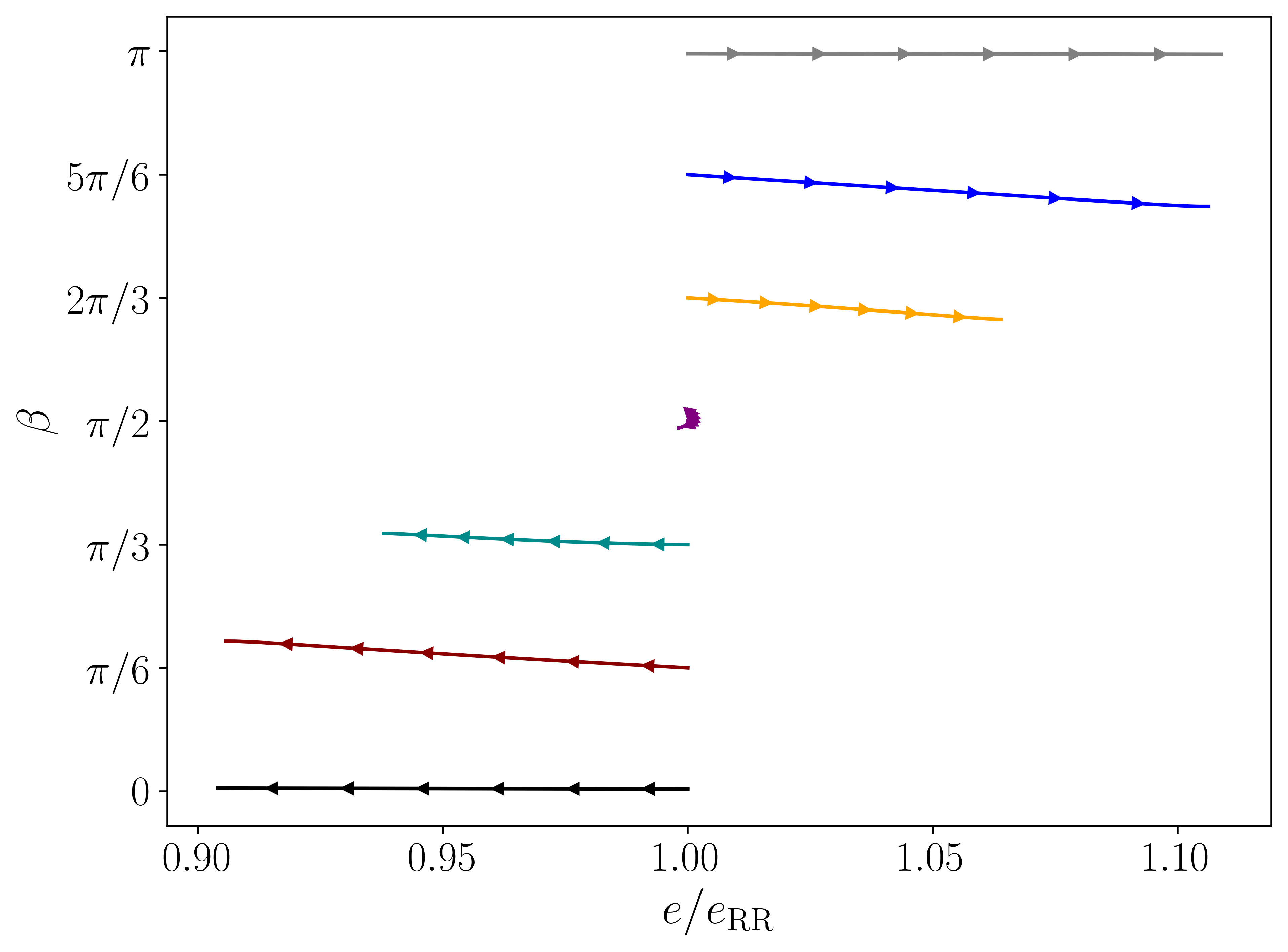}%
  \hspace{0pt}
  \includegraphics[width=.5\textwidth]{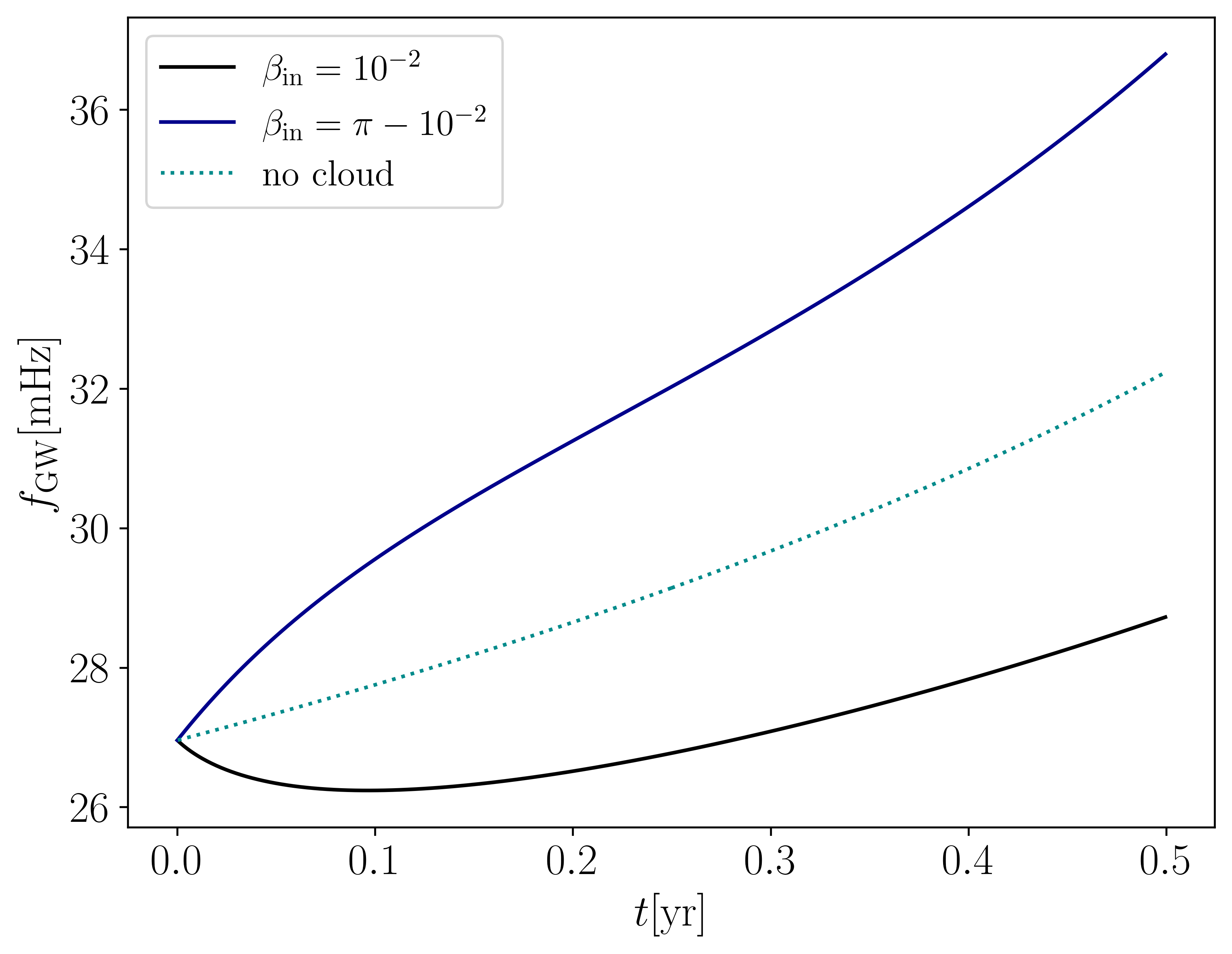}%
}
\caption{Evolution of the (normalized) eccentricity and obliquity [\textit{left}] due to the wide mixing $\ket{322} \to \ket{300}$ in the \textit{(ii)} scenario (see main text); with $\alpha=0.2$, $q=0.1$, $e_{\rm in}=0.3$, and seven different values for $\beta_{\rm in}$. Evolution of the peak GW frequency over the timescale of the LISA mission in the \textit{(iii)} scenario [\textit{right}]; for a $(50+25)M_\odot$ binary on near equatorial co- (black) and counter-rotating (blue) orbits, with $\Omega_\mathrm{in}=0.12 \Omega^{(322,300)}_0$, $e_\mathrm{in}=0.05$, $\alpha=0.28$.}
\label{fig:wide_322}
\end{figure*}

In the {(\it ii)} scenario the cloud is instead formed after the $\mathcal{H}$ regime. In this case we find that the cloud  depletes through perturbative mixing with the wide $\ket{300}$ $\mathcal{F}$-transition, much before entering the realm of narrower $\mathcal{F}$-resonances. This holds irrespective of the values for $\{e,\beta\}_{\rm in}$. We find that the obliquity changes only moderately from its initial value, while eccentricity decreases slightly more (less) compared to vacuum evolution for $0\lesssim \beta \lesssim \pi/2$ ($\pi/2 \lesssim \beta \lesssim \pi$), see Fig.~\ref{fig:wide_322}.\vskip 4pt  The \textit{(iii)} possibility, on the other hand, is interesting for the case of in-band transitions. Although the strong {\it pull} of the $\ket{300}$ state over the narrower $\ket{31m}$ (and early $\mathcal{B}$) transitions renders the latter ineffective, deviations from the standard scenarios may reveal the presence of a boson cloud, see Fig.~\ref{fig:wide_322}.\footnote{The closer the binary system lies to the $\Omega^{(322,300)}_0$ resonant frequency, the more pronounced the backreaction from the  transition becomes. For $\beta_\mathrm{in} \lesssim \pi/2$, this can even induce a transient outspiral that remains observable in band, whereas for $\beta_\mathrm{in} \gtrsim \pi/2$ the binary can undergo a temporary growth of eccentricity.}

\vskip 4pt In summary, across all scenarios the cloud depletes well before---or at the latest at---$\Omega^{(322,300)}_0$ across all parameter space. Binaries undergoing $\mathcal{H}$-transitions attain eccentricities somewhat above the cloud-free baseline, with obliquity for initially off-equatorial orbits being driven towards the counter-rotating limit, see~Fig.~\ref{fig:322_hyperfine}.  At the same time, we find that strong perturbative mixing with $\ket{300}$ prevents comparable growth in the $\mathcal{F}$ regime. Nevertheless, as shown in Fig.~\ref{fig:wide_322}, broad in-band transitions can still drive noticeable departures from standard vacuum evolution.\vskip 4pt   

{\it \underline{Chronology of the $\ket{433}$ state.}} The presence of older BHs allows for higher excited states to form. For example, for \(M \simeq 70\,M_\odot\) and \(\alpha\simeq 0.25\), the \(\ket{433}\) level overtakes \(\ket{322}\) at \(t_{\rm age}\simeq 10^{9}\,\mathrm{yr}\) and for $\alpha \gtrsim 0.3$, already at \(t_{\rm age} \gtrsim 10^{8}\,\mathrm{yr}\), within typical stellar BH ages. For $\mathcal{F}$-transitions to spherical states $\ket{n_a l_a m_a} \to \ket{n_a 0 0}$, the selection rule enforces $(l,m)=(l_a,m_a)$, such that the width decreases rapidly with increasing values of $l_a$. As discussed above, wide transitions can mitigate the impact of narrower \(\mathcal{F}\)-resonances. In the \(\ket{322}\) case, perturbative mixing with \(\ket{300}\) suppressed eccentricity growth in the region of parameter space considered in the Letter. This obstruction is lifted, to a degree, for higher excitations, such as \(\ket{433}\). Although the quantitative details slightly differ, the full picture developed in the Letter remains qualitatively robust under general conditions.\vskip 4pt

\begin{figure*}[t!]
\centering


\resizebox{\textwidth}{!}{%
  \includegraphics[height=6cm]{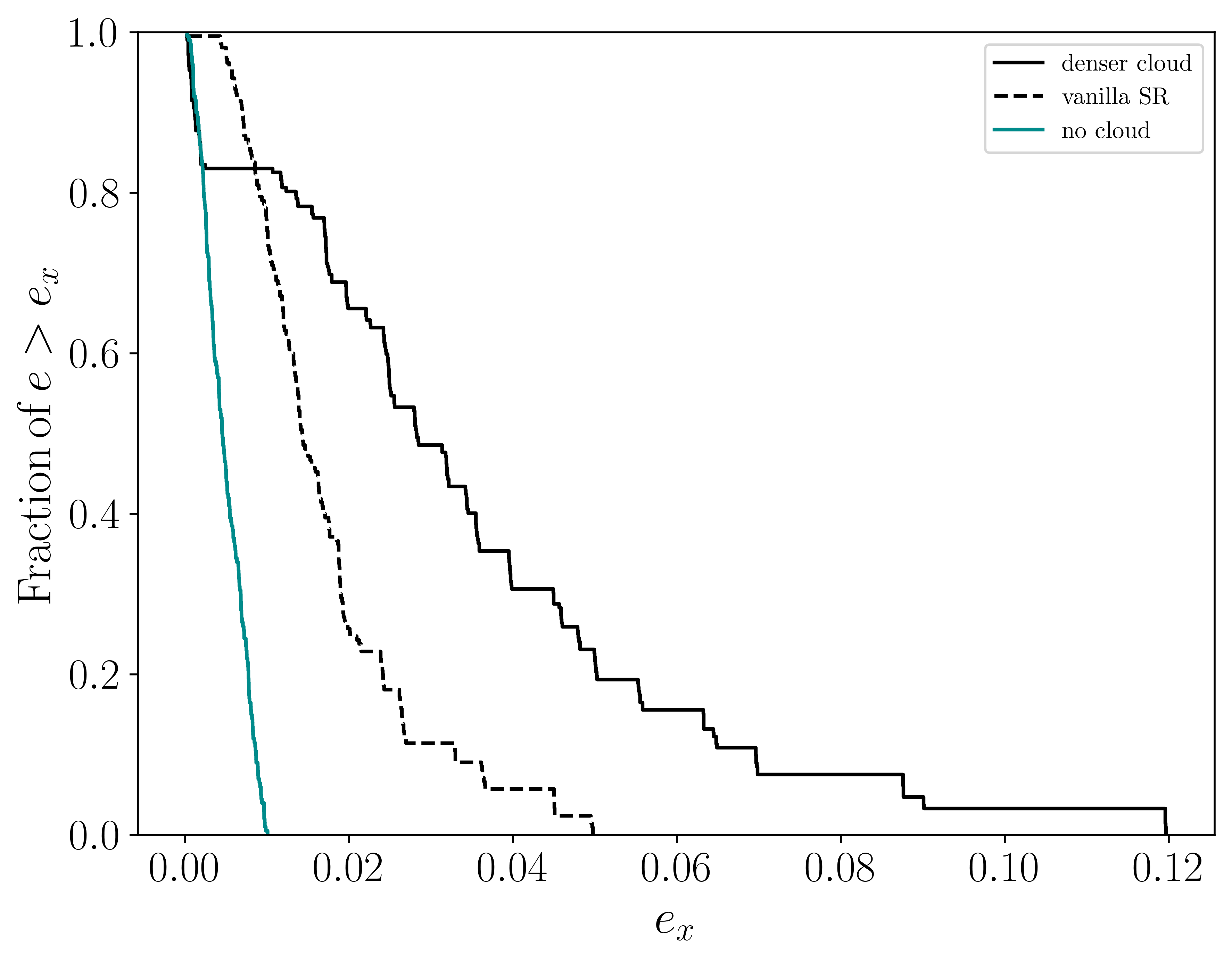}\qquad
  \includegraphics[height=6cm]{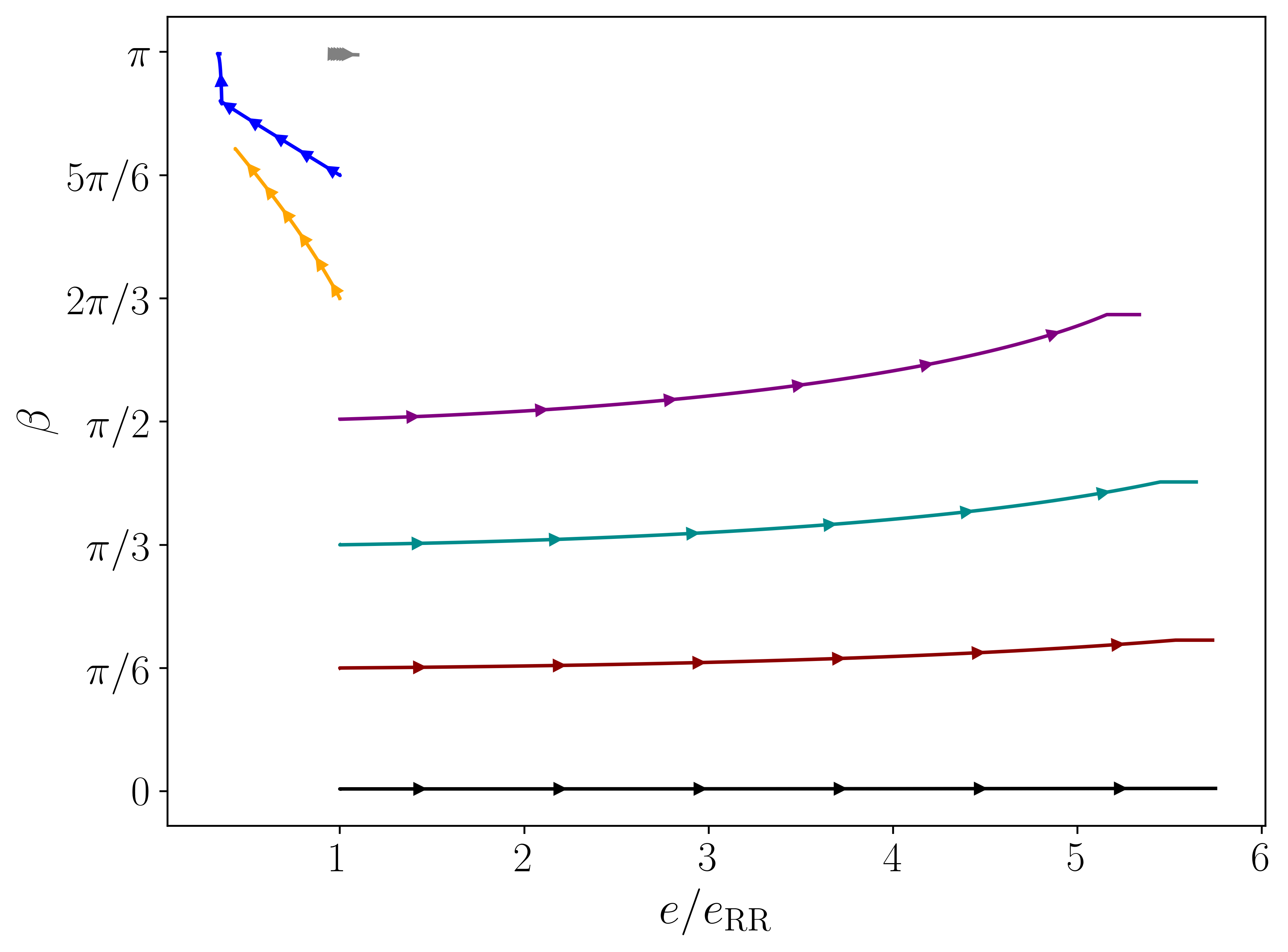}%
}

\caption{Features of the cloud-orbit co-evolution for $\ket{433}$ in the $\mathcal{F}$ regime.
Ratio of binaries observed at $f_{\rm GW} = 10^{-2}\,\mathrm{Hz}$
with eccentricities above a given value $e_x$, for
$\beta_\mathrm{in} \simeq 0$, and both for vanilla and denser clouds [\textit{left}].
Flow of (normalized) eccentricity and obliquity for the specific case $\alpha=0.25$, $q=0.1$, $e_\mathrm{in}=0.3$ and six different $\beta_{\rm in}$ values, considering only vanilla-type clouds [\textit{right}].}
\label{fig:433_fine}
\end{figure*}

For illustrative purposes, we consider clouds that form after the $\mathcal{H}$ regime, taking the same mock distribution as in the \textit{(ii)} scenario for $\ket{322}$. Chronologically, the first overtone band is $\{k \leq 0, g=3\}$ of $(l=3)$-mediated transitions to $\{\ket{420},\ket{421},\ket{422}\}$. As~anticipated, and consistent with our findings in the Letter, sufficiently large initial values drive the increase of eccentricity for co-rotating planar orbits ($\beta_{\rm in} \simeq 0$) toward the fixed points, with negligible change in obliquity. The eccentricity distribution is shown on the left in~Fig.~\ref{fig:433_fine}, for vanilla as well as for denser clouds. As illustrated in the plot, an even greater fraction of binaries (compared to  $\ket{322}$) may achieve large values of in-band eccentricities at the heart of the LISA band. Remarkably, a similar distribution arises when $\beta_\mathrm{in} \lesssim \pi/2$, with somewhat smaller gains in the eccentricity, but with a more prominent change in the obliquity, displayed on the right. For cases with $\pi/2 \lesssim \beta_\mathrm{in} \lesssim \pi$, the main ($k=0$) resonances are activated, yet they typically shut off before the cloud is exhausted, enabling the evolution toward the next band of $g=1$ transitions.
 The eccentricity then depletes faster than in vacuum, while the obliquity grows towards the counter-rotating fixed point. \vskip 4pt

\begin{figure*}[t!]

\resizebox{\textwidth}{!}{%
  \includegraphics[height=6cm]{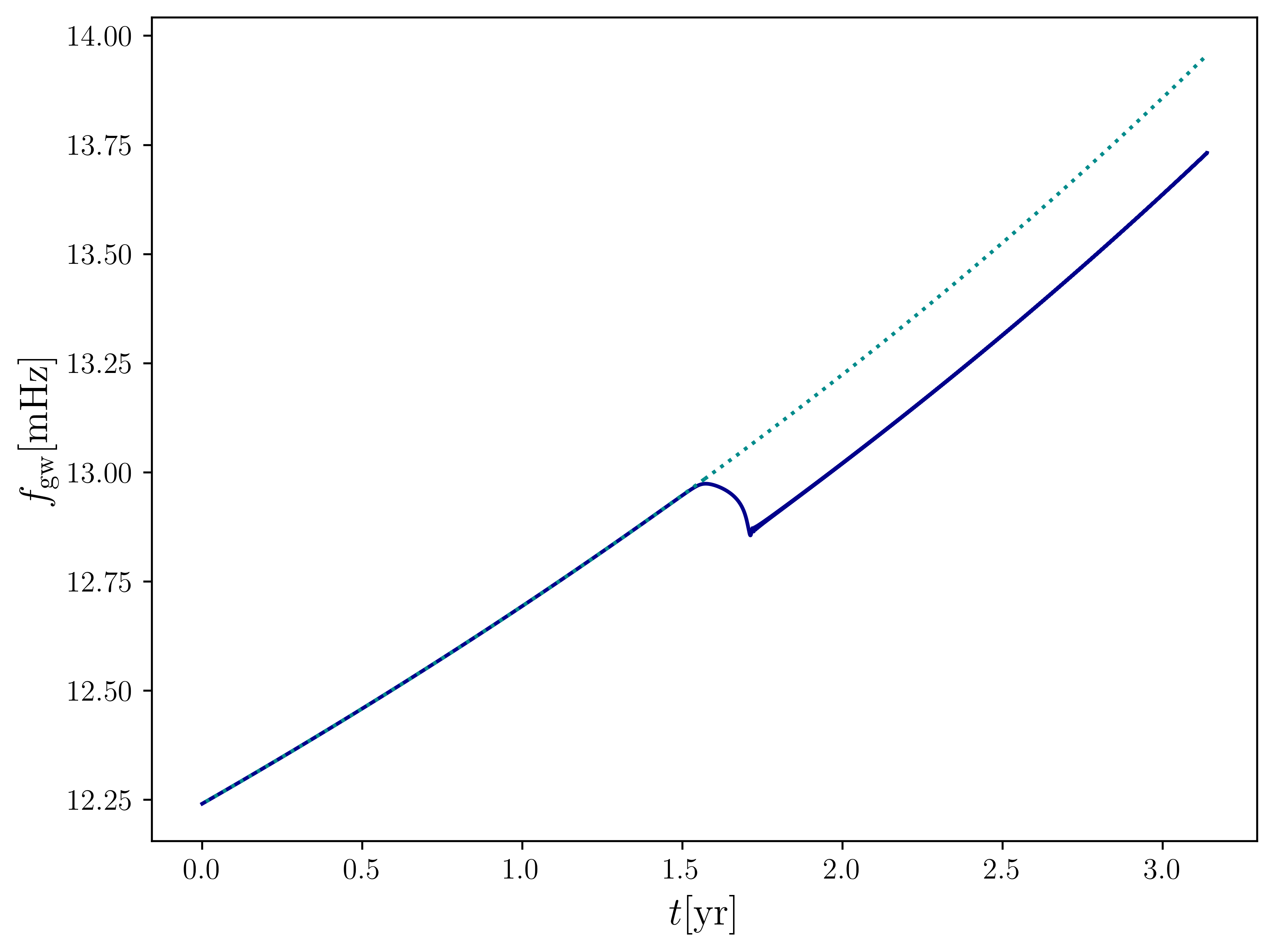}\qquad
  \includegraphics[height=6cm]{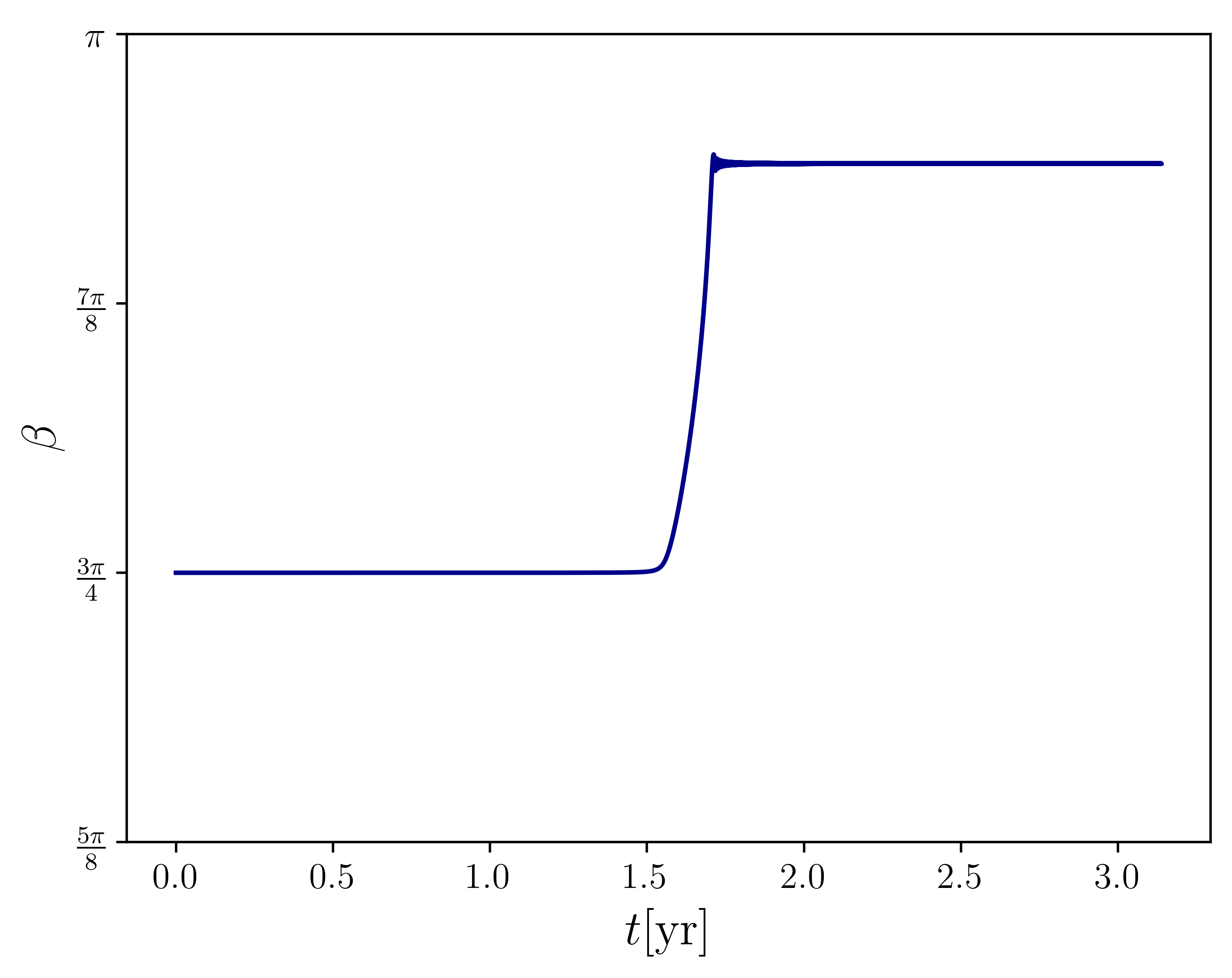}%
}

    \caption{Evolution of the peak GW
    frequency $f_\mathrm{GW}$ [\textit{left}] and the obliquity [\textit{right}] over the nominal LISA
    mission lifetime for the $\ket{433} \to \ket{421}$ $\mathcal{F}$-transition, with $\alpha = 0.28$,
    $(50+30)M_\odot$, $e_{\rm in}=0.1$, and $\beta_{\rm in} = 3\pi/4$.}
    \label{fig:433_band}
\end{figure*}

Finally, the most phenomenologically relevant scenarios for the $\ket{433}$ state are those in which resonance transitions occur within the detector band. As a representative example, consider a binary that forms close to the $g=1$ resonance band with $\beta_\mathrm{in} = 3\pi/4$, where it encounters the strongest transition, $\ket{433} \to \ket{421}$. In Fig.~\ref{fig:433_band} we show the evolution of the peak frequency  and obliquity, with the dotted curve on the left panel corresponding to the standard vacuum case. Strikingly, the right panel illustrates how the cloud-driven evolution of the obliquity is orders of magnitude larger than relativistic effects. (For instance, even assuming a non-negligible spin for the companion, which for simplicity we have ignored in this paper, the relativistic evolution of the obliquity, $\dot \beta_{\rm GR}$, would first enter at 2PN order.)

\subsection{EMRI/IMRI binaries} \label{sec:pheno_emri}

\begin{figure*}[t!]
\begin{tabular}{cc}
\includegraphics[width=.5\textwidth]{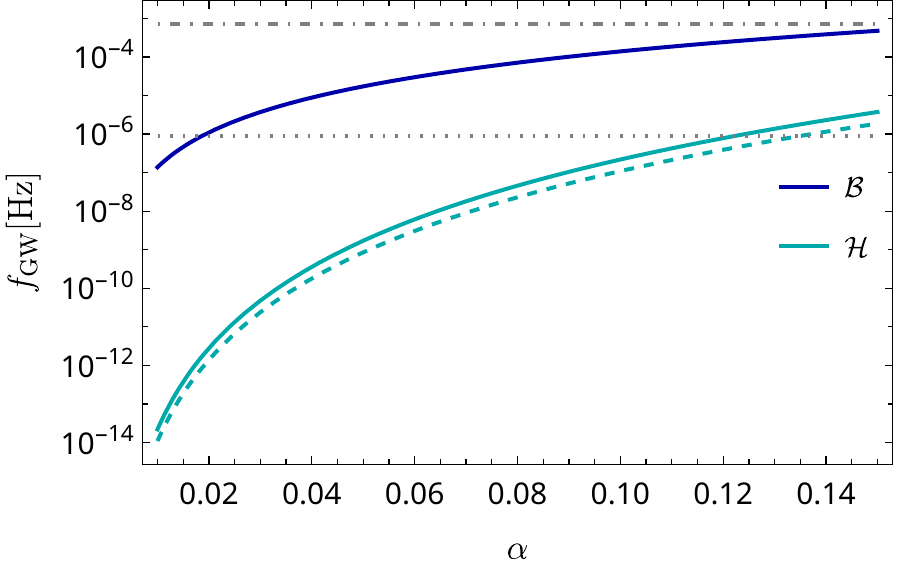}
\includegraphics[width=.5\textwidth]{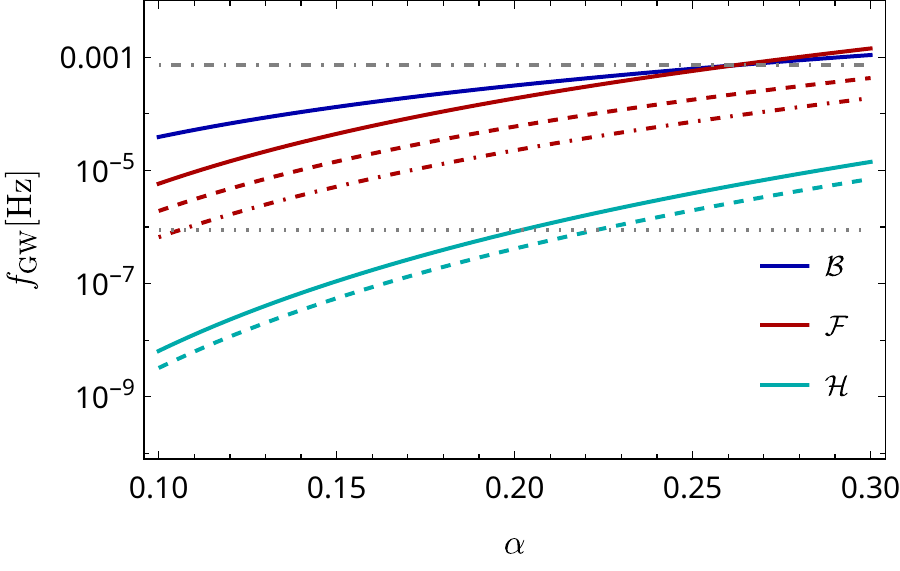}
\end{tabular}
    \caption{Position of the resonance transitions for $M =  10^4 M_\odot$. Same conventions as in Fig.~\ref{fig:freq}, and assuming $q\simeq 10^{-3}$.}
    \label{fig:freq_emri}
\end{figure*}

We organize the discussion similarly to the stellar case. However, in contrast to stellar binaries, for the $\alpha \lesssim 0.3$ range that we focus on in this paper, the $\ket{211}$ and $\ket{322}$ states are the only relevant ones for very massive BHs carrying a boson cloud (see Fig.~\ref{fig:freq_emri}). \vskip 4pt

 {\it \underline{Chronology of the $\ket{211}$ state.}} The window of support for the first excited state is $0.015 \lesssim \alpha \lesssim 0.11$ for the intermediate mass case, $M \simeq 10^4 M_\odot$, and $0.03 \lesssim \alpha \lesssim 0.15$ for bigger BHs, $M \simeq 10^6 M_\odot$, respectively. Generically, $\ket{211}$ states which are sensitive to the $\mathcal{H}$ regime do not coalesce within a Hubble time, and therefore are not relevant for our purposes here (see Fig.~\ref{fig:freq_emri} and App.~\ref{app:lines_res}). Likewise, early (sinking) $\mathcal{B}$-resonances, which do not lead to the disruption of the cloud, also lie  outside the band of present and future GW detectors. The fingerprints of the $\ket{211}$ state on GW observables are thus encoded in transitions occurring deeper into the $\mathcal{B}$ regime, where ionization effects become relevant. We do not discuss this possibility in this paper, see, e.g.,~\cite{Tomaselli:2024bdd, Dyson:2025dlj} for preliminary results.\vskip 4pt

{\it \underline{Chronology of the $\ket{322}$ state.}} 
Outside the $\alpha$-range relevant for the $\ket{211}$ state, and for $M \gtrsim 10^3\,M_\odot$, the $\ket{322}$ state remains stable throughout the $\alpha \lesssim 0.3$ interval considered here. In the window $0.1 \lesssim \alpha \lesssim 0.2$ (for the fiducial values in Fig.~\ref{fig:freq_emri}), binaries that may merge on timescales shorter than a Hubble time are primarily influenced by the $\mathcal{F}$ regime, which is typically encountered before the system enters the LISA band. For larger $\alpha$, binaries may additionally excite $\mathcal{H}$-resonances, depending on their birth frequency and on astrophysical constraints.\vskip 4pt 

\begin{figure*}[t!]
\begin{tabular}{cc}
\includegraphics[width=.48\textwidth]{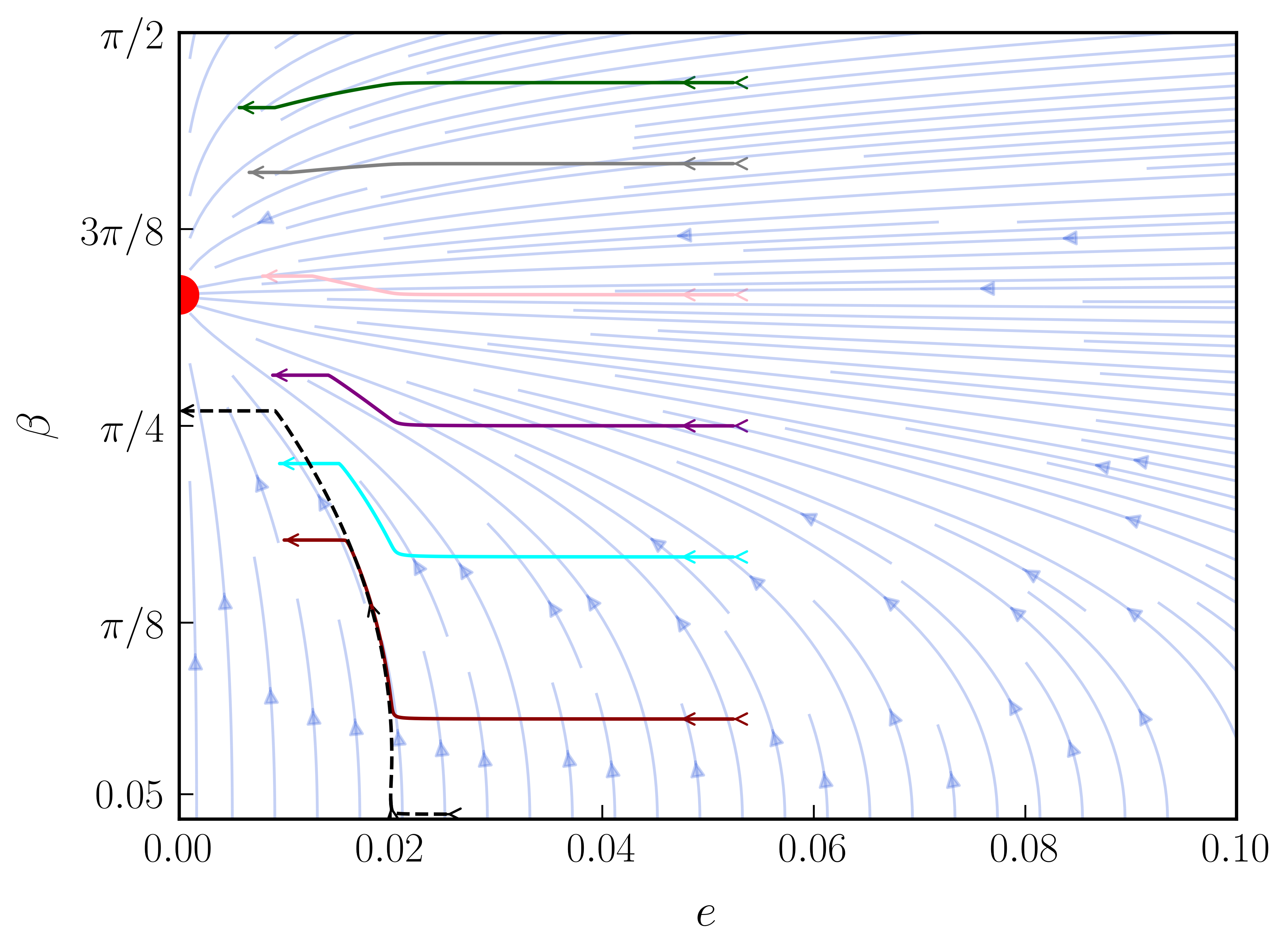}
\includegraphics[width=.48\textwidth]{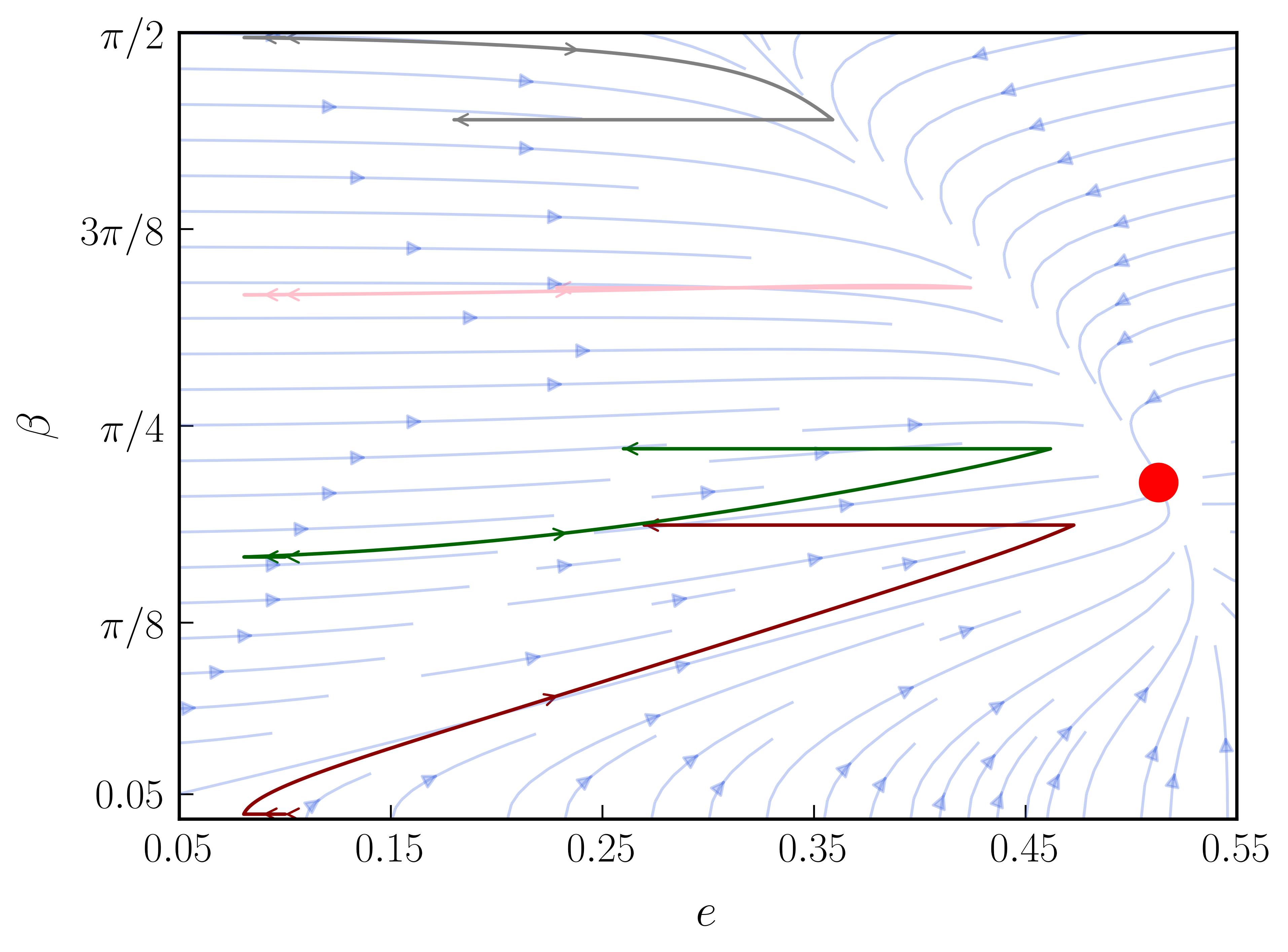}
\end{tabular}
   \caption{Eccentricity/obliquity flow for $(\alpha,q)=(0.3,10^{-3})$. We show on the left trajectories initialized near $\Omega^{(322,321)}_{2,0}$, for $e_{\rm in}=0.05$ and $\beta_{\rm in}\in\{0.2,\pi/6,\pi/4,\pi/3,5\pi/12,\pi/2-10^{-1}\}$ (solid), together with a near--co-rotating configuration with $e_{\rm in}\simeq 0.025$ and $\beta_{\rm in}=10^{-2}$ (black dashed). The numerical curves are overlaid on the analytic flow obtained under idealized floating conditions (lower-left quadrant of Fig.~\ref{fig:e_i_flow}). On the right we use initial conditions chosen near the  overtone $\Omega^{(322,321)}_{2,-1}$. We take $e_{\rm in}=0.1$ and $\beta_{\rm in}\in\{10^{-2},\pi/6,\pi/3,\pi/2-10^{-2}\}$, and superpose the resulting trajectories on the analytic prediction (lower-right quadrant of Fig.~\ref{fig:e_i_flow}).}
    \label{fig:emri_hf}
\end{figure*}

We examine two broad dynamical regimes, distinguished by whether the host binary is either \textit{(i)} shaped by the cloud's $\mathcal{H}$-transition sequence, or \textit{(ii)} born in the radiation-reaction--driven phase after the $\mathcal{H}$ regime has passed. Unlike in the stellar-mass case, scenario~\textit{(i)} is only weakly affected by the broad transition to $\ket{300}$, so that earlier overtones can proceed essentially uninterrupted. Yet, these transitions typically occur well before the LISA band, so any transient eccentricity growth is largely erased by the time the system becomes observable. The situation is markedly different for the obliquity. Because of the presence of fixed points, and the fact that a cloud-free inspiral exhibits no comparable changes, gravitational atoms can leave a pronounced and qualitatively distinct imprint on the observed values of~$\beta$.\vskip 4pt 
We consider two paradigmatic examples in Fig.~\ref{fig:emri_hf}.\footnote{For the cases shown in~Fig.~\ref{fig:emri_hf}, the condition that the binary reaches the LISA band within a  Hubble time requires $M \lesssim 5 \cdot 10^4 M_\odot$. For smaller $\alpha \simeq 0.2 - 0.25$, the backreaction is even larger (and thus the floating time). Consequently, the BH mass range that will not be filtered out from coalescing within a Hubble time is shifted downwards (see also App.~\ref{app:lines_res}).} Each trajectory exhibits three clearly separated stages. We initialize the system with orbital parameters chosen close to a resonant transition; the early evolution therefore proceeds as a standard, radiation-reaction--driven inspiral. At a later time, the orbit sweeps into the relevant resonance and the binary transitions into a quasi-floating phase. Finally, once the cloud has been sufficiently depleted, the inspiral resumes its vacuum-driven evolution, which we follow for a further interval up to the point of (almost) complete depletion.\vskip 4pt 

The left panel illustrates a configuration in which the binary enters the $\mathcal{H}$ regime with initial eccentricity $e_{\rm in}\ll 1$ and obliquity $\beta_{\rm in}<\pi/2$. In this case, the first (adiabatic) transition encountered is $\ket{322}\to\ket{321}$ with $g-k=2$ (since $e_{\rm in}\ll 1$ renders the early overtones ineffective), featuring a fixed point at $\{e_{\rm cr},\beta_{\rm cr}\}\simeq (0,\pi/3)$. We find that the semi-analytical (idealized) floating approximation in~\eqref{eq:e_i_emri} captures the dynamics remarkably well away from the fixed point (see the discussion in \S\ref{sec:incl}). For the cases with larger $\beta_\mathrm{in}$, initialized at $\Omega_{\rm in}=0.4\,\Omega^{(322,321)}_{2,0}$, GW emission from the cloud reduces the occupancy by a factor of $\simeq \mathcal{O}(10)$ before floating begins; nevertheless, the strong drive toward the fixed point remains clearly visible (solid curves). For a configuration with dynamical capture closer to resonance, $\Omega_{\rm in}=0.8\,\Omega^{(322,321)}_{2,0}$, we find that the push toward the fixed point is unavoidable even for a quasi-equatorial setup with $\beta_{\rm in}\simeq 10^{-2}$ (dashed curve).\footnote{For smaller values of $\beta_{\mathrm{in}}$, closer to the co-rotating case, the binary may be dominated by the 
$\ket{322} \to \ket{320}$ transition, whose fixed point lies at $\beta_\mathrm{cr}=0$ [cf.~\S\ref{sec:incl}].}
In the plot to the right, we consider a moderately eccentric binary initialized prior to the $g-k=3$ overtone, with $\Omega_{\rm in}=0.9\,\Omega^{(322,321)}_{2,-1}$. We observe the same eccentricity growth as in previous examples in \S\ref{sec:stellar}. However, these occur much before the LISA band, and can only leave a significant remnant value for $M\simeq 10^{3}\,M_\odot$. Crucially, the growth of obliquity toward the off-equatorial fixed point is a generic feature that persists even for larger BHs.\vskip 4pt

The features shown in Fig.~\ref{fig:emri_hf} correspond to configurations with initial obliquities $\beta_{\rm in}\lesssim \pi/2$. The qualitative picture remains unchanged in the complementary regime $\beta_{\rm in}\gtrsim \pi/2$, where trajectories typically alternate between phases of standard radiation-reaction--driven evolution and intervals in which the dynamics closely follows the analytic flow in Fig.~\ref{fig:e_i_flow} toward the fixed points. While resonant effects weaken as $\beta_{\rm in}$ approaches the counter-rotating limit (particularly for $e_{\rm in}\ll 1$), we nonetheless find that, through the combined action of narrow resonances, cloud GW emission, and the broad $\ket{322}\!\to\!\ket{300}$ transition, most binaries---including quasi-circular systems with $\beta_{\rm in}\simeq \pi$---undergo substantial depletion, reaching $N_\mathrm{c}/N_\mathrm{c}^{\rm sat} \lesssim 10^{-6}$, before the orbital frequency reaches $\Omega^{(322,300)}_0$.\vskip 4pt

Let us now turn to the scenario in \textit{(ii)}. As in the stellar-mass case, the presence of the wide transition $\ket{322} \to \ket{300}$, on top of the narrower $\mathcal{F}$-resonances, plays a crucial role. On the one hand, the effective width of the transitions is narrower than in the comparable-mass limit [cf.~\eqref{eq:nc_pert}]. On the other hand, the larger orbital backreaction slows down (speeds up) the inspiral for $\beta_\mathrm{in} \lesssim \pi/2$ ($\beta_\mathrm{in} \gtrsim \pi/2$), thus also prolonging (shortening) the depletion. At the same time, and unlike the stellar case, the stronger orbital backreaction enables narrower transitions $\ket{322} \to \ket{31m}$, $m \in [-1,1]$, to have a more prominent role. This complicates a straightforward repetition of our previous analyses.  
For illustrative purposes, we therefore highlight below a few representative phenomenological examples, and defer a more comprehensive investigation of the interplay between wide and narrow $\mathcal{F}$-transitions in the $q \ll 1$ limit to future work.
\vskip 4pt

\begin{figure*}[t!]
\begin{tabular}{cc}
\includegraphics[width=.48\textwidth]{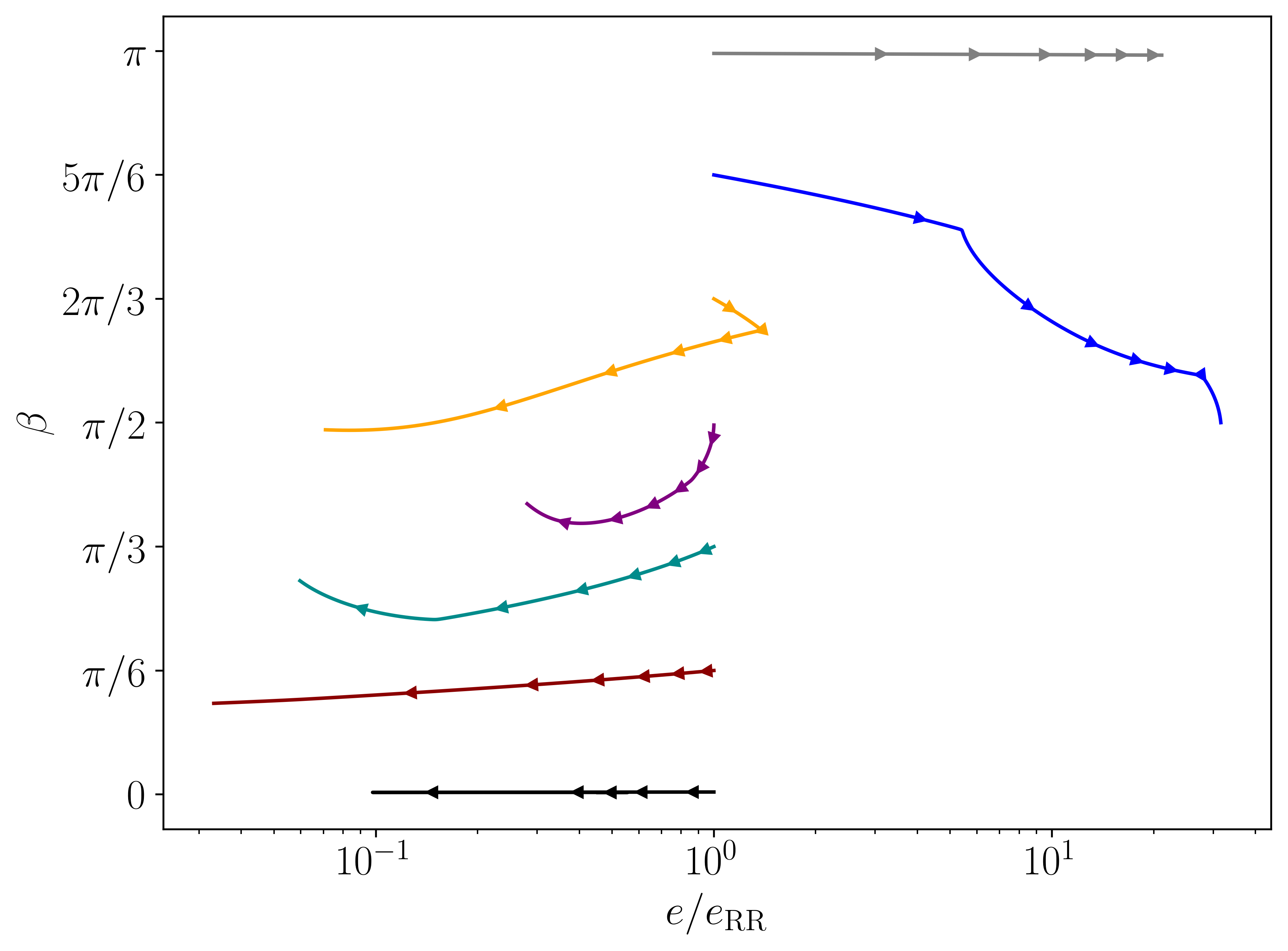}
\includegraphics[width=.48\textwidth]{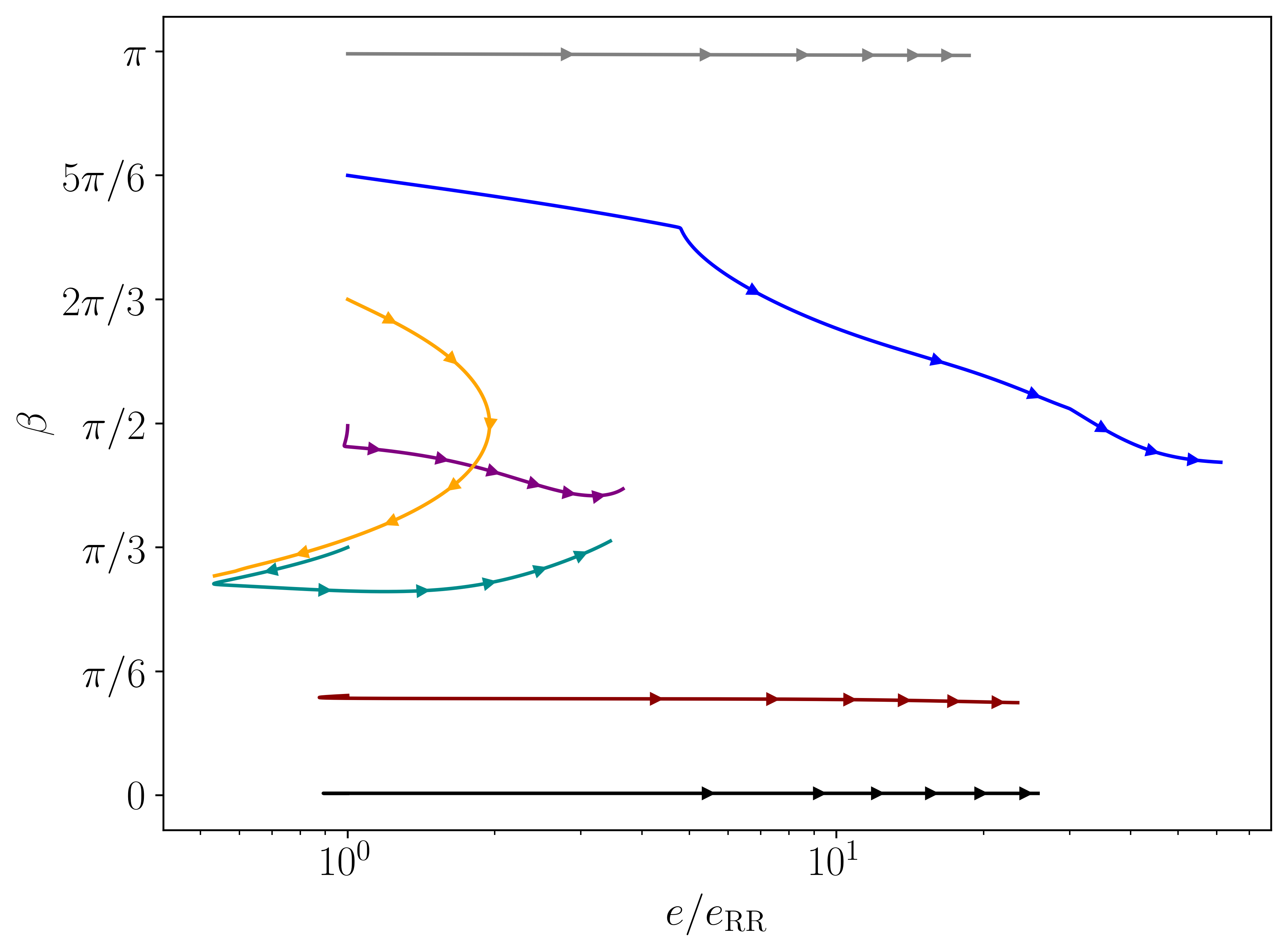}
\end{tabular}
   \caption{ Flow of $\{e/e_\mathrm{RR},\beta\}$ (relative to vacuum) due to $\mathcal{F}$-resonances combined with wide mixing. We set $\alpha=0.3$, $q=10^{-2}$, $\Omega_\mathrm{in}=5\times 10^{-2}\,\Omega^{(322,300)}_{0}$ and $e_\mathrm{in}=0.05$ [{\it left}], $e_\mathrm{in}=0.2$ [{\it right}]. (For illustration, black and red curves on the right are initialized at the onset of the narrow resonance.)}
    \label{fig:emri_fine}
\end{figure*}

Let us consider a binary with a BH of mass $M \simeq 10^3- 10^4 M_\odot$, carrying a boson cloud in an IMRI with $q = 10^{-2}$.  We initialize the system after the $\mathcal{H}$ regime, with $\Omega_\mathrm{in} = 5 \cdot 10^{-2} \, \Omega^{(322,300)}_{0}$, and evolve until the vicinity of $\Omega^{(322,300)}_0$. The obliquity/eccentricity flow (relative to radiation-reaction in vacuum) is shown in Fig.~\ref{fig:emri_fine} for two values of the initial eccentricity, $e_\mathrm{in}=0.05$ (left) and $e_\mathrm{in}=0.2$ (right).\vskip 4pt For the case of low eccentricity, and smallest values of $\beta_\mathrm{in}$ (black, red), the wide transition accelerates the depletion of both the cloud occupancy and the orbital eccentricity. The first adiabatic narrow resonance is $\ket{322} \to \ket{31-1}$, $(g,k)=(3,-1)$ ($(g,k)=(3,0)$) for black (red),\footnote{Although the $k=-1$ overtone is triggered for $\beta_\mathrm{in} \simeq 0$, the growth of eccentricity is only moderate.} yielding $N_\mathrm{c}/N_\mathrm{c}^{\rm sat}\simeq 10^{-7}$ at the end of the evolution. For the next two values of $\beta_\mathrm{in} \lesssim \pi/2$ (green, purple), the depletion of the cloud and changes in the eccentricity are comparatively slower. For these  ranges of the obliquity, narrow resonances are typically triggered, details of which (growth/decay of $e$ and $\beta$), are somewhat sensitive to the initial conditions. In all cases, we find that the cloud at the end of the evolution satisfies $N_\mathrm{c}/N_\mathrm{c}^{\rm sat} \lesssim 10^{-3}$. For the second-largest $\beta_{\rm in}$ (blue),  the interplay between degenerate overtones leads to significant eccentricity growth,\footnote{By analysing the idealised flows in \S\ref{sec:incl}, we find the horizontal asymptote (from above) $\{e \to 1 , \beta \to \pi/2\}$.} 
while the cloud is approximately depleted by $N_\mathrm{c}/N_\mathrm{c}^{\rm sat}\lesssim 10^{-2}$. Finally, for the counter-rotating case (gray), no strong narrow resonances are triggered, and the system is dominated by perturbative mixing. Near orbital frequencies of the order of $\Omega \simeq \Omega^{(322,300)}_0$ we find $N_\mathrm{c}/N_\mathrm{c}^{\rm sat} \simeq 1/2$. The strong  backreaction near $\Omega^{(322,300)}_0$ induces the growth of eccentricity shown in the plot (left). For this trajectory, the cloud can proceed to a sequence of early sinking-$\mathcal{B}$ transitions (which again do not deplete the cloud), and subsequently onward to the deep-$\mathcal{B}$ regime.\vskip 4pt

Increasing the initial eccentricity to $e_\mathrm{in}=0.2$ activates additional early-time overtones, including for near-equatorial configurations. The resulting changes in the system are reflected in the  flow chart shown in the right panel. Notably, binaries with small obliquity (black, red) can experience a pronounced growth in orbital eccentricity, outpacing the standard radiation--reaction evolution. While the behaviour at other obliquities depends more sensitively on the initial conditions, the $\beta \gtrsim \pi/2$ region exhibits a clear increase in eccentricity---due to either resonant-induced growth off the equatorial plane (blue), or wide-transition effects for counter-rotating orbits (gray).
 \vskip 4pt

\begin{figure*}[t!]
\begin{tabular}{cc}
\includegraphics[width=.48\textwidth]{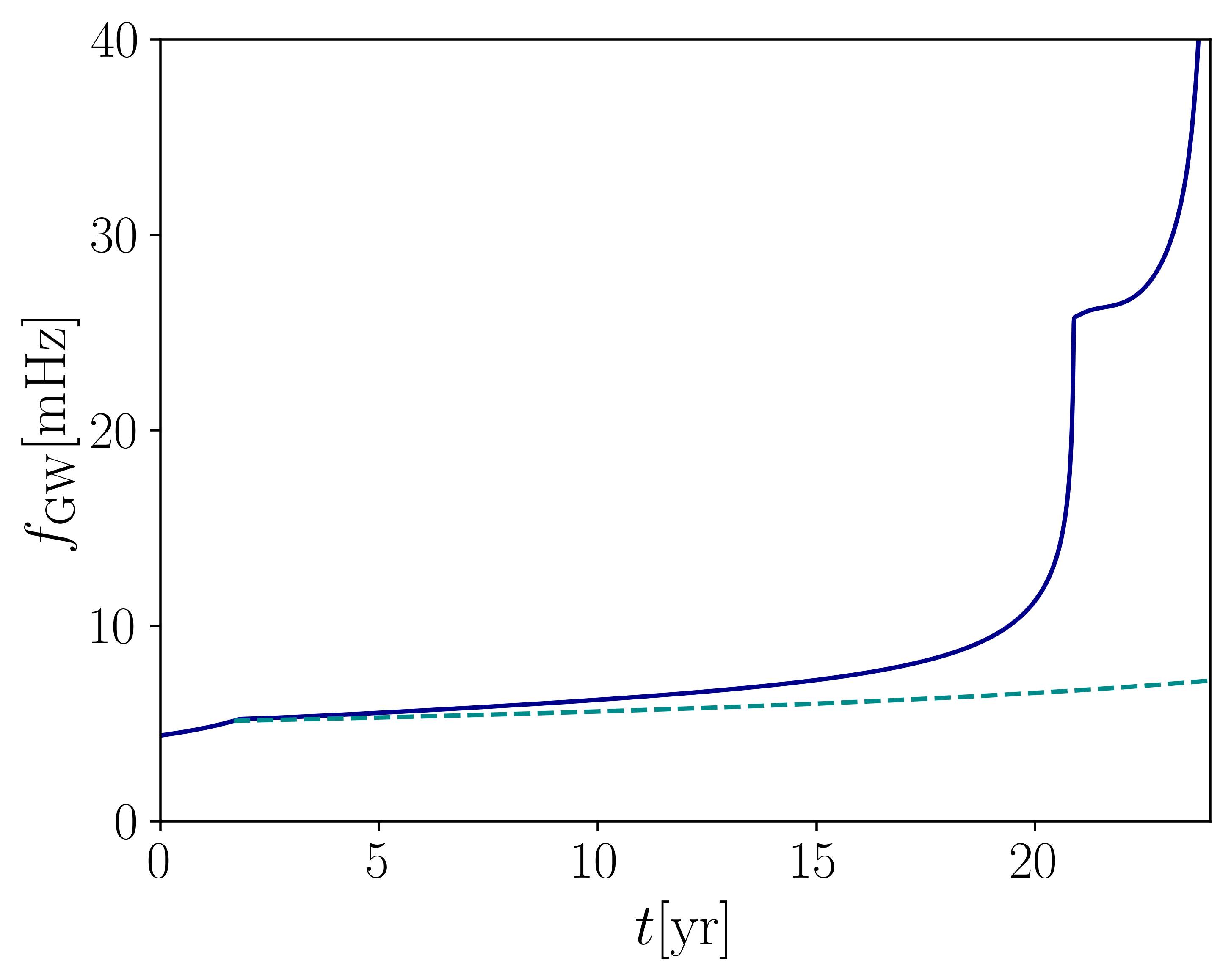}
\includegraphics[width=.48\textwidth]{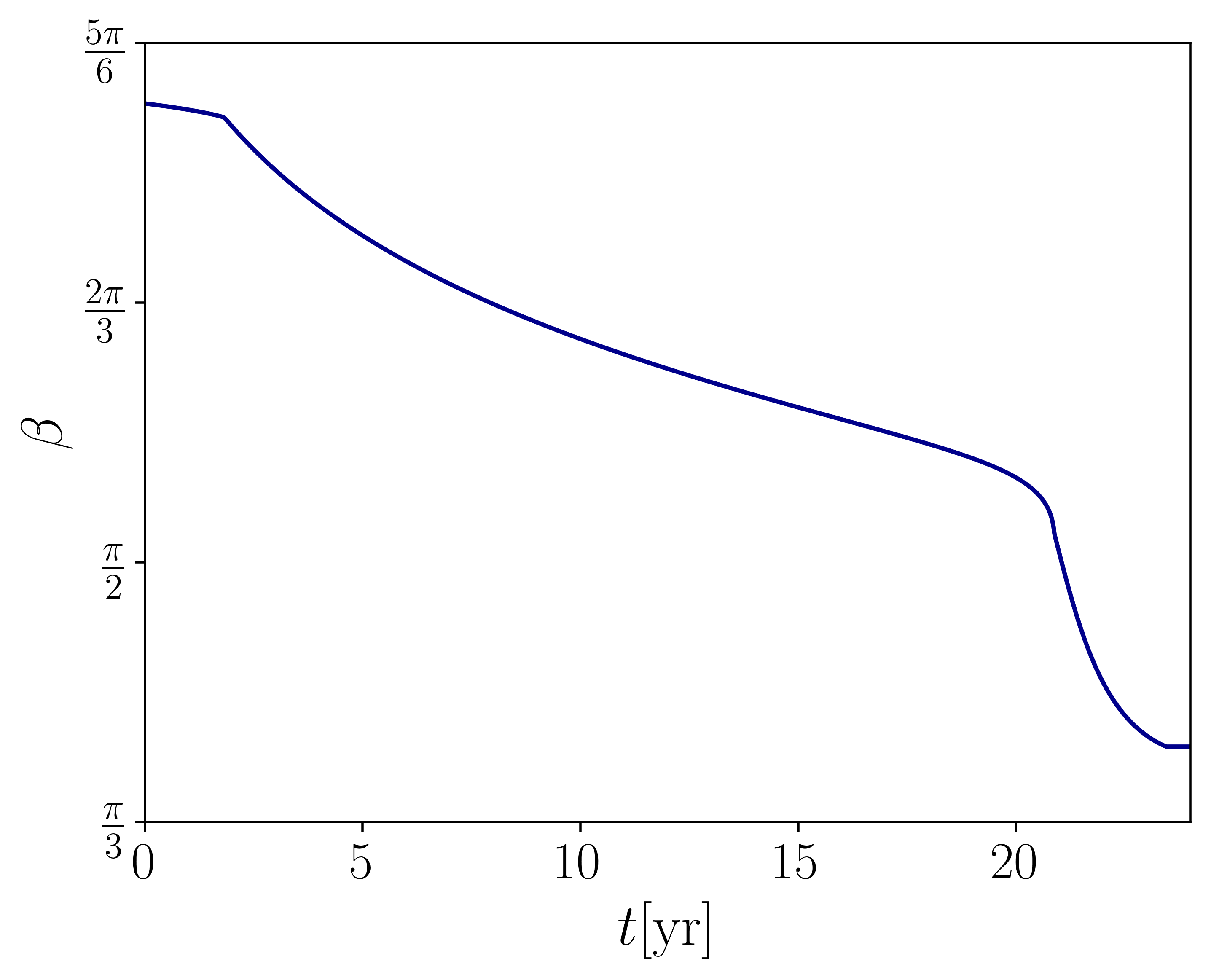}
\end{tabular}
   \caption{Peak GW
    frequency $f_\mathrm{GW}$ [\textit{left}] and obliquity [\textit{right}]  for the $\{\ket{322},\ket{311},\ket{300}\}$ mixing, with $\alpha = 0.3$,
    $(10^3+10)M_\odot$, $\Omega_{\rm in}=0.1 \Omega^{(322,300)}_0$, $\{e_{\rm in},\beta_{\rm in}\}= \{0.1, 5\pi/6\}$.}
    \label{fig:fine_emri_time}
\end{figure*}

The effects illustrated above become increasingly pronounced as the cloud--hosting binary starts its radiation-reaction--driven phase closer to the resonance $\Omega^{(322,300)}_0$, as well as for smaller mass ratios. The qualitative impact of a given transition also depends sensitively on the obliquity. For $\beta \lesssim \pi/2$, the wide transition alone can in principle induce a temporary outspiral, followed by a quasi-floating phase. In these situations, although comparatively stronger than in the stellar-mass case, the influence of narrower transitions is suppressed  and the coalescence time is prolonged, yielding an almost monochromatic signal in the LISA band.\footnote{In the IMRI regime, the typical frequency drift may still exceed $t_\mathrm{LISA}^{-1}$. See~\cite{Kim:2025wwj} for a dedicated study of the detectability of circular co-rotating orbits in the lower–mass IMRI regime.} In contrast, for $\beta \gtrsim \pi/2$, the wide transition accelerates the inspiral, opening up additional in-band observational opportunities. In Fig.~\ref{fig:fine_emri_time}, we show an example featuring a segment of the evolution of the peak frequency (left) and the obliquity (right), starting at the onset of a narrow ($g-k=1$) resonance and extending to the end of the (quasi-)floating period, shortly after which the occupancy has dropped to $N_\mathrm{c}/N^{\mathrm{sat}}_\mathrm{c} \lesssim 10^{-4}$. We observe a substantial growth of eccentricity accompanied by a decrease in obliquity, on timescales compatible with the LISA mission. This growth of eccentricity makes the peak frequency evolve much faster than in vacuum (cyan, dotted). The large final value of the eccentricity at the end of the transition, $e \simeq 0.7$, further accelerates the inspiral, causing the IMRI to exit the LISA band on a timescale of days.\footnote{Note that at the end of the resonance, the periastron distance remains parametrically separated from the innermost stable circular orbit, $[a(1-e)]_{\mathrm{res}^+}/r_\mathrm{ISCO} \simeq 25$, and thus our PN-based calculation remains valid. This may not hold in more extreme examples, where  counter-rotating resonances could precipitate the plunge.}

\section{Summary and Outlook} \label{sec:concl}

Provided they exist in nature, clouds of ultralight bosons surrounding BHs in binaries are inevitably transient phenomena---the question is not whether they fade away, but rather {\it when}~and~{\it how}. The timing determines if we witness their demise unfold within the frequency band of GW detectors, or whether they imprint measurable changes into the binary's orbital parameters relative to (cloud-free) vacuum evolution. The mechanism, in turn, fixes the character and strength of the observable signatures. Building on the worldline EFT approach \cite{Goldberger:2004jt,Goldberger:2005cd,Porto:2005ac,Porto:2007qi,Porto:2016pyg}, in this paper we developed a unified formalism that tracks the nonlinear dynamics of the gravitational atom and its host binary on generic orbits. Armed with this powerful framework, we demonstrated the existence of fixed points in the evolution of the eccentricity and obliquity---both for co- and counter-rotating (planar) orbits as well as off the equatorial plane---which allowed us to identify several orbital fingerprints and in-band GW signatures of gravitational atoms in binary systems.\vskip 4pt

{\it \underline{Novel features.}} We established the following key points for stellar binaries: 

\begin{itemize}
\item  Boson clouds in stellar binaries that form at low orbital frequencies---prior to the $\mathcal{H}$/$\mathcal{F}$ regime but merging within a Hubble time---are generically disrupted by resonance transitions, essentially independent of the initial eccentricity and obliquity.\footnote{Due to the absence of $\mathcal{F}$-transitions, clouds in the $\ket{211}$ state with $0.05 \lesssim \alpha \lesssim 0.1$ can survive the early inspiral, provided the orbit is nearly counter-rotating ($\beta_\mathrm{in} \simeq \pi$) and the initial eccentricity is small ($e_\mathrm{in} \ll 1$). More generally, if dynamical captures at higher orbital frequencies can be realised, boson clouds can also persist beyond the $\mathcal{H}/\mathcal{F}$ region and into the $\mathcal{B}$ region, irrespective of the initial state. In particular, for $\alpha < 0.05$, the boson cloud does not encounter adiabatic resonances prior to the $\mathcal{B}$ regime, see~Fig.~\ref{fig:freq}.}
This induces either a substantial modification of the in-band GW phase evolution or, if it occurs outside the direct observational window, leaves behind a characteristic \textit{trail} in the distribution of orbital parameters that may be observed at later times. See Figs.~\ref{fig:322_hyperfine}-\ref{fig:433_band}.

\item In-band transitions---whether broad or narrow---drive pronounced departures from the standard (cloud-free) GW phase evolution of stellar binaries. These arise not only from shifts in the peak frequency and the relevance of higher harmonics, but also from the evolution of the obliquity, which can outpace relativistic effects. 

\item Larger-than-expected values for the eccentricity---$e(10^{-2}\,\mathrm{Hz}) \gtrsim 0.01$---are observed in a significant fraction of co-rotating $(\beta_{\rm in}\simeq 0)$ stellar binaries with $q \lesssim 0.1$ (consistent with our findings in the Letter). Even greater values may be observed in the presence of dense clouds. Similar increase of the eccentricity develops for a portion of stellar binaries with $\beta_{\rm in} \lesssim  \pi/2$, $q\lesssim 0.1$, with growth of eccentricity correlated with an increase in the obliquity.

\end{itemize}
Whereas, for IMRI/EMRI binaries:
\begin{itemize}

\item Boson clouds around supermassive black holes, with $M \gtrsim 10^{6} M_\odot$, that are part of an EMRI at (very) low orbital frequencies, i.e., prior to the $\mathcal{H}/\mathcal{F}$-transitions, typically do not enter the detector band within a Hubble time. This remains true whether the cloud is dominated by the $\ket{211}$ state at small $\alpha$, or by $\ket{322}$ at larger values.\footnote{The (non-)resonant transitions prolong the inspiral for the later, whereas for the former the dominant effect arises from the orbital influence due to the cloud's own GW emission.} In contrast, for clouds in binaries that are born at higher frequencies, i.e., already past the $\mathcal{H}/\mathcal{F}$ regime, these may remain undisrupted until well into the $\mathcal{B}$ regime.

\item Binaries in which an intermediate-to-supermassive BH (with mass in the range
$M \simeq 10^3$--$10^6\,M_\odot$) is surrounded by a boson cloud during the early inspiral---sensitive to the $\mathcal{H}/\mathcal{F}$ regime---may ultimately reach the detector band. This is especially true at large $\alpha$, and, at the upper end of the mass range, provided $q \gtrsim 10^{-3}$. In most cases, the cloud is disrupted outside the direct observational window, but it may still leave behind observable signatures of its prior existence.\footnote{A significant portion of the cloud may however survive if the binary enters a radiation-reaction--driven phase right before the $\mathcal{F}$ regime, and with $\beta_\mathrm{in} \simeq \pi$ on a quasi-circular ($e_\mathrm{in} \ll 1$) orbit.}

\item The obliquity distribution exhibits strong resonance-driven structural changes with respect to stellar-mass binaries, together with a more direct correlation with orbital eccentricity. We identify two characteristic features in the resulting populations:
\begin{itemize}
    \item[$\star$] Binaries clustering on quasi-circular orbits---due to the large separation between the $\mathcal{H}$ transitions and the LISA band---with obliquities in the range $\beta \in [3\pi/8,\,\pi/4]$ (Fig.~\ref{fig:emri_hf}).
    \item[$\star$] Moderate-to-highly eccentric binaries grouped near $\beta \simeq \{0,\,\pi/2,\,\pi\}$ (Fig.~\ref{fig:emri_fine}). The increase in eccentricity over vacuum evolution is more robust for $\beta \gtrsim \pi/2$, whereas the $\beta < \pi/2$ region is more sensitive to wide mixing, which tends to lower the eccentricity on long timescales.

\end{itemize}

\item In-band transitions may occur in the IMRI regime. For cases with $\beta_{\rm in} \gtrsim \pi/2$, depending on the initial value, $e_{\rm in} \gtrsim 10^{-2}$, the resonance can trigger a substantial increase of the eccentricity that can speed up the binary's progression towards coalescence. The increase in eccentricity is correlated with a decrease in the obliquity~ (Fig.~\ref{fig:fine_emri_time}).

\end{itemize}

Our results provide a basis for turning both detections and null results into constraints on putative ultralight particles via precision~data, both for stellar and EMRI/IMRI binaries.\vskip 4pt

{\it \underline{Comparison with previous work.}} 
A first step toward characterizing the orbital dynamics of misaligned configurations was taken in~\cite{Tomaselli:2024bdd}, albeit restricted to the $q\ll 1$ limit. While we concur with some of the qualitative results---such as the disruption through  resonances for boson clouds formed in the early binary inspiral (at low frequencies), phenomena which we show here for the first time persists into the comparable-mass regime---the analysis in~\cite{Tomaselli:2024bdd} 
does not include all effects required for a comprehensive and self-consistent description of the system's full dynamics on generic orbits. In addition, we find that certain assumptions and simplifications, both  in~\cite{Tomaselli:2024bdd} and in our Letter~\cite{Boskovic:2024fga}, are not always justified. Moreover, although extremely useful as a guiding principle~\cite{Baumann:2019ztm}, the complete reliance on balance laws can hinder the proper description of the evolution of the system at all times. We~identify the following key issues:

\begin{itemize}
\item Unlike previous work, our present analysis incorporates the combined effects of GW emission from the cloud---qualitatively significant in the $\mathcal{H}$ regime---the dynamical evolution of the background BH parameters---which leads to a widening of the floating regime---and the introduction of `degenerate overtones', a concept that is crucial for capturing consistent resonant dynamics on generic orbits.
\item The use of balance laws---as originally introduced in~\cite{Baumann:2019ztm} and implemented in \cite{Boskovic:2024fga,Tomaselli:2024bdd}---is well justified for complete \emph{in--out} processes and for (very) narrow transitions. However, capturing the full dynamics requires the inclusion of the \emph{interaction} terms encoded in the Hamiltonian~\eqref{eq:hamiltonian_psi}. While these terms are less relevant for the $\mathcal{H}/\mathcal{F}$ transitions considered in~\cite{Boskovic:2024fga}, their influence grows as the width increases, leading to significant deviations, particularly for non-resonant mixing (see App.~\ref{app:flux_balance}).
\item As we have demonstrated here, wide transitions have a profound impact on the dynamics. Yet they were largely overlooked in the previous literature~\cite{Baumann:2019ztm,Boskovic:2024fga,Tomaselli:2024bdd} and only recently considered~\cite{Kim:2025wwj}, albeit restricted to co-rotating circular orbits. An attempt to include their effects for more generic configurations appeared in~\cite{Tomaselli:2025jfo}, although it still omits the crucial interaction term in the balance laws.
\item The coupled evolution of the spin and orbital angular momenta plays a central role in determining the dynamical equations governing the eccentricity and obliquity. These effects, which are overlooked in~\cite{Tomaselli:2024bdd}, modify the flow equations and position of the fixed points in the $\{e,\beta\}$ plane, away from $\beta=\{0,\pi\}$.~(See Fig.~\ref{fig:e_i_flow}.) Furthermore, although not the focus of our present work, precession significantly influences the position and duration of resonances deep in the $\mathcal{B}$ regime, even for planar orbits [cf.~\eqref{eq:Dchi_1pn}-\eqref{eq:Dchi_qm0}].
\end{itemize}These considerations show that the chronological sequence of resonances proposed in~\cite{Tomaselli:2024bdd} is altered in a non-trivial manner. First, while clouds are typically disrupted near co-rotating orbits, analogous disruption can also occur in the counter-rotating limit. Second, the presence of non-equatorial fixed points in the obliquity, together with the broadening of the transitions, substantially reshapes the resonance sequence proposed in~\cite{Tomaselli:2024bdd}. Consequently, as demonstrated throughout this work, the resulting in-band and off-band observational signatures of ultralight boson clouds around BHs in binaries are considerably richer than previously anticipated. \vskip 4pt

{\it \underline{Outstanding questions \& Prospects.}}  Notwithstanding the advances achieved in this work, several important questions remain open that deserve further investigation:
\begin{itemize}
\item We considered only gravitationally-bound clouds. In principle, self-interactions can influence both the bound-state spectrum and the superradiant growth~\cite{Gruzinov:2016hcq, Baryakhtar:2020gao, Witte:2024drg}. For~instance, assuming a standard axion-like potential, $V = \mu^2 f^2_\mathrm{a} \left(1 - \cos{\Psi/f_\mathrm{a}} \right)$, the effect of self-interactions on the superradiant evolution is subdominant for $f_\mathrm{a} \gtrsim (10^{-2} - 10^{-3}) m_\mathrm{Pl}$. For~smaller decay constants, on the other hand, self-interactions may dilute the cloud. Although a more systematic treatment is in principle needed, as we have shown here even  diluted clouds can dominate over the quadrupole moment of the parent BH [c.f.~\eqref{eq:Q_c_vs_bh}]. This motivates the extension of our framework to include additional couplings beyond the minimal models we studied here. 
\item We have ignored relativistic corrections to the bound-state spectrum, which become important for $\alpha\gtrsim 0.3$~\cite{Baumann:2019eav}. As discussed in~\cite{Cannizzaro:2023jle}, one must incorporate relativistic effects not only in the eigenvalues but also in the scalar eigenfunctions and the inner product. Although we do not expect these to modify our qualitative results, they may impact quantitative values. As shown in \S\ref{sec:stellar},  stellar binaries also populate higher-$l_a$ states~\cite{Arvanitaki:2014wva}, and relativistic effects may ultimately have a sizeable impact for large values of $\alpha$.
\item  In general, a full description of the dynamics requires accounting for the simultaneous participation of multiple levels. The resulting transition patterns are often intricate, potentially spanning several channels. In our examples, for each initial eccentricity and obliquity, we have restricted attention to cases where level overlap is not expected to introduce significant error (see App.~\ref{app:lines_res}). A comprehensive treatment that evolves all levels concurrently may ultimately be required to capture the system's full evolution and to validate the approximations employed here.
\item We have ignored self-gravity effects. As shown in~\cite{Kim:2025wwj}, depending on the density of the cloud, self-gravity can shift its energy levels and, for some $\mathcal{H}$-transitions, it can  even reverse the sign of the level splitting. So far this effect has been quantified only for  two (fastest growing) states, $l_a=\{1,2\}$, on co-rotating, equatorial orbits~\cite{Kim:2025wwj}. Let us emphasize, however, that prior to the $\mathcal{H}$-transitions, the inspiral is slow enough for the cloud's GW emission to become relevant, reducing its mass by up to an order of magnitude and thus suppressing self-gravity effects. In addition, because the backreaction parameters scale as $1/q$ in the $q \ll 1$ limit, even a diluted cloud can still leave a notable imprint on the orbital dynamics. This motivates a more systematic analysis that robustly characterises the $\mathcal{H}$ regime for generic cases.

\item In general, the expansion in~\eqref{eq:psi_expansion} is incomplete: a tidal perturber also sources scalar radiation~\cite{Baumann:2021fkf,Tomaselli:2023ysb,Tomaselli:2024bdd,Tomaselli:2025jfo}. Although this channel does not significantly affect the $\mathcal{H}$ and $\mathcal{F}$ regimes, it is amplified in the deep $\mathcal{B}$ regime, inducing strong backreaction analogous to tidal disruption~\cite{Zhou:2025lzg}. 
This has been explored using numerical relativity in~\cite{Guo:2025pea,Roy:2025qaa,Cheng:2025wac}, and self-force methods in~\cite{Brito:2023pyl,Duque:2023seg,Dyson:2025dlj}. Recent advances within the EFT approach~\cite{Modrekiladze:2024htc} suggest a path to analytic control that merits further study.
\item Searching for imprints of scalar clouds in binary BH orbital parameter distributions demands going beyond the mock population study explored here. Building on well-motivated astrophysical priors, one must contrast the orbital evolution across three key scenarios: vacuum binaries, binaries embedded in baryonic environments (where relevant), and systems hosting scalar clouds. Such an analysis would sharpen our ability to disentangle features uniquely induced by ultralight degrees of freedom from a dark sector, with particular emphasis on the emergence and structure of fixed points in the joint evolution of eccentricity and obliquity.
\item Dedicated waveform models encompassing all the relevant physical phenomena are thererfore essential to bring our results to ready-to-use form. (For instance, as in~\cite{Chia:2020psj,Shterenberg:2024tmo}, by incorporating systems with large quadrupole moments.) As we have demonstrated here, the worldline EFT framework offers a natural (and efficient) way to include finite-size and environmental effects, which will be accessible with next-generation GW detectors. 
\end{itemize}

We will return to these issues in greater detail in future work.
\vskip 4pt {\bf Acknowledgements.} During the 
long gestation of this paper, we have benefitted from discussions with Hyungjin Kim, Alessandro Lenoci and Giovanni Tomaselli. MB~is grateful to Nikola Savić and Clemente Smarra for discussions on related topics. The work of MB and MK was supported in part by the Deutsche Forschungsgemeinschaft (DFG, German Research Foundation) under Germany's Excellence
Strategy – EXC 2121 ``Quantum Universe" – 390833306. MB and RAP were supported in part by the ERC-CoG ``Precision Gravity: From the LHC to LISA" provided by the European Research Council under the European Union's H2020 research and innovation program (grant No. 817791). 

\appendix

\section{Flow of orbital elements} \label{app:cel_mech}

{\it \underline{Phase space.}} 
In celestial mechanics it is customary to encode the orbital state in phase space through a set of (generally non-canonical) orbital elements $\mathbb{E}$, which possess a direct geometric interpretation.  
Alternatively, one may employ canonical action-angle variables, such as the Delaunay pairs $\mathbb{D} \equiv \{(\vartheta,\Lambda), (\chi,L), (\Upsilon,L_z)\}$, where
\begin{equation} \label{eq:delaunay_lambda}
\Lambda \equiv \frac{\mathcal{M}^{3/2}q}{\sqrt{1+q}} \sqrt{a}
\end{equation}
is the canonical action associated with the semi-major axis.  
In this language, the symplectic structure appearing in~\eqref{eq:orb_evo_H} reads
\begin{equation}
\hat{\mathbb{M}}
   = \hat{\mathbb{G}}^{\mathrm T} \hat{\mathbb{J}} \hat{\mathbb{G}}
   \,,\qquad
   \hat{\mathbb{G}} \equiv \frac{\partial \mathbb{E}}{\partial \mathbb{D}}
   \,,\qquad
   \hat{\mathbb{J}} \equiv 
   \begin{bmatrix}
      0 & \hat{\mathbb{I}}_n \\
      -\hat{\mathbb{I}}_n & 0
   \end{bmatrix},
   \label{eq:jacobian}
\end{equation}
where $\hat{\mathbb{G}}$ is the Jacobian for the transformation $\mathbb{D}\!\to\!\mathbb{E}$, 
$\hat{\mathbb{J}}$ is the canonical symplectic form, 
$\hat{\mathbb{I}}_n$ the $n$-dimensional identity matrix, 
and $n=3$ is the number of canonical pairs~\cite{Tremaine_Dynamics}.   
Hence, if the effective mass parameter ${\cal M}(1+q)$ evolves only adiabatically in time, $\Lambda$ remains an adiabatic invariant, $\dot{\Lambda}\simeq0$. \vskip 4pt 

If the perturbation enters through the Hamiltonian, the evolution of the orbital elements follows from Lagrange’s planetary equations~\cite{Tremaine_Dynamics} \,[cf.~\eqref{eq:a_gen}–\eqref{eq:ups_gen}]. On the other hand, 
for dissipative (or non-Hamiltonian) perturbations, we can instead express $\dot{\mathbb{E}}$ in terms of the perturbing force via Gauss variational equations~\cite{Tremaine_Dynamics,1976AmJPh..44..944B}.  
Since $\dot{\mathbb{E}}$ depends \emph{linearly} on the non-Keplerian perturbation, the total evolution is simply the superposition of all perturbations acting on the orbit. This description adopts the \emph{osculating} viewpoint that if the perturbation were switched off at time $t$, the trajectory would instantaneously revert to a Keplerian orbit whose initial conditions are given by the instantaneous values $\mathbb{E}(t)$. While the perturbation is active, however, the motion is in general not Keplerian, and all orbital elements evolve non-trivially---including the mean anomaly, such that $\dot{\vartheta}\neq\Omega$.\footnote{Notice that the angular elements $\chi$ (argument of pericenter) and $\Upsilon$ (longitude of ascending node) become ill-defined in the limits $e\to0$ and $\iota\to0$, respectively.   
In these regimes one may switch to alternative sets of (non-)canonical elements~\cite{Tremaine_Dynamics}, or work directly with the Keplerian conserved quantities (see App.~\ref{app:flux_balance}).}\vskip 4pt

\begin{figure}
    \centering
    \includegraphics[width=0.7\linewidth]{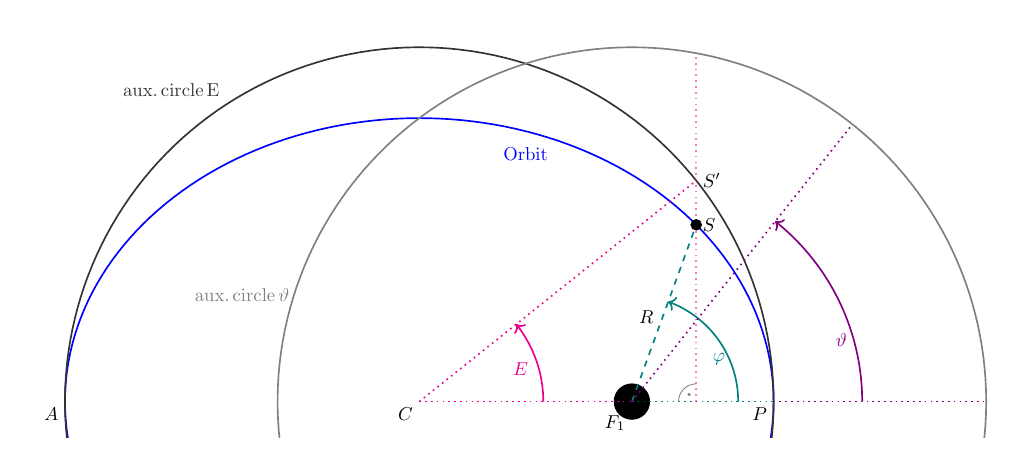}
    \caption{The three anomalies in an eccentric orbit.  
    The central object is at $F_1$ and the companion at $S$, moving along the blue orbit.  
    The periapsis is $P$ and the apoapsis $A$.  
    The true anomaly $\varphi$ is the angle between $\overline{F_1P}$ and $\overline{F_1S}$.  
    The auxiliary circle used to define the eccentric anomaly $E$ is centered at $C$ and has radius $a$.  
    The eccentric anomaly $E$ is the angle between $\overline{CP}$ and $\overline{CS'}$, where $S'$ is the point on the auxiliary circle lying on the line through $S$ that is perpendicular to $\overline{CP}$.}
    \label{fig:anomalies}
\end{figure}

{\it \underline{Eccentric anomaly.}} For general Keplerian orbits it is often useful, in addition to the mean ($\vartheta$) and true ($\varphi$) anomaly, to introduce $E$, the eccentric one~\cite{Tremaine_Dynamics}. The latter is connected to the orbital radius through the relation $R = a(1 - e \cos E)$, and to the mean anomaly via (Kepler's equation)
\begin{equation} \label{eq:Kepler_eq}
\vartheta = E - e \sin E \,.
\end{equation}
The relationship between $E$ and $\varphi$ follows from the identities
\begin{equation} \label{eq:phi_via_E}
\sin \varphi = \frac{\sqrt{1-e^2}\,\sin E}{1 - e \cos E}\,,
\qquad 
\cos \varphi = \frac{\cos E - e}{1 - e \cos E} \,.
\end{equation}
We illustrate the connection between true, eccentric, and mean anomaly in Fig.~\ref{fig:anomalies}.

\vskip 4pt

\vskip 4pt

{\it \underline{Adiabatic approximation.}} 
Typically, we are interested in the secular evolution on time scales $\gg 2\pi/\Omega$, i.e.~much longer than the orbital period.  
On such time scales, Lagrange’s planetary equations may be averaged over the fast orbital motion.  
Equivalently, one may first perform an orbit-average of the non-Keplerian part of the Hamiltonian and then insert the result into Lagrange’s equations~\cite{Tremaine_Dynamics,Nabet:2014kva}:
\begin{equation}
 \big\langle H_{\mathrm I} \big\rangle
 \;\equiv\;
 \frac{1}{2\pi} \int_{0}^{2\pi} \! d\vartheta \;
 H_{\mathrm I}\big(\vartheta;\,\mathbb{E}\setminus\{\vartheta\}\big) \,,
 \label{eq:H_I_avg}
\end{equation}
where we treat the remaining orbital elements $\mathbb{E}\setminus\{\vartheta\}$ as constant over a single orbital period.  
A successful orbital averaging of $H_{\mathrm I}$ immediately implies 
$\langle \dot{a} \rangle = 0$ for conservative perturbations, since the explicit dependence on the mean anomaly $\vartheta$ is removed; cf.~\eqref{eq:a_gen}.

In principle, the presence of an additional dynamical timescale with characteristic frequency comparable to or higher than $\Omega$ can obstruct this averaging procedure (for instance, near resonances or when the perturbation varies rapidly in time).  
Throughout this work we implement orbital averaging wherever it is justified, without always making the notation $\langle \cdots \rangle$ explicit.  
For approaches that go systematically beyond the adiabatic approximation---particularly relevant for highly eccentric or nearly plunging orbits with $1-e \lesssim 0$---see, e.g.,~\cite{Damour:2004bz,Pound:2007th,Galley:2016zee,Fumagalli:2025rhc}.\vskip 4pt

\section{Spin and positional elements} \label{app:spin}
{\it \underline{Euler angles.}} 
To track the orbital and spin dynamics with respect to an inertial reference frame, specified by the unit vector $\bm{n}$, we recast the parameters $\{\kappa,\beta,\xi\}$ [cf.~\eqref{eq:wigner}] in terms of the Euler angles associated with the rotations
$\frak{R}(\bm{n},\bm{L}) \equiv \{\Upsilon,\iota,\chi\}$ and 
$\frak{R}(\bm{n},\bm{{\cal S}}) \equiv \{\altupsilon,\altiota,\altchi\}$,
through the composition rule
$\frak{R}(\bm{{\cal S}},\bm{L})=\frak{R}(\bm{n},\bm{L})\,[\frak{R}(\bm{n},\bm{{\cal S}})]^{-1}$.
Without loss of generality, we fix the remaining gauge freedom by choosing $\altchi=0$.

More explicitly, to pass from the frame aligned with $\bm{n}$ to the frames aligned with $\bm{L}$ or $\bm{{\cal S}}$, we use a passive $zyz$ Euler rotation (in that order),
\begin{equation}
\frak{R}(\frak{a},\frak{b},\frak{c}) \equiv
\begin{pmatrix}
 \cos \frak{c}  & \sin \frak{c}  & 0 \\
 -\sin \frak{c} & \cos \frak{c}  & 0 \\
 0              & 0              & 1
\end{pmatrix}
\begin{pmatrix}
 \cos \frak{b}  & 0 & -\sin \frak{b} \\
 0              & 1 & 0              \\
 \sin \frak{b}  & 0 & \cos \frak{b}
\end{pmatrix}
\begin{pmatrix}
 \cos \frak{a}  & \sin \frak{a} & 0 \\
 -\sin \frak{a} & \cos \frak{a} & 0 \\
 0              & 0             & 1
\end{pmatrix} ,
\end{equation}
where $(\frak{a},\frak{b},\frak{c})$ denote, respectively, the Euler angles for
$\{(\bm{n},\bm{L}),\,(\bm{n},\bm{{\cal S}}),\,(\bm{L},\bm{{\cal S}})\}$ introduced above.
We caution the reader that, in the $zyz$ convention for $(\bm{n},\bm{L})$, the final $\chi$-rotation does not land on the periastron direction, but rather on a direction shifted by $\pi/2$.
Since this is a constant offset, it plays no role in the dynamical content of the main text and we suppress it throughout.\footnote{This is the reason why the $zxz$ convention is often adopted in celestial mechanics, where $\chi\big|_{zyz}=\chi\big|_{zxz}+\pi/2$ and $\Upsilon\big|_{zyz}=\Upsilon\big|_{zxz}-\pi/2$. 
On the other hand, the $zyz$ convention is particularly convenient here, as it renders the Wigner small-$d$ matrices purely real~\cite{1987AuJPh..40..465M}. 
Finally, we caution the reader that our convention for Wigner's small-$d$ matrix, $d^{(l)}_{mg}$, differs from the one implemented in \texttt{Mathematica}~\cite{Mathematica}, which corresponds to $\mathrm{\texttt{WignerD}}[\{l,-m,-g\},\frak{a},\frak{b},\frak{c}]$.}
From $\frak{R}(\bm{{\cal S}},\bm{L})=\frak{R}(\bm{n},\bm{L})\,[\frak{R}(\bm{n},\bm{{\cal S}})]^{-1}$ we may then express the obliquity in terms of the $\{(\bm{n},\bm{L}),\,(\bm{n},\bm{{\cal S}})\}$ angles as
\begin{equation} \label{eq:obliquity}
\cos\beta = \cos \iota \cos \altiota + \cos\!\left( \Upsilon - \altupsilon \right)\sin \iota \sin \altiota \,,
\end{equation}
with analogous, albeit more cumbersome, relations for $\{\kappa,\xi\}$.\vskip 4pt 

{\it \underline{Spin dynamics.}} 
The multipolar couplings encoded in $V_{\mathcal{Q}}$ induce the following evolution equation for the spin of the gravitational atom~\cite{poissonwillbook,Goldberger:2020fot}:
\begin{eqnarray}
\dot{{\cal S}}^j_{\cal Q} &=& \left( \frak{R}(\bm{n},\bm{{\cal S}})^{-1} \right)^j_a \,
\epsilon^{abc} \sum^\infty_{l=1} \frac{M_\star}{l!} \,
{\cal Q}^{\langle b L \rangle}\,
\partial_{\langle c L \rangle} \left( \frac{1}{R} \right) \,.
\end{eqnarray}
Upon shifting $l\to l+1$, and using~\eqref{eq:ir_multi} together with standard STF identities,\footnote{One such non-trivial identity is
\[
{\hat s}^{\langle aL\rangle} n^{\langle bL\rangle}
= \frac{l!}{(2l+1)\,(2l+1)!!}
\Big[
  K_1 \delta^{ab}
  + K_2 \big( {\hat s}^a {\hat s}^b + n^a n^b \big)
  + K_3\, {\hat s}^{(b} n^{a)}
  + K_4\, {\hat s}^{[b} n^{a]}
\Big] \,,
\]
where $\{\bm{\hat s},\bm{n}\}$ are unit vectors and $K_1 = d_x P_l$, $K_2 = - d_x^2 P_{l+1}$, $K_3 = 2 d_x^2 P_l + (2l+1) d_x P_{l+1}$, $K_4 = (2l+1) d_x P_{l+1}$, with $x = {\bm{\hat s}}\cdot \bm{n}$. See e.g.~\cite{poissonwillbook} for a review.}
we recover the compact form reported in~\eqref{eq:spin_dynamics}.

For illustrative purposes, consider the gravitational atom in a given state (without mixing)[cf.~\S\ref{sec:1l_atom}]. 
In this case the body is axisymmetric, ${\cal Q}^{\langle aL \rangle}=q_l\,{\hat s}^{\langle aL \rangle}$, with
$q_l=\sqrt{\tfrac{4\pi}{2l+1}}\,{\cal Q}_{l0}$, and $\bm{\hat s}$ denotes the unit vector along the symmetry axis.
The integral in~\eqref{eq:spin_dynamics} can then be carried out explicitly, yielding
\begin{eqnarray}
\frac{d\bm{\mathcal{S}}}{dt} 
= \sum_{l \geq 2} \frac{M_\star}{R^{l+1}}\,(-1)^l\, q_l \,
\frac{d P_l(x)}{d x} \Big|_{x=\bm{\hat{s}} \cdot \bm{\hat{R}}} \,.
\end{eqnarray}
We may further expose the qualitative content of this equation by performing an adiabatic average. %
Retaining, for simplicity, only the leading $l=2$ contribution, and after straightforward manipulations, we obtain
\begin{eqnarray} \label{eq:spin_1l}
\dot{\bm{\mathcal{S}}} = \bm{\varOmega}_{{\cal Q}{\cal S}} \cross \bm{\mathcal{S}} \,,\quad  
\bm{\varOmega}_{{\cal Q}{\cal S}} \equiv  \frac{3 M_\star}{2 a^3 \sqrt{1-e^2}^3}\, q_2 \cos{\beta}\, \frac{\bm{J}}{L \mathcal{S}} \,, \quad 
\bm{J} \equiv \bm{L}+ \bm{{\cal S}} \,,
\end{eqnarray}
in agreement with the standard spin dynamics of axisymmetric celestial bodies; see, e.g.,~\cite{Tremaine_Dynamics,poissonwillbook}. 
(Notice that the same result also follows from~\eqref{eq:spin_balance}.)\vskip 4pt

\underline{\it{Precession.}} As shown in~\eqref{eq:spin_1l}, precession effects are an important part of the spin dynamics, even within a single-state atomic description. The same parametric scaling, however, continues to apply in a multi-level setting. This will allow us to generalize  scaling estimates to assess the impact of precession effects in more realistic, multi-level scenarios. We begin by comparing the relevant timescales. Using~\eqref{eq:spin_1l} combined with~\eqref{eq:RR_a}, and evaluating at orbital separations near resonance transitions [c.f.~\eqref{eq:res_positions}], we obtain
\begin{eqnarray}
\frac{t_\mathrm{RR}}{t_{{\cal Q}{\cal S}}} &\sim&
\frac{\alpha^{(-2p-9)/3}}{(1+q)^{2/3}}
\frac{1-e^2}{1+\tfrac{7}{8}e^2} \,,
\hspace{2.8cm} (\mathcal{S} \gg L) \,, \nonumber \\
&\sim&
\frac{\alpha^{-3-p/3}}{q(1+q)^{1/3}}
\frac{\sqrt{1-e^2}}{1+\tfrac{7}{8}e^2}
\left(\frac{M_\mathrm{c}/M}{\alpha}\right) \,,
\qquad (\mathcal{S} \ll L) \,,
\label{eq:tRRvsQ}
\end{eqnarray}
where $t_\mathrm{RR} \equiv L/[\dot{L}]_\mathrm{RR}$ and $t_{{\cal Q}{\cal S}} \equiv \mathcal{S}/[\dot{\mathcal{S}}]$ are the radiation-reaction and precession scales, respectively. In both limiting cases $(\mathcal{S} \ll L\,,  L\gg \mathcal{S})$ we find that $t_\mathrm{RR}\gg t_\mathcal{{\cal Q}{\cal S}}$ throughout the parameter space of interest. A similar estimate (for $e \lesssim 0.6$) yields $t_\mathrm{RR} \gg t_\mathrm{SO}$~\cite{Fumagalli:2023hde}, with $t_{\rm SO}$ the radiation-reaction scale associated with the standard spin-orbit coupling (which contributes to precession at 1.5PN order, e.g.,~\cite{Porto:2005ac,Nabet:2014kva}).\vskip 4pt

In general, over the long timescales associated with resonant transitions we find
$t_{\mathrm{float}} \gg t_{\mathrm{RR}}$, which precludes the use of the total angular
momentum as a fixed reference axis. (Indeed, this is one of our main points of
disagreement with the treatment in~\cite{Tomaselli:2024bdd}.) Nevertheless, continuing with the one-level system---where resonant transitions are absent---as a paradigmatic example,  it is then convenient to adopt a
frame aligned with the total angular momentum, $\bm{n}=\hat{\bm J}$, which streamlines
the derivation, yielding
$\dot{\Upsilon}\big|_{\mathcal Q}=\dot{\altupsilon}\big|_{\mathcal Q}=\Omega_{{\cal Q}{\cal S}}$.\footnote{Moreover, unlike in a generic $\bm n$-frame, aligning with $\bm J$ implies
$\partial_t \iota\big|_{\mathcal Q}=\partial_t \altiota\big|_{\mathcal Q}=0$; whereas in
general one would have $\partial_t \iota\big|_{\mathcal Q}\neq 0$ and
$\partial_t \altiota\big|_{\mathcal Q}\neq 0$. In both cases we find
$\partial_t \beta\big|_{\mathcal Q}=0$, since here we consider only conservative
interactions that keep the obliquity constant.} The relative strength of the precession effects discussed in the following remain parametrically valid even in multi-state scenarios. \vskip 4pt

Two observations follow immediately. First, in the EMRI limit (${\cal S}\gg L$) we obtain
$\dot{\Upsilon}\big|_{\mathcal Q}/\dot{\Upsilon}\big|_{\mathrm{SO}}\sim \alpha^{3} q$,
whereas for comparable masses (${\cal S}\ll L$) we find instead
$\dot{\Upsilon}\big|_{\mathcal Q}/\dot{\Upsilon}\big|_{\mathrm{SO}}\sim \alpha^{p/3-2}$.
Hence, quadrupole-induced precession can be parametrically important across the relevant
parameter space for comparable-mass binaries, while it is suppressed in the EMRI regime
(in particular for $q\lesssim \alpha^{3}$).\vskip 4pt

Second, the rates
$\{\dot{\Upsilon}\big|_{\mathcal Q},\dot{\altupsilon}\big|_{\mathcal Q}\}$ enter the
evolution equations for $\{\dot{\kappa},\dot{\xi}\}$ and could, in principle, shift the
resonance condition in~\eqref{eq:resonance}. However, one can show that the ratio
$\dot{\Upsilon}\big|_{\mathcal Q}/\Omega^{(ab)}_{0}$ scales with $\alpha$ and $q$ as in~\eqref{eq:energy_pert_shift} for ${\cal S}\gg L$, while in the opposite regime it
follows the scaling of~\eqref{eq:Dchi_qm0}. We therefore conclude [see also the
discussion around~\eqref{eq:Dchi_qm0}] that, within the parameter ranges of interest
here, neither apsidal nor nodal precession appreciably alters the location of the
resonances.\footnote{We emphasize, however, an important caveat. Even for $g=k=0$ 
the resonance condition~\eqref{eq:resonance} can still be met on misaligned orbits, provided
$m\,\dot{\kappa}+\Delta {\cal E}_{ab}=0$. For such \emph{shifted} resonances, the required
precession rate is realized only at very large eccentricities, or deep into the ${\cal B}$
regime.}

\vskip 4pt
\underline{{\it Spin-orbit misalignment.}} In extending the analysis to the fully general $n$-state case, with $n\geq 2$, we find that
the integral in~\eqref{eq:spin_dynamics} becomes considerably less transparent. Furthermore, its closed form is not especially illuminating. Nevertheless, one may verify term-by-term that, to the PN order relevant here, the resulting equations of motion are fully equivalent to imposing the constraint~\eqref{eq:spin_balance}.\vskip 4pt 
The relation in~\eqref{eq:spin_balance} also enables us to follow the evolution of the obliquity, cf.~\eqref{eq:obliquity}. 
To this end, it is convenient to express the spin kinematics in the inertial $\bm{n}$-frame via the identities below [in direct analogy with the corresponding relations for the orbital angular momentum; see~\eqref{eq:orbital_via_fl}]:
\begin{eqnarray}
\frac{d\mathcal{S}}{dt} &=& 
\big(\dot{\mathcal{S}}_x \cos \altupsilon + \dot{\mathcal{S}}_y \sin \altupsilon\big)\,\sin \altiota 
+ \dot{\mathcal{S}}_z \cos \altiota \,, 
\label{eq:spin_angle_S} \\
\frac{d\altiota}{dt} &=& 
\frac{1}{\mathcal{S}} \left[
\big(\dot{\mathcal{S}}_x \cos \altupsilon + \dot{\mathcal{S}}_y \sin \altupsilon\big)\cos \altiota 
- \dot{\mathcal{S}}_z \sin \altiota 
\right] \,, 
\label{eq:spin_angle_I} \\
\frac{d\altupsilon}{dt} &=& 
\frac{\cos \altupsilon}{\mathcal{S} \sin \altiota}
\left(-\dot{\mathcal{S}}_x \tan \altupsilon + \dot{\mathcal{S}}_y\right) \,,
\label{eq:spin_angle_Y}
\end{eqnarray}
with ${\cal S}\equiv |\bm{\mathcal{S}}|$. 
After a sequence of algebraic manipulations, combining~\eqref{eq:spin_angle_S}-\eqref{eq:spin_angle_Y} with~\eqref{eq:a_gen}-\eqref{eq:ups_gen}, and using the explicit structure of the interaction terms in $V_{{\rm \cal Q}}$, we arrive at~\eqref{eq:i_Vq}. 
Crucially, when working in the inertial $\bm{n}$-frame, the final result receives contributions from the dynamics of both $\bm{L}$ and $\bm{\mathcal{S}}$.\vskip 4pt

\section{Microphysics of gravitational atoms} \label{app:atom}

{\it \underline{Hamiltonian and Bloch sphere}.} Starting from the Klein-Gordon equation on a Kerr background and carrying out the non-relativistic expansion in $\alpha$, one can obtain the cloud Lagrangian density (see App.~A of~\cite{Baumann:2018vus}). From here, and implementing the field redefinition in~\eqref{eq:sgm_delta}, we find (neglecting dissipative effects) 
\begin{eqnarray} \label{eq:cloud_lagra}
\frac{L_\mathrm{c}}{N_\mathrm{c}}
= 2\dot{\delta}\,(\sigma-1) + 2\Delta\mathcal{E}_{ab}\,\sigma
- \sqrt{1-\sigma^2}\sum_{l,m,g,k}\eta^{(ab)}_{l,m,g,k}
\cos\!\big(\delta-\Sigma^{(ab)}_{g,k}\big)\,.
\end{eqnarray}
The phase space of the system becomes $\mathbb{E}\cup\{\sigma,\delta\}$.
Performing the Legendre transformation on $L_\mathrm{c}$ we obtain the interacting Hamiltonian~\eqref{eq:hamiltonian_psi}, and the equations of motion in~\eqref{eq:d_sgm_decay}-\eqref{eq:del_Vq} (with $\bar{\Gamma}^{-}_{ab}= 0$).\vskip 4pt

The decay widths are incorporated by promoting the two-level evolution to
\begin{eqnarray} \label{eq:EoM_occup}
i \partial_t 
\begin{pmatrix}
c_a \\[4pt]
c_b
\end{pmatrix}
&=& \sum_{l,m,g,k}
\left(
\begin{array}{cc}
 -i \Gamma_a 
 & \eta^{(ab)}_{l,m,g,k}\, e^{-i \mathcal{E}_{ab} t - i \Sigma^{(ab)}_{g,k}} 
\\[6pt]
 \eta^{(ab)}_{l,m,g,k}\, e^{ i \mathcal{E}_{ab} t + i \Sigma^{(ab)}_{g,k}} 
 & -i \Gamma_b
\end{array}
\right)
\begin{pmatrix}
c_a \\[4pt]
c_b
\end{pmatrix}\,,
\end{eqnarray}
which generates the non-symplectic contribution in~\eqref{eq:d_sgm_decay} and yields the occupancy evolution in~\eqref{eq:d_occup}. In the single-overtone $(g,k)$ limit, one may perform a unitary transformation to the dressed frame of~\eqref{eq:EoM_occup} (see, e.g., the analysis in~\cite{Baumann:2019ztm}). This makes it natural to recast the dynamics in terms of Bloch-sphere variables, defined by
\begin{eqnarray} \label{eq:FVH_ab}
u^{(ab)}_{g,k} = \frac{1}{N_\mathrm{c}}\!\left( c_a c^\ast_b + c_a^\ast c_b \right)\,, 
\qquad 
\nu^{(ab)}_{g,k} = \frac{-i}{N_\mathrm{c}}\!\left( c_a c^\ast_b - c_a^\ast c_b \right)\,.
\end{eqnarray}
Differentiating~\eqref{eq:FVH_ab} and substituting~\eqref{eq:EoM_occup}  leads directly to~\eqref{eq:du_decay}-\eqref{eq:dv_decay}. The extension to degenerate overtones is immediate; see the discussion surrounding~\eqref{eq:v_flt_inc}.\vskip 4pt 

{\it \underline{Perturbative mixing}.} The Bloch-sphere formulation allows us to implement, whenever $|\sigma - 1| \ll 1$, a perturbative mixing approximation. To leading order in $\epsilon$ the constraint equation $[u^{(ab)}_{g,k}]^2+[\nu^{(ab)}_{g,k}]^2 = 1-\sigma^2$ gives 
\begin{eqnarray}
&& \left[ \frac{\left(3 \bar{\Gamma}^{-}_{ab}\partial_t F^{(2)}-\partial_t^2 F^{(2)}\right)}{[\eta^{(ab)}_{l,m,g,k}]^2} + \partial_t F^{(2)} \frac{\partial_t \eta^{(ab)}_{l,m,g,k}}{[\eta^{(ab)}_{l,m,g,k}]^3}  -2 \bar{\Gamma}^{-}_{ab} \frac{F^{(2)}}{ [\eta^{(ab)}_{l,m,g,k}]^2} \left(\bar{\Gamma}^{-}_{ab} + \frac{\partial_t \eta^{(ab)}_{l,m,g,k}}{\eta^{(ab)}_{l,m,g,k}} \right)+4 \right]^2 + \nonumber \\
&& \left( \frac{\Delta^{(ab)}_{g,k}}{\eta^{(ab)}_{l,m,g,k}}\right)^2 \left[ -8F^{(2)}+ \frac{\left(\partial_t F^{(2)}-2 \bar{\Gamma}^{-}_{ab}F^{(2)} \right)^2}{[\eta^{(ab)}_{l,m,g,k}]^2}   \right]= 0\,.  \label{eq:pert_mix_app}
\end{eqnarray}
Let us now address the adiabaticity assumption that leads to~\eqref{eq:pert_mix}. As this assumption doesn't hold in general, let us first consider the decoupling limit, where  away from the highly-eccentric limit we have $\partial_{\tau^{(ab)}_{g,k}} \sim [\frak{f}^{(ab)}]^{8/3}/w_{g,k}$  [via~\eqref{eq:RR_a}]. 
We can use this scaling to check the self-consistency of dropping derivatives in~\eqref{eq:pert_mix_app}.\vskip 4pt

Consider further, for simplicity, a special case $ \bar{\Gamma}^{-}_{ab} = 0$, so that $\Delta^{(ab)}_{g,k} /\eta^{(ab)}_{l,m,g,k} \gg 1$ for the perturbative mixing approximation to be valid. Thus the term in the second line of~\eqref{eq:pert_mix_app} dominates, where [via~\eqref{eq:pert_mix}]:  
\[
F^{(2)}/([\partial_t F^{(2)}]^2/ [\eta^{(ab)}_{l,m,g,k}]^2) \sim w_{g,k}^4 [\frak{f}^{(ab)}]^{-16/3}.
\]
Hence, as long as $[w^{(ab)}]^4_{g,k} \gg [\frak{f}^{(ab)}]^{4/3}$, which is satisfied in the regime of our interest (as $w^{(ab)}_{g,k} \sim \alpha^{-5p/6}$ ), we can drop the derivative term.  Similar reasoning applies in the case of a strong-decay resonance.\footnote{More explicitly, since in this regime $\Delta^{(ab)}_{g,k}\simeq 0$ and
$\bar{\Gamma}^{-}_{ab}/\eta^{(ab)}_{l,m,g,k}\gg 1$, the second line of
Eq.~\eqref{eq:pert_mix_app} vanishes. Moreover, from the first line of
Eq.~\eqref{eq:pert_mix_app} we can verify that the derivative terms are parametrically
suppressed, in the sense that
$\bar{\Gamma}^{-}_{ab}/\!\left[(\partial_t \eta^{(ab)}_{l,m,g,k})/\eta^{(ab)}_{l,m,g,k}\right]
\sim (vw)_{g,k}$.
Now, for the transitions of interest $w^{(ab)}_{g,k} \gg 1$, and,
by construction---from the requirement of an adiabatic strong-decay resonance---we have
$v^{(ab)}_{g,k}\gg \sqrt{z}^{(ab)}_{g,k}\gg 1$, implying $(vw)_{g,k}^{(ab)} \gg 1$. We thus recover the
self-consistency of the adiabatic approximation.
}\vskip 4pt 

\begin{figure*}[t!]
\includegraphics[width=.5\textwidth]{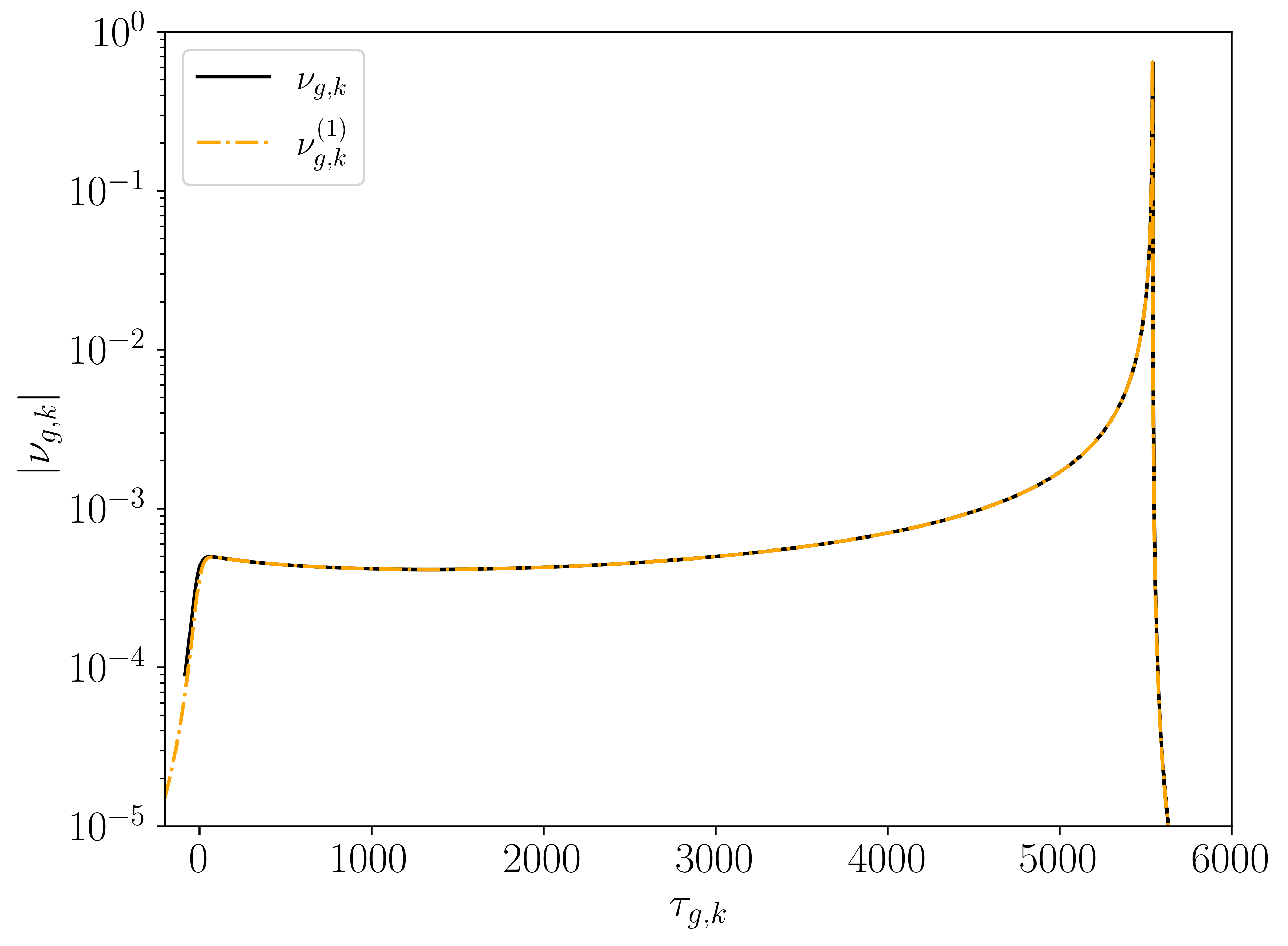}
    \caption{Numerical evolution of the Bloch variable $\nu^{(ab)}_{g,k}$ (black) for the transition in Fig.~\ref{fig:quasi_flt}, with the perturbative mixing~\eqref{eq:pert_mix} superposed in orange dot-dashed.}
    \label{fig:bloch_nu}
\end{figure*}

The preceding consistency check assumed an inspiral driven by radiation reaction. However,
(quasi-)floating further enhances the adiabaticity of the evolution, most notably by
stalling the orbital-frequency (cf.~\S\ref{sec:equatorial}). This, in turn, suppresses
the relevant (time) derivatives even more. We illustrate the robustness of the
perturbative-mixing treatment by direct comparison with a numerical calculation during
quasi-floating in Fig.~\ref{fig:bloch_nu}.\vskip 4pt

Throughout our perturbative-mixing treatment we have used the bookkeeping assignment
$\eta^{(ab)}_{l,m,g,k}\sim \mathcal{O}(\varepsilon)$, although the correct expansion parameter is given by the ratio 
$\eta^{(ab)}_{l,m,g,k}/\sqrt{\big[\Delta^{(ab)}_{g,k}\big]^2+\big(\bar{\Gamma}^{-}_{ab}\big)^2}$.
The latter is the reason why the approximation remains valid even for resonances in the strong-decay regime. In other words,  although the cloud decay is resonantly enhanced, the \emph{relative} mixing between states is still perturbative. That said, the derivation of~\eqref{eq:pert_mix} relies on the assumption of adiabatic evolution of the orbital
parameters. This, however, fails for \emph{sinking} resonances, for which the mixing itself is
perturbative but the evolution is not adiabatic, and~\eqref{eq:pert_mix} ceases to be a
reliable approximation (in agreement with our numerical validation).

\vskip 4pt \underline{\textit{Beyond Bloch: Three-state mixing.}} In most scenarios we have considered a two-state (at a time) mixing approach. However, we often must grapple with a more general situation, involving an $n$-state system with $n>2$ (see also App.~\ref{app:lines_res}). In this case, unitarity and phase-space invariance imply that there are $2n-2$ degrees of freedom in the occupancy dynamics. For all practical purposes in this paper, we focus on scenarios with $n \leq 3$.\vskip 4pt

For three-state mixing, the representation in~\eqref{eq:sgm_delta} can be extended as
\begin{eqnarray} \label{eq:sgm_delta_3s}
c_a &=& \sqrt{\frac{1+\sigma}{2}}\,
\exp\!\left[-\frac{i}{2}\Big(\Delta \mathcal{E}_{ab} + \Delta \mathcal{E}_{ac}\Big)t \right] \,, \\[2pt]
c_b &=& \sqrt{\frac{(1-\sigma)(1+\varrho)}{4}}\,
\exp\!\left[i\left(\delta + \frac{1}{2}\Big(\Delta \mathcal{E}_{ab} - \Delta \mathcal{E}_{ac}\Big)t\right)\right] \,, \\[2pt]
c_c &=& \sqrt{\frac{(1-\sigma)(1-\varrho)}{4}}\,
\exp\!\left[i\left(\delta_\varrho + \frac{1}{2}\Big(\Delta \mathcal{E}_{ac}- \Delta \mathcal{E}_{ab}\Big)t\right)\right] \,,
\end{eqnarray}
including two additional degrees of freedom $\{\varrho,\delta_\varrho\}$, such that in the limit $\varrho \to \pm 1$ the system reduces to the familiar $n=2$ case. The dynamical system simplifies dramatically if the overlap between the decaying mode and the additional state is suppressed, i.e.,
$|\bra{b} V_\star \ket{c}_{lm}| \ll \mathrm{min}\left\{|\bra{a} V_\star \ket{b}_{lm}|,|\bra{a} V_\star \ket{c}_{lm}|\right\}$, allowing us to perform the same unitary transformation to the dressed frame as in the two-state scenario. This turns out to be an excellent approximation throughout parameter space.\footnote{This is the case since the multi-state overlap occurs mostly in the strong-decay transitions, thus inhibiting the $b \leftrightarrow c$ transitions, due to the fast decaying nature of the $\ket{b}$ state. In contrast, in the weak-decay regime overlaps are typically narrower, in which case an $n=2$ approximation suffices.} We can now introduce the Feynman--Vernon--Hellwarth (FVH) representation,\footnote{In principle, the Bloch representation (which relies only on the Hermiticity and unit-trace properties of the density matrix) introduces $n^2-1$ variables. In contrast, the FVH representation introduces $2n-1$ variables. Only for $n=2$ do they coincide.} both for the $(ab)$ and $(ac)$ sectors, i.e.~\eqref{eq:FVH_ab} and $\{u^{(ac)}_{g,k},\nu^{(ac)}_{g,k}\}$, defined by substituting $c_b \to c_c$ in~\eqref{eq:FVH_ab}. Both sectors satisfy constraints that depend on $\sigma$ and $\varrho$.\vskip 4pt

We  proceed to analyze the perturbative mixing as in the two-state case. Using~\eqref{eq:eps}, we introduce the same ansatz as in~\eqref{eq:pert_mix} and expand in $\varepsilon$, assuming the same power-counting rules:
$\eta^{(ab)}_{l,m,g,k} \sim \eta^{(ac)}_{l,m,g,k} \sim \mathcal{O}(\varepsilon)$, and suppressing time derivatives (see the previous discussion). After straightforward, albeit tedious, calculations we arrive at
\begin{eqnarray}
F^{(2)} &=& \frac{2 \big[\eta^{(ab)}_{l,m,g,k}\big]^2}{\big[\Delta^{(ab)}_{g,k}\big]^2 + \big(\bar{\Gamma}^{-}_{ab}\big)^2}
\;+\;
\frac{2 \big[\eta^{(ac)}_{l,m,g,k}\big]^2}{\big[\Delta^{(ac)}_{g,k}\big]^2 + \big(\bar{\Gamma}^{-}_{ac}\big)^2}
\,, \nonumber
\end{eqnarray}
and likewise to an expression for $\varrho$, which asymptotes to $\varrho \to 1$ for $\eta^{(ac)}_{l,m,g,k} \to 0$, and $\varrho \to -1$ for $\eta^{(ab)}_{l,m,g,k} \to 0$. We can then proceed to implement an extension of the analysis leading to~\eqref{eq:a_Vq}--\eqref{eq:i_Vq}, now applicable to a $3$-state system.

\section{Overtone expansion for eccentric orbits} \label{app:ecc}

 {\it \underline{Eccentric overtones.}} In the Letter we performed an expansion in overtones in the small-eccentricity limit~\cite{Boskovic:2024fga}. In principle, this can be generalized  to arbitrary powers of $e$. We start by noticing that, for each $g$-overtone, we find [cf.~\eqref{eq:eta}]
\begin{eqnarray}
\bra{a} V_\star\ket{b}_{lmg} \supset \E^{-i g \varphi} \left(1-e \cos{E} \right)^{-(l+1)}  \,,
\end{eqnarray}
where $E$ and $\varphi$ can be expressed in terms of $\vartheta$ as a series in $e$ (see~\cite{Tremaine_Dynamics}). For instance, for the exponential factor, %
\begin{eqnarray}
\E^{-i g \varphi} = \E^{-i g \vartheta } \prod^\infty_{j=1} \E^{-i g \alpha_j \sin (j \vartheta)} \,,
\end{eqnarray}
where $\alpha_j$ are coefficients that can be calculated to arbitrary order in $e$ (e.g., $\alpha_1 = 2e + \mathcal{O}(e^3)$)  by the procedure explained in~\cite{Tremaine_Dynamics}. We can then further expand each product factor using the Jacobi–Anger identity, and rewrite the full expression as
\begin{eqnarray} \label{eq:ecc_first_term}
\E^{-i g \varphi} = \sum^{\infty}_{k=-\infty} \mathcal{C}_k \E^{-i (g-k) \vartheta} \,, \quad \mathcal{C}_k \equiv \sum_{\{m_j|k\}}\, \prod^\infty_{j=1} J_{m_j} (-g \alpha_j) \,,
\end{eqnarray}
with $J_k$ are the Bessel functions obeying $J_k \sim e^{|k|}$ for $e\ll1$, and $\{m_j|k\}$ indicates a sum over $\{m_1, \ldots, m_j\}$, obeying the constraint $\sum_j j m_j = k$. For the term depending on the distance, on the other hand, we find both harmonic and non-harmonic contributions,
\begin{eqnarray}
\left(1-e \cos{E} \right)^{-(l+1)} = \left[1 - e \left(-\frac{e}{2} + 2 \sum^\infty_{\mathsf{m}=1} \frac{J'_\mathsf{m} (\mathsf{m} e)}{\mathsf{m}} \cos{\mathsf{m}\vartheta} \right) \right]^{-(l+1)} \,.
\end{eqnarray}
Notice that each $\mathsf{m}$-term, multiplying~\eqref{eq:ecc_first_term}, can be absorbed into the $\mathcal{C}_k$'s by shifting the value of $k$. Finally, we have
\begin{eqnarray}
\left(1-e \cos{E} \right)^{-(l+1)} \E^{-i g \varphi} = \sum^{\infty}_{k=-\infty} \tilde{\mathcal{C}}_k \E^{-i(g-k) \vartheta} \,, \quad \tilde{\mathcal{C}}_k = \sum_{n = |k|} f_n (l, g,k) e^{n} \,,
\end{eqnarray}
where $f_n(l,g,k)$ are coefficients that can be calculated using the procedure outlined above.\vskip 4pt

Extending the earlier work in the Letter, in this paper we implemented the overtone expansion up to $\mathcal{O}(e^6)$. This is sufficient for all the examples studied in \S\ref{sec:dynamics} and \S\ref{sec:pheno}.\footnote{We note that in $\lesssim 20\%$ of cases, $k<-5$ overtones are triggered in the $\mathcal{H}$ regime studied in \S\ref{sec:pheno}. In these cases, we take $f_n(l,g,k) \simeq 10$, motivated by the trend observed for lower overtones, since $\mathcal{O}(1)$ variations in $\eta^{(ab)}_{l,m,g,k}$ do not affect the results shown in Fig.~\ref{fig:322_hyperfine}. In particular, {\it (i)} the excess eccentricity relative to vacuum evolution is driven by (relatively) later overtones, and {\it (ii)} the earliest overtone that may be triggered depends on the initial orbital frequency. As a result, a large portion of the binary systems do not have access to very early overtones.} The value of the $f_n(l,g,k)$'s can be found in the ancillary file accompanying the arXiv submission.

{\it \underline{Arbitrarily-eccentric orbit.}} The above representation is particularly well suited to isolating resonant transitions and estimating their excitation through the coefficients $\eta^{(ab)}_{l,m,g,k}$. However, even for modest initial conditions, floating can induce a rapid growth of the eccentricity toward fixed points---sometimes reaching large values (see \S\ref{sec:equatorial}). In this regime, the overtone decomposition may cease to be under control, as terms nominally suppressed by powers of $e^{|k|}$ can become appreciable. To describe these scenarios, we turn to a formulation of the mixing potential $V_{\cal Q}$ in~\eqref{eq:Vq_mixing} that does not explicitly rely on a converging $k$-overtone expansion, i.e.,
\begin{eqnarray} \label{eq:VQ_notexp}
V_{\cal Q}&=& - M_\star \sum_{l,m, g}\frac{\mu  N_\mathrm{c} r^{l}_c }{R^{l+1}}  \frac{4\pi}{2l+1}  Y_{lg}\left(\frac{\pi}{2},0\right)  (I_r   I_\Omega)^{(ab|lm)}  d^{(l)}_{mg}(\beta) \sqrt{1-\sigma^2}   \cos{\left[\delta- \Sigma^{\varphi,(ab)}_{g}\right]}  \,, \nonumber \\
\Sigma^{\varphi,(ab)}_{g} &\equiv& g\varphi + g\xi + m \kappa. 
\end{eqnarray}
We next change variables from $(\vartheta,e)$ to $(E,e)$ using~\eqref{eq:Kepler_eq}. This induces a Jacobian $\hat{\mathbb{G}}'$ on the subspace spanned by $(\vartheta,e)$, which can be composed with~\eqref{eq:jacobian}. Hence, under the transformation
\[
\hat{\mathbb{M}}' = \hat{\mathbb{G}}'^{\mathrm T}\,\hat{\mathbb{M}}\,\hat{\mathbb{G}}'\,,
\]
we obtain a new representation of Lagrange's planetary equations for $\mathbb{E}'=\mathbb{E}\big|_{\vartheta\to E}$. For illustrative purposes, we present below only the evolution equations for the orbital frequency and eccentricity,
\begin{eqnarray}
\frac{d\Omega}{dt} \Big|_{\cal Q}  &=& - \sum_{l,m,g}\frac{(\lim_{e \to 0}\eta_{l,m,g,0}) \sqrt{\gamma_0} b_0}{\left( 1 - e \cos{E} \right)^{l+3}} [\frak{f}^{(ab)}]^{4/3} \times \\ && \left[\sqrt{1-e^2} g \nu^{\varphi,(ab)}_{g} -  e (l+1) \sin{E}  u^{\varphi,(ab)}_{g} \right] \, , \nonumber\\
\frac{de^2}{dt} \Big|_{\cal Q}  &=& - \sum_{l,m,g}\frac{(\lim_{e \to 0}\eta_{l,m,g,0}) \sqrt{\gamma_0} b_0}{ [\Omega^{(ab)}_0]^\mathrm{sat} \left( 1 - e \cos{E} \right)^{l+3}} [\frak{f}^{(ab)}]^{1/3}\sqrt{1-e^2} \times \nonumber\\
&& \left[  e \sqrt{1-e^2} (l+1) \sin E  \, u^{\varphi,(ab)}_{g} +  \left[\left(1-e^2\right) g- g (e \cos E -1)^2\right] \nu^{\varphi,(ab)}_{g} \right]  \nonumber\,,\\
\nu^{\varphi,(ab)}_{g} &\equiv& -\sqrt{1-\sigma^2} \sin{(\delta - \Sigma^{\varphi,(ab)}_{g})} \,, \quad u^{\varphi,(ab)}_{g} \equiv \sqrt{1-\sigma^2} \cos{(\delta - \Sigma^{\varphi,(ab)}_{g})} \,,\nonumber
\end{eqnarray}
where the trigonometric functions (of $g\varphi$) entering in $\{\nu^{\varphi,(ab)}_{g}, u^{\varphi,(ab)}_{g}\}$ can be expressed  using~\eqref{eq:phi_via_E} in terms of the eccentric anomaly and orbital eccentricity.\vskip 4pt

\begin{figure*}[t!]
\includegraphics[width=.5\textwidth]{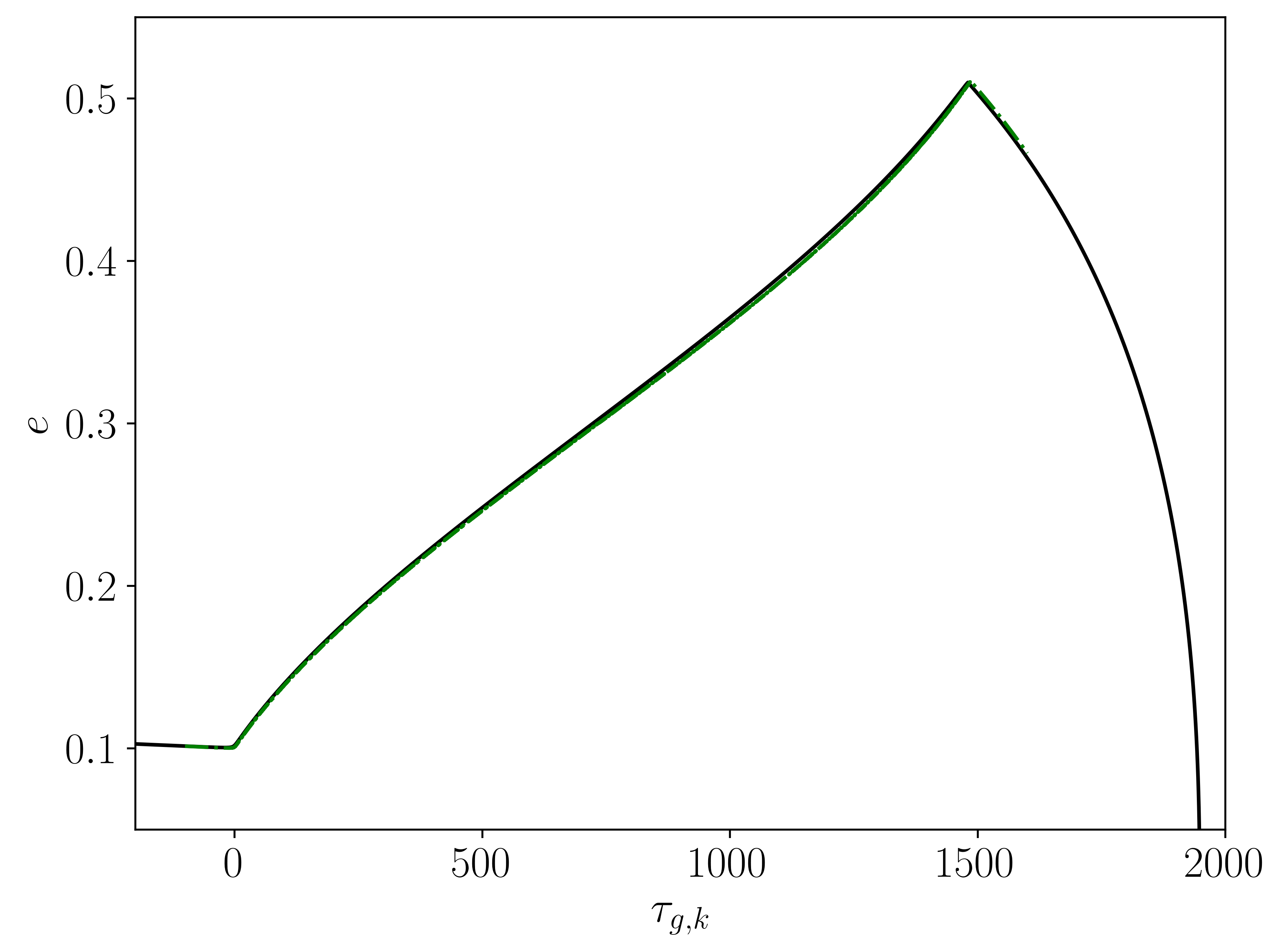}
    \caption{Evolution of the eccentricity for the example in Fig.~\ref{fig:quasi_flt}; using the truncated overtone-based expansion (black) and the complete dynamics [c.f.~\eqref{eq:VQ_notexp}] 
    (green).}
    \label{fig:ecc_comparsion}
\end{figure*}

In all numerical evolutions presented here we employ the overtone expansion. Over the region of parameter space of interest, the error remains under control. A direct comparison between the overtone expansion and the full evolution at arbitrary eccentricity is shown in Fig.~\ref{fig:ecc_comparsion}. Moreover, as demonstrated in App.~\ref{app:flux_balance}, in the narrow-resonance limit the full dynamics reduces to the analysis in~\cite{Boskovic:2024fga}. In such regime, and during floating, the cloud may be integrated out in a $k$-resummed form, yielding fixed points that remain valid for arbitrary $e<1$.

\section{Flux-balance approach} \label{app:flux_balance}

{\it \underline{Balance laws.}} In many situations of interest, we may be able to read off the change of the orbital elements via the balance laws associated with the {\it would-be} conserved quantities of the Keplerian problem \cite{Baumann:2019ztm}: the orbital energy $H_\mathrm{K}$, angular momentum $\bm{L}$ and the Laplace-Runge-Lenz vector~\cite{Tremaine_Dynamics,Nabet:2014kva},\footnote{We  can also use $H_\mathrm{K}$ and $\bm{L}$ combined with kinematic constraints, see e.g., ~\cite{Damour:2004bz,Pound:2007th}.}
\begin{eqnarray}
\bm{e} = \frac{\bm{\dot{r}} \cross (\bm{r} \cross \bm{\dot{r}})}{{\cal M}(1+q)} - \frac{\bm{r}}{r}\,,
\end{eqnarray}
due to the losses induced by ``fluxes'' from orbital and spin perturbations. (On top of the spin dynamics already discussed in App.~\ref{app:spin}).\vskip 4pt
There are in total  $1+3+3=7$ components. However, the constraints $\bm{L} \cdot \bm{e}=0$ and $(L/(\text{\textmu}\Lambda))^2 + e^2 = 1$ reduce them to $5$ equations for the $\mathbb{E} \setminus \{ \vartheta \}$ variables. For instance, for the $\mathbb{E} \setminus \{ \vartheta, \chi \}$ variables, we have 
\begin{eqnarray} \label{eq:orbital_via_fl}
\frac{da}{dt} &=& \frac{2  a^2}{M^2 q} \dot{H}_\mathrm{K} \,,  \\
\frac{de}{dt} &=& \frac{1}{a e M^2  q} \Big[ a^2 \dot{H}_\mathrm{K} \left(1-e^2\right) - \sqrt{a \left(1 -e^2\right) M (q+1)} \times  \nonumber \\
&&  \left(\sin \iota (\dot{L}_x \cos \Upsilon +\dot{L}_y  \sin \Upsilon)+\dot{L}_z  \cos  \iota \right) \Big] \,, \nonumber\\
\frac{d\iota}{dt} &=& \frac{1}{q} \sqrt{\frac{1+q}{aM^3(1-e^2)}} \times  \, \nonumber \\
&& \left((\dot{L}_x \cos{\Upsilon}+\dot{L}_y \sin{\Upsilon}) \cos{\iota} - \dot{L}_z  \sin{\iota} \right)\,,  \nonumber\\
\frac{d\Upsilon}{dt} &=& \frac{1}{q} \sqrt{\frac{1+q}{aM^3(1-e^2)}} \frac{1}{\sin \iota}  \left(- \dot{L}_x \sin \Upsilon + \dot{L}_y \cos \Upsilon \right)\,. \nonumber
\end{eqnarray}
These are kinematic identities. The specific evolution equations will then depend on prescribing the values for the given fluxes of $\dot{H}_\mathrm{K}$ and $\dot{\bm{L}}$. In the case of GW emission, upon adiabatic averaging (see App.~\ref{app:cel_mech}) we  can then recover the standard evolution equations, e.g, for $\{\dot{a}, \dot{e}\}$ first derived in ~\cite{Peters:1963ux, Peters:1964zz}.\vskip 4pt

{\it \underline{Isolated-atom approximation.}}  From the full interacting Hamiltonian, $H_{\mathrm I}$, and using Milankovi\'c's equations~\cite{Allan_Ward_1963, Tremaine_Dynamics}, we obtain the fluxes of $\dot{\bm e}$ and $\dot{\bm L}$, and from there the evolution of the orbital elements through the relations in~\eqref{eq:orbital_via_fl}. By construction, this procedure reproduces the equations of motion derived directly from the Hamiltonian in~\S\ref{sec:overview_orbit}. By contrast, the earlier analyses in our Letter~\cite{Boskovic:2024fga} and in~\cite{Tomaselli:2024bdd} adopted a more direct---but, as we will see, generally incomplete---identification. For simplicity, we restrict to a two-state system and neglect decay-width effects:
\begin{eqnarray}
 \dot{H}_\mathrm{K} &=& - \dot{E}^\mathrm{IA}_\mathrm{c}  = N_\mathrm{c}\,\Delta\mathcal{E}_{ab}\,\dot{\sigma}/2 \,,\qquad 
E^\mathrm{IA}_\mathrm{c} = N_\mathrm{c}\left(\mathcal{E}_a |c_a|^2 + \mathcal{E}_b |c_b|^2\right) \,, \label{eq:iso_at_E} \\
\dot{L}_z &=& - \dot{S}_\mathrm{c} \,,\qquad 
S_\mathrm{c} = N_\mathrm{c}\left(m_a |c_a|^2 + m_b |c_b|^2\right)
\label{eq:iso_at_L}  \,,
\end{eqnarray}
where we used~\eqref{eq:sgm_delta} to rewrite the energy flux as
$\dot{E}^\mathrm{IA}_\mathrm{c} = - N_\mathrm{c}\,\Delta\mathcal{E}_{ab}\,\dot{\sigma}/2$.
We refer to this as the \emph{Isolated-Atom} (IA) approximation. A direct comparison with~\eqref{eq:hamiltonian_psi} makes it immediately apparent that the interaction term is missing in the IA prescription.
\vskip 4pt

The IA approximation was originally introduced in~\cite{Baumann:2019ztm} for use within an $S$-matrix framework, where interaction terms can be neglected because the \emph{in} and \emph{out} states are, by construction, non-interacting. In the present context, however, such terms can play an important role. To illustrate the impact of the missing contribution in the IA results, consider, for simplicity, a resonant transition on a co-rotating circular orbit. From~\eqref{eq:hamiltonian_psi} we find
\begin{eqnarray} \label{eq:int_IAAvsH}
\frac{2}{N_\mathrm{c}} \frac{d H_\mathrm{c}}{dt} 
= - \Delta\mathcal{E}_{ab} \dot{\sigma} 
- 2 \eta^{(ab)}_{l,m,g,k} \Big\{ (\Delta\mathcal{E} - \Delta m \,\Omega)\, \nu
+ \Big[ \frac{\partial_t\eta}{\eta} \Big|_{l,m}\, u  + \Delta m\,(\dot{\vartheta}-\Omega)\, \nu\Big] \Big\}^{(ab)}_{g,k} .
\end{eqnarray}
Using~\eqref{eq:tht_gen}, one can show that the last term, in square brackets ($[\ldots]$), does not contribute.
The remaining piece, proportional to $\nu^{(ab)}_{l,m,g,k}$, becomes negligible only in the limit of narrow resonances---as expected in an $S$-matrix treatment. This is indeed the case for several of the scenarios studied in~\cite{Boskovic:2024fga, Tomaselli:2024bdd}, for which the IA approximation is therefore justified. 
However, in general, the evolution of the full system during a resonance (see Fig.~\ref{fig:quasi_flt}) tends to broaden the transition, thereby promoting the interaction term to a phenomenologically relevant role (cf.~\S\ref{sec:dynamics}, \S\ref{sec:pheno}). For instance, comparing the IA approximation against the dynamics implied by~\eqref{eq:a_Vq} and~\eqref{eq:e_Vq} for equatorial orbits and under the single-overtone dominance, we obtain
\begin{eqnarray}
\frac{d\Omega_\mathrm{IA}}{d\Omega} \Big|^{(ab)}_{l,m,g,k} &=&
\frac{\frak{f}^{(ab)}_{g,k}}{\frak{f}^{(ab)}(t)} \,, \\
\frac{d(e^2)_\mathrm{IA}}{d(e^2)} \Big|^{(ab)}_{l,m,g,k} &=&
\frac{\frak{f}^{(ab)}_{g,k}}{\frak{f}^{(ab)}(t)}
\frac{\sqrt{1-e^2} - \frak{f}^{(ab)}(t)}{\sqrt{1-e^2} - \frak{f}^{(ab)}_{g,k}} \,,
\end{eqnarray}
which provides a direct quantification of the error incurred by the IA prescription. Crucially, this discrepancy is not merely quantitative, but can also be qualitative. To see this, notice that in the circular limit one may attempt to determine the orbital backreaction either by balancing the cloud's energy or by balancing its spin. However, only the latter option (as employed in~\cite{Baumann:2019ztm}) yields the correct result, since $S_\mathrm{c}$ (effectively) carries no interaction term. Energy balance via the IA approximation, instead, leads to inconsistent predictions away from the exact floating case.
\vskip 4pt

Another regime, for which the IA approximation induces a significant qualitative error, is that of wide mixing. As a concrete example, consider circular counter-rotating orbits [cf.\S\ref{sec:pheno}]. From~\eqref{eq:a_Vq} one finds $\dot{\Omega} \sim g\,\dot{\sigma} > 0$ (for $g>0$), while the IA approximation incorrectly predicts $\dot{\Omega} \sim - \Delta \mathcal{E}_{ab}\,\dot{\sigma} < 0$. Physically, as the cloud spin decreases, the orbital angular momentum must also decrease, and the orbital frequency therefore accelerates.\footnote{Floating resonances on counter-rotating eccentric orbits are then possible, because the loss of angular momentum can be accommodated by a combination of stalling the orbital frequency and increasing the eccentricity.}\vskip 4pt

Finally, an additional subtlety arises for non-equatorial orbits. In the analysis of~\cite{Tomaselli:2024bdd}, the balance between $\bm L$ and $\bm S_\mathrm{c}$ was enforced in a frame aligned with $\bm n=\bm S_\mathrm{c}/S_\mathrm{c}$, while imposing the conditions $\dot L_x=0$ and $\dot L_y=0$. Substituting these assumptions into~\eqref{eq:orbital_via_fl} then yields the flow equations for $\{\dot e,\dot\beta\}$ adopted in~\cite{Tomaselli:2024bdd}. However, spin precession contributes to the energy-momentum balance, even in the EMRI limit, rendering the chosen $\bm n$-frame non-inertial (see App.~\ref{app:spin}). As a result, the procedure in~\cite{Tomaselli:2024bdd} leads to incorrect predictions for inclined orbits, even for narrow resonances. (See Sec.~3.4 of~\cite{Tomaselli:2024bdd} and compare with our results in~\S\ref{sec:incl}.)

\section{Remarks on multi-level phenomenology} \label{app:lines_res}

{\it \underline{Superradiant growth.}} We implement the following minimal (``vanilla'') evolution model. The boson field initially populates a single hydrogenic level,
$\ket{a}\equiv \ket{m_a\!+\!1\,\, m_a\,\, m_a}$ (e.g.\ $\ket{211}$, and so on), and remains in this state throughout the superradiant growth phase until saturation. At saturation, the BH spin parameter is driven to
\begin{eqnarray}\label{eq:cloud_sat}
\tilde{a}_\mathrm{sat} = \frac{4 m_a \,\alpha_\mathrm{sat}}{m_a^2 + 4 \alpha_\mathrm{sat}^2} \,.
\end{eqnarray}
After the occupation number reaches its maximum, the $\ket{a}$ state depletes via GW emission~\cite{Arvanitaki:2010sy,Brito:2014wla,Arvanitaki:2014wva,East:2018glu}. While the cloud may lose up to an $\mathcal{O}(1)$ fraction of its mass, a substantial overdensity can persist over timescales of phenomenological relevance. As the next level, $\ket{m_a\!+\!2\,\, m_a\!+\!1\,\, m_a\!+\!1}$, begins to grow and extracts additional angular momentum, the original mode $\ket{m_a\!+\!1\,\, m_a\,\, m_a}$ is pushed across the superradiant threshold and becomes decaying. Provided the requisite conditions are maintained, the system proceeds through the familiar ladder
$\ket{211}\to\ket{322}\to\ket{433}\to\cdots$.

Denoting by $(\alpha_n,\tilde{a}_n)$ the parameters at the onset of the $n$-th superradiant cycle, and neglecting $\mathcal{O}(\alpha^2)$ corrections to the bound-state energies~\cite{Baryakhtar:2020gao,Khalaf:2024nwc}, we obtain
\begin{eqnarray}
\alpha_\mathrm{sat} &=& \alpha_n \bigl(1 - \alpha_n\, n_\mathrm{c,sat}\bigr) \,, \\
\tilde{a}_\mathrm{sat} &=& \frac{\tilde{a}_n - m_a\, n_\mathrm{c,sat}}{\bigl(1 - \alpha_n\, n_\mathrm{c,sat}\bigr)^2} \,,
\label{eq:SR_sat}
\end{eqnarray}
where $n_\mathrm{c}\equiv N_\mathrm{c}/M_n^2$ is the (dimensionless) cloud occupancy. As long as $\tilde{a}>\tilde{a}_\mathrm{sat}$, the equations in~\eqref{eq:cloud_sat}--\eqref{eq:SR_sat} form a closed algebraic system that determines $(n_\mathrm{c},\alpha,\tilde{a})|_\mathrm{sat}$ from the parameters at the beginning of the cycle. Expanding to leading order in $\alpha_n$ yields
$n_\mathrm{c,sat}\simeq (\tilde{a}_n-\tilde{a}_\mathrm{sat})/m_a$.
Fixing the boson mass $\mu$, we then propagate from an initial spin $\tilde{a}_\mathrm{in}=0.995$ and a distribution of BH masses (equivalently set by $\alpha_\mathrm{in}$), obtaining the (saturated) values  which we use as initial data for the numerical evolution of the binary dynamics shown in \S\ref{sec:pheno}.\vskip 4pt

Let us emphasize that, within the superradiant-growth scenario considered above, the cloud can carry at most ${\cal O}(10\%)$ of the original BH mass~\cite{Herdeiro:2021znw,Hui:2022sri}. Accretion onto the BH---from either baryonic or dark-matter environments---may nevertheless yield substantially denser configurations, reaching up to $\sim 1/3$ of the original mass~\cite{Brito:2014wla,Hui:2022sri} (see also~\cite{Budker:2023sex}). Incorporating matter accretion into our cloud-evolution model is beyond the scope of this work. We have, however, probed the impact of higher masses in a few representative cases (see Fig.~\ref{fig:322_hyperfine} and Fig.~\ref{fig:433_fine}), where we find a significant enhancement of backreaction effects.\vskip 4pt

{\it \underline{Loss of the cloud's mass.}} As noted above, the cloud may deplete through GW emission well before the binary encounters any resonant regime. In addition, level mixing can also drain the cloud mass. To assess the impact of this secular mass evolution on the orbital dynamics---through the time dependence of the mass parameters $\{\mathcal{M},q\}$---we must include an additional contribution to the orbital flow. Assuming the cloud-mass evolution is adiabatic, we may follow the evolution of the semi-major axis $a(t)$ from the adiabatic invariance of the action variable $\Lambda$ [cf.~\eqref{eq:delaunay_lambda}], which implies
\begin{eqnarray}\label{eq:a_dotMc}
\frac{da}{dt}\Big|_{\mathcal{M}(t)} = -\,a\,\frac{1+2q}{1+q} \frac{\dot{M}_\mathrm{c}}{\mathcal{M}} \,.
\end{eqnarray}

The associated loss of cloud mass generically drives an {\it outspiral} of the orbit. Consider first level mixing, for which $\Delta {\cal M}/{\cal M}\sim \mathcal{O}(\alpha^p)$ with $p=3,5,7$. Comparing the resulting drift against the resonant evolution of the orbital parameters [cf.~\eqref{eq:a_Vq}], we obtain
\begin{eqnarray}
\frac{\dot{a}\big|_\mathcal{Q}}{\dot{a}\big|_{\mathcal{M}(t)}} \sim \frac{1}{q(1+q)^{2/3}}\,\alpha^{-2p/3}\,.
\end{eqnarray}
For small-to-moderate values of $\alpha$, this ratio is parametrically large, implying that cloud mass loss from level mixing does not materially modify the resonant imprint on the binary evolution. We therefore set $\mathcal{G}\simeq 1$ in~\eqref{eq:a_Vq}, and do so throughout the numerical evaluations in \S\ref{sec:dynamics} and \S\ref{sec:pheno}.\vskip 4pt

By contrast, GW-driven cloud depletion can remove an $\mathcal{O}(1)$ fraction of the cloud mass, $\Delta {\cal M}/{\cal M}\sim \mathcal{O}(\alpha^0)$, and can therefore have a significantly larger impact on the orbital evolution~\cite{Cao:2023fyv}. For instance, comparing with  standard radiation-reaction effects evaluated near a resonance, $M \Omega \simeq \alpha^p$, we find [via~\eqref{eq:gw_from_cloud}]\footnote{We ignore here the level-mixing channel of GW emission, which can be shown to be suppressed for the transitions of interest, most of which lie in the moderate-to-strong decay regime (see also~\cite{Kyriazis:2025fis}).}
\begin{eqnarray}\label{eq:gwc_outspiral}
\frac{\dot{a}\big|_\mathrm{RR}}{\dot{a}\big|_{\mathcal{M}(t)}} \simeq
- \frac{1}{G_{n_a l_a m_a}} \, q\, (1+q)^{-1/3}\,
\alpha^{8p/3-4(m_a+3)}
\left(\frac{M_\mathrm{c}/M}{\alpha}\right)^{-2} \,.
\end{eqnarray}
From here it follows that mass-loss effects are most relevant---and can even become dominant---for $\mathcal{H}$ transitions of the $\ket{211}$ state with $q \lesssim 0.1$. In this regime, GW emission from the cloud may induce a quasi-floating phase, persisting until an $\mathcal{O}(1)$ fraction of the cloud mass has been radiated away. In practice, the resulting delay can push the system out of the detector band, with effective timescales that may exceed a Hubble time. Hence, whether by substantially postponing the inspiral, depleting the cloud, or both, $\mathcal{H}$ transitions of $\ket{211}$ become ineffective for $q \lesssim 0.1$, and can further prevent the cloud from surviving long enough to re-enter the GW band at later stages of the inspiral. By contrast, for $\mathcal{F}/\mathcal{B}$ transitions or for higher-$m_a$ states, the extra suppression in GW emission rates---together with the longer timescales ($p=3,5$)---renders this effect negligible.

\vskip 4pt {\it \underline{Wide transitions.}}  
As emphasized in \S\ref{sec:pheno}, the width of a transition plays a central role in the interaction between the cloud and the orbit~\cite{Tong:2022bbl,Tomaselli:2025jfo,Kim:2025wwj}. To quantify its impact, we can use~\eqref{eq:nc_pert} as a measure of the effective width of the transition
\begin{eqnarray} \label{eq:width}
\frac{1}{N_\mathrm{c}}\frac{dN_\mathrm{c}}{d\frak{f}^{(ab)}} [\frak{f}^{(ab)}]^{-(4l-7)/3} \simeq - \Bigg(\hat{z}_{l,m} \frac{\Omega_0}{\Gamma_b} \left[ 1 + (1-\frak{f})^2 \left( \frac{\Omega_0}{\Gamma_b} \right)^2 \right]^{-1} \Bigg)^{(ab)}_{g,0}\,. 
\end{eqnarray}
Notice that $(\Omega_0/\Gamma_b)^{(ab)} \sim \alpha^{p-6-4 l_b} \gg 1$, which tends to suppress the width of strong-decay transitions for $l_b \geq 1$. This suppression is absent, however, for transitions into spherical states with $l=l_a$ and $l_b = 0$. For instance, for $\alpha \gtrsim 0.1$ we have $(\Omega/\Gamma_b)^{(322,300)} \sim \mathcal{O}(1)$.\vskip 4pt

\begin{figure*}[t!]
\begin{tabular}{cc}
\includegraphics[width=.5\textwidth]{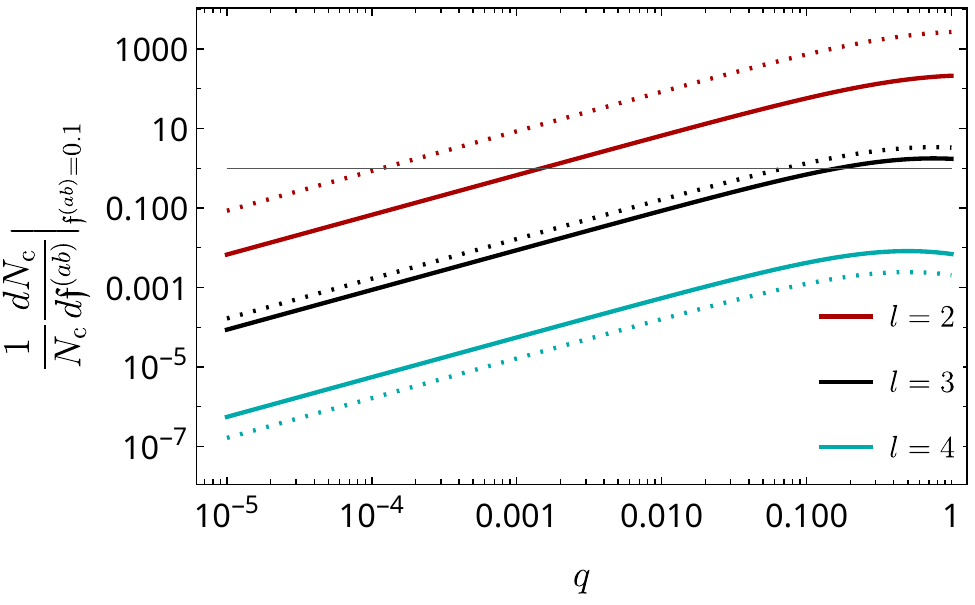}
\qquad
\includegraphics[width=.5\textwidth]{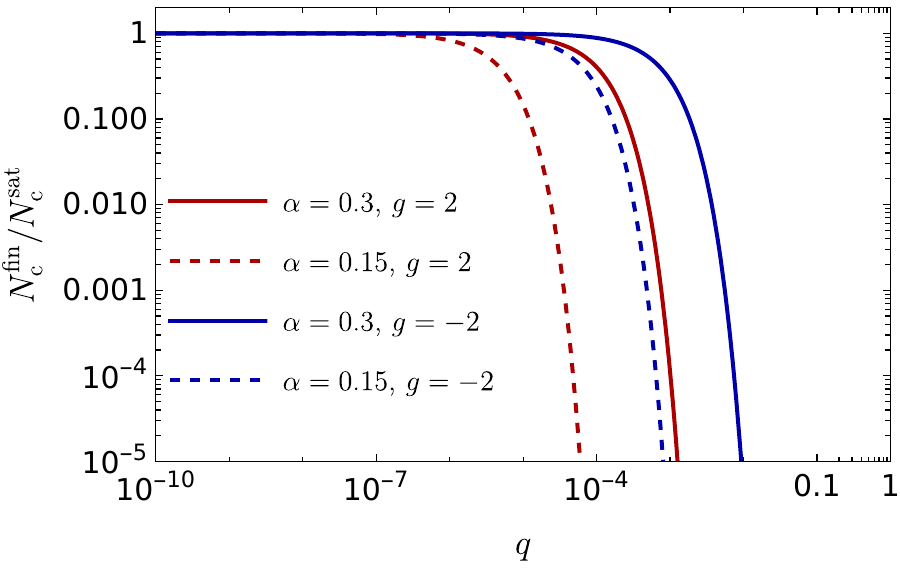}
\end{tabular}
    \caption{Width of fine transitions to spherical states on co-rotating circular orbits [via~\eqref{eq:width}, evaluated at $\frak{f}^{(ab)}=0.1$]: 
    $\ket{322} \to \ket{300}$, $\ket{433} \to \ket{400}$, and $\ket{544} \to \ket{500}$ [\textit{left}],
    shown in red, black, and cyan, respectively, for $\alpha=0.3$ (solid) and 
    $\alpha=0.15$ (dotted). Leftover occupancy of the cloud in the decoupling limit [Eq.~\eqref{eq:nc_pert}], due to the $l=2$ wide transition, normalized to the value at saturation $N^{\mathrm{sat}}_\mathrm{c}$ [\textit{right}],  for $\alpha=0.3$ (full line) and $\alpha=0.15$ (dashed), on co-rotating (red) and counter-rotating (blue) orbits, respectively.}
    \label{fig:wide_app}
\end{figure*}

For the $\ket{211}$ state, aside from the dipole-driven $\mathcal{F}$-type  transition $\ket{211} \to \ket{200}$ that occurs {\it inside} the cloud\footnote{This occurs for $\alpha \gtrsim 0.2$~\cite{Tomaselli:2024bdd}. Note, however, that for such high values of $\alpha$, the lifetime of the $\ket{211}$ state shortens due to the growth of the $\ket{322}$ state. For instance, for $M=10 M_\odot$, we have $t_{\ket{322}} \simeq 10^3 \mathrm{yr}$, while for $M=10^6 M_\odot$, $t_{\ket{322}} \simeq 10^8 \mathrm{yr}$.}, $\mathcal{H}$ transitions have a moderate width, while early $\mathcal{B}$ transitions are narrow. In contrast, $\ket{322}$ exhibits strong quadrupolar mixing with the $\ket{300}$ state, acting {\it outside} the cloud. Comparatively, for subsequent excited states, the effect of wide mixing becomes increasingly suppressed as $l_a$ increases. We plot in Fig.~\ref{fig:wide_app} the width defined in~\eqref{eq:width}, for $l_a = 2,3,4$. This demonstrates that, even in the $q\ll 1$ limit, the width of the $\ket{322} \to \ket{300}$ mixing remains non-negligible, whereas for $\ket{544}$ and higher-$l_a$ states multipolar suppression narrows the effect, also for comparable masses.\vskip 4pt

Let us now turn to the $l_a=2$ case. In the decoupling regime [cf.~\eqref{eq:nc_pert}], reducing the mass ratio suppresses the cloud-depletion rate. We illustrate this behavior in Fig.~\ref{fig:wide_app} (left), where we consider quasi-circular orbits and integrate Eq.~\eqref{eq:nc_pert} for orbital frequencies lying in the interval $[10^{-2}, 2]$ of the main resonance $\Omega^{(322,300)}_0$.
 This trend, however, should be viewed as a leading-order baseline rather than a robust prediction. Once orbital backreaction is included, the picture becomes qualitatively richer. The effective backreaction, $b_{g,k}/w_{g,k}\sim q^{-1}$, is enhanced for $q \ll 1$. As a result, the inspiral is prolonged (accelerated) for $\beta$ close to the co-rotating (counter-rotating) equatorial limit, yielding correspondingly more (less) cloud depletion relative to the baseline, as discussed in \S\ref{sec:pheno_emri}.

 \begin{figure*}[t!]
\begin{tabular}{cc}
\includegraphics[width=.5\textwidth]{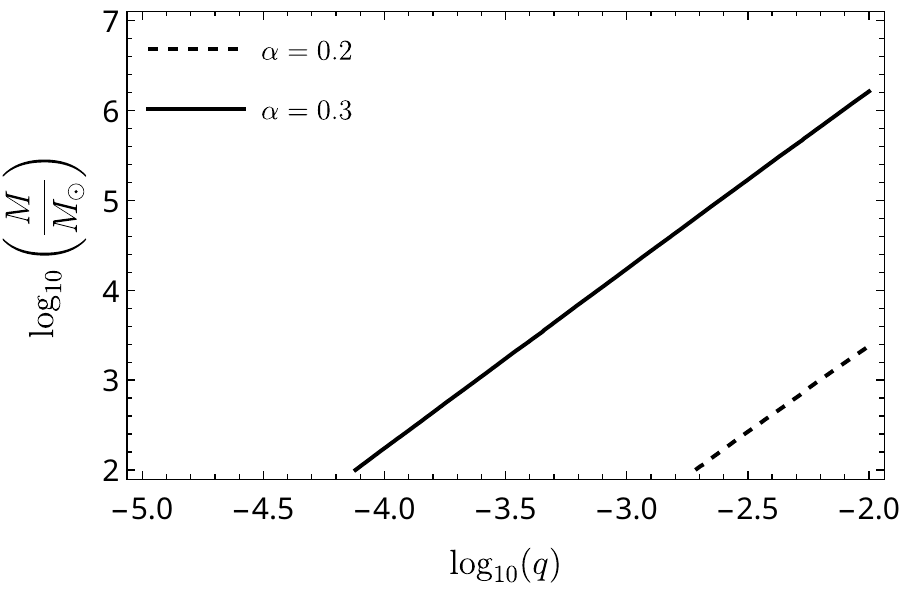}
\qquad
\includegraphics[width=.5\textwidth]{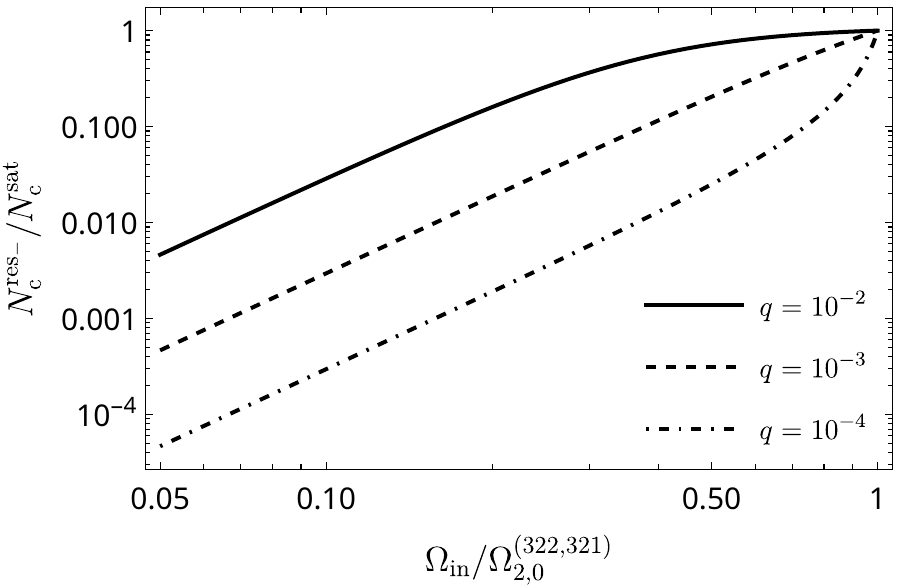}
\end{tabular}
    \caption{The curves on the left delineate the region in which $(t_{\rm col}+t_{\rm fl})/t_{\rm H}<1$, where $t_{\rm col}$ is the radiation-reaction--driven coalescence time starting \emph{after} the resonance $\Omega^{(322,321)}_{2,0}$ (with $e_{\rm in}\ll 1$), $t_{\rm fl}$ is the floating timescale for the transition under consideration, and $t_{\rm H}$ is the Hubble timescale. The criterion is satisfied to the right of each curve. We plot on the right the cloud occupancy immediately before the $\ket{322}\to\ket{321}$ resonance, $N^{\mathrm{res}_{-}}_{\rm c}$, shown relative to the saturation occupancy in the vanilla superradiant-evolution model, as a function of the orbital frequency at which superradiance saturates and the binary evolution is initialized. All curves are for $\alpha=0.3$.}

    \label{fig:emri_hf_time}
\end{figure*}

\vskip 4pt {\it \underline{Time vs. frequency evolution.}} Quasi-floating can substantially prolong the inspiral when the relevant transition is encountered very early. In many cases we find that a time-domain evolution yields a quasi-floating duration that differs by a factor of $\simeq \mathcal{O}(1)$ from the estimate in~\eqref{eq:flt_time}, while the orbital frequency drifts away from the resonant value by $\mathcal{O}(10\%)$ over the course of the quasi-floating phase (as occurs, for example, for the dashed trajectory in Fig.~\ref{fig:wide_app}, right). With this in mind, a practical criterion is obtained by simply adding the radiation-reaction--driven time to coalescence to the ideal floating timescale in~\eqref{eq:flt_time}. This provides a reliable estimate of whether a given IMRI/EMRI that encounters an $\mathcal{H}$ resonance will be ``filtered out'' before ever reaching the detector band. We demonstrate this in Fig.~\ref{fig:emri_hf_time} (left), assuming that the companion is captured immediately prior to the resonant transition.  Initializing the evolution earlier, on the other hand, can partially deplete the cloud through GW emission, weakening the ensuing quasi-floating phase (and, conversely, shortening it). We quantify these effects in more detail using~\eqref{eq:gw_from_cloud} and~\eqref{eq:d_occup}, and illustrate them in Fig.~\ref{fig:emri_hf_time} (right).\vskip 4pt

Formulating an analogous estimate in the $\mathcal{F}$ regime is more subtle, owing to 
the interplay between narrow and wide transitions. Nevertheless, a rough assessment 
can be obtained by adopting the floating timescale associated with the strongest 
transition, i.e., $\ket{322} \to \ket{300}$. Doing so relaxes the curves in 
Fig.~\ref{fig:emri_hf_time}. In particular, focusing on the $\alpha=0.3$ curve, binary inspirals with $q \gtrsim (10 M_\odot/M_\mathrm{sat})$ $\left( q \gtrsim 10^{-3} \right)$ will traverse the $\mathcal{F}$ regime within Hubble time, in the range 
$M \simeq 10^3-10^5\, M_\odot$ $\left( M \simeq 10^6\, M_\odot \right)$ of BH masses.

\vskip 4pt {\it \underline{Multi-state overlap.}} 
Since we have $\Delta \epsilon_{ab} \sim [h_{nl}\Delta m]_{ab}$ [cf.~\eqref{eq:spectrum_R}], all $\mathcal{H}$ transitions share the same fundamental frequency $\Omega^{(ab)}_0$. Consequently, for the dominant $l=2$ transitions considered here, any resonant overtone occurring at the same fractional frequency $\mathfrak{f}^{(ab)}_{g,k} = 1/\mathbb{N}^+$ of $\Omega^{(ab)}_0$ may activate both the $m=1$ and $m=2$ channels simultaneously, whereas only the $m=2$ degenerate overtones can be triggered at $\mathfrak{f}^{(ab)}_{g,k}=2/(2\mathbb{N}^0+1)$. Although this degeneracy is lifted by $\mathcal{O}(\alpha^6)$ corrections, the quasi-floating character of the transitions implies that mutual overlap may still play a role. To streamline the discussion in \S\ref{sec:pheno}, we  initialized most of our examples---both for $q\simeq 1$ binaries and in the IMRI/EMRI regime---at values away from $\beta_{
\rm in}=\{0,\pi\}$, and prior to the (chronologically first) $m=1$ transition.\vskip 4pt For the case of quasi-equatorial orbits, we determined instead which transition the system encounters first by evaluating the quantity $\sum_{(g,k|d)} \eta^{(ab)}_{2,m,g,k}(\beta_\mathrm{in})/\Gamma_{32\, 2-m}$, and selecting the transition that maximizes it. In Fig.~\ref{fig:emri_hf} (left) we further restricted our analysis to the $e_\mathrm{in}\ll 1$ regime, so that the $(g,k)=(2,-3)$ overtone of the $m=2$ transition---which coincides with the $(g,k)=(2,0)$ overtone of the $m=1$ transition---remains ineffective.\vskip 4pt 

\begin{figure*}[t!]
\begin{tabular}{cc}
\includegraphics[width=.5\textwidth]{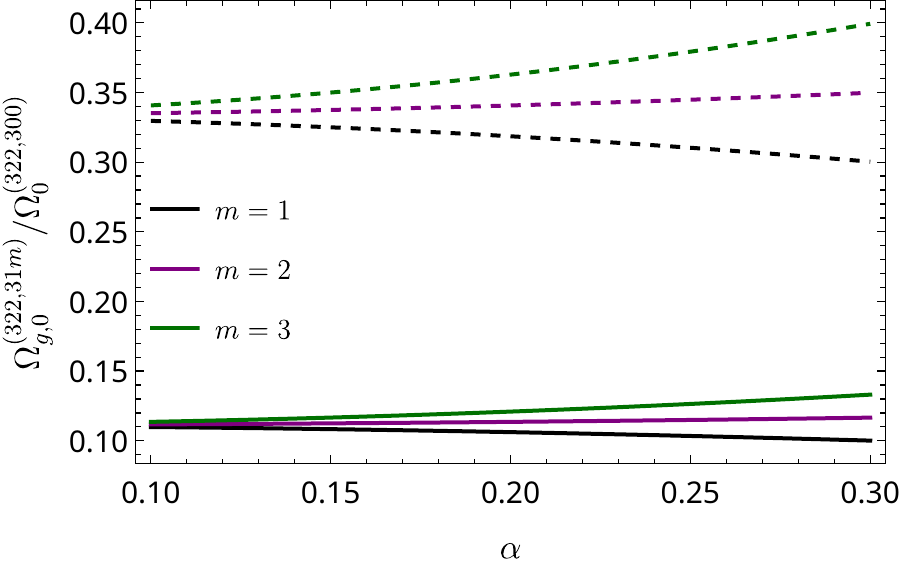}
\qquad
\includegraphics[width=.5\textwidth]{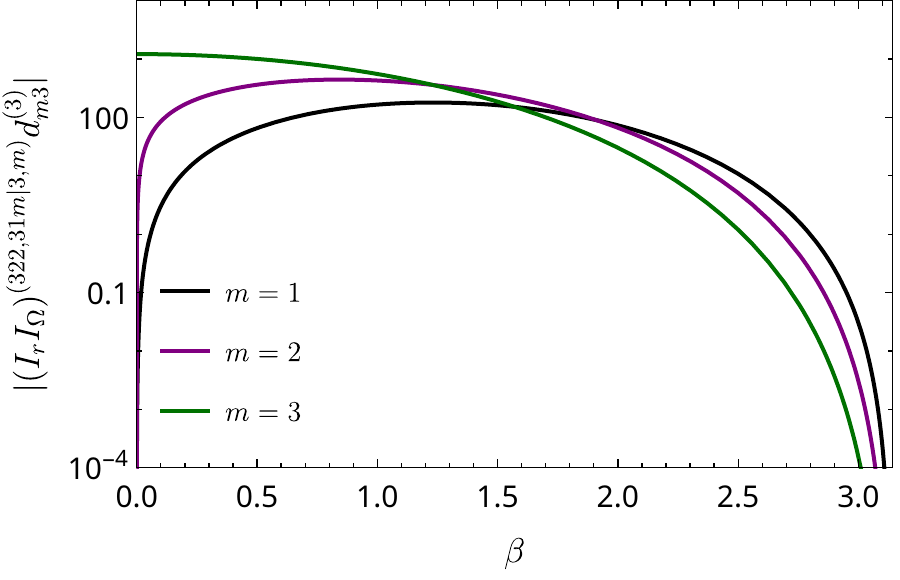}
\end{tabular}
    \caption{Position of the $(g,0)$ resonant overtones relative to $\Omega^{(322,300)}_0$, with $g=3$ (full line) and $g=1$ (dashed) [\textit{left}]. Wigner $d$-matrix of the transitions, weighted by $(I_r I_\Omega)^{(322,31m|3,m)}$ [\textit{right}].}
    \label{fig:322_split_wigner}
\end{figure*}

For $\mathcal{F}$-transitions, the narrower $l=3$ bands of the $\ket{322}$ level cluster into two groups, $g=\{1,3\}$, as shown in Fig.~\ref{fig:322_split_wigner} (left). Although the individual transitions within each band lie closer together than in the $\mathcal{H}$ regime, we apply criteria analogous to those used above. As illustrated in Fig.~\ref{fig:322_split_wigner}, this yields a good approximation, except within a small range of $\alpha$ and in the vicinity of $\beta_{\rm in}\simeq \pi/2$.
\vskip 4pt

Finally, the full transition structure of $\ket{433}$ contains a richer hierarchy of $g$-bands, spanning $l=5$ down to $l=2$–mediated transitions. Chronologically, very weak and largely ineffective $l=5$ transitions occur first, followed by the $g=3$ band of $l=3$ transitions on which we concentrate. We apply the same criteria as above to identify the dominant resonance. In a few cases with $\beta_{\rm in}\gtrsim 3\pi/4$, where these $g=3$ transitions are insufficiently adiabatic to deplete the cloud, we continue the evolution to the next strongest $g=1$ band, see~Fig.~\ref{fig:433_fine} (right).

\newpage


\newpage
\bibliography{references.bib}
\bibliographystyle{apsrev4-1}
\end{document}